\documentclass[11pt]{article}

\usepackage{graphicx,graphics}
\usepackage{mathtools}
\usepackage{amsfonts,amssymb,amsmath,mathbbol}
\usepackage{amsthm,amsbsy}
\usepackage{stmaryrd}
\usepackage{stackrel}
\usepackage{fixmath}
\usepackage{mathabx}
\usepackage{algorithm}
\usepackage{algorithmicx}
\usepackage{algpseudocode}
\usepackage[square, numbers, comma, sort&compress]{natbib}
\usepackage[margin=0.75in]{geometry}
\usepackage{url}
\newcommand{\pkg}[1]
{\texttt{#1}}

\newcommand{\Lower}[1]{\smash{\lower 1.5ex \hbox{#1}}}

\def\spacingset#1{\renewcommand{\baselinestretch}{#1}}

\makeatletter
    \newcommand{\thefigurename}{Figure}
    \def\fnum@figure{\footnotesize{\thefigurename\ \thefigure}}
\makeatother

\makeatletter
    \newcommand{\thetablename}{Table}
    \def\fnum@table{\footnotesize{\thetablename\ \thetable}}
\makeatother

\begin{document}

%============== Title ====================================================================
\title{\textbf{Cross-validation and Peeling Strategies \\for Survival Bump Hunting \\using Recursive Peeling Methods}}

%============== Authorship ===============================================================
\author{Jean-Eudes Dazard \thanks{Center for Proteomics and Bioinformatics, Case Western Reserve University. Cleveland, OH 44106, USA. Corresponding author Email (JED): jxd101@case.edu} \and  Michael Choe \thanks{Center for Proteomics and Bioinformatics, Case Western Reserve University. Cleveland, OH 44106, USA.} \and  Michael LeBlanc \thanks{Department of Biostatistics, School of Public Health, University of Washington, Seattle, WA 98195, USA; Public Health Sciences, Fred Hutchinson Cancer Research Center, Seattle, WA 98109.} \and J. Sunil Rao \thanks{Division of Biostatistics, Dept. of Epidemiology and Public Health, The University of Miami. Miami, FL 33136, USA.}}

%============== Date ==================================================================
\date{\today}
%\date{}

%============== With page numbers ==========================================================
\pagestyle{plain}

\maketitle

\begin{abstract}
  We introduce a framework to build a survival/risk bump hunting model with a censored time-to-event response. Our Survival Bump Hunting (SBH) method is based on a recursive peeling procedure that uses a specific survival peeling criterion derived from non/semi-parametric statistics such as the hazards-ratio, the log-rank test or the Nelson--Aalen estimator. To optimize the tuning parameter of the model and validate it, we introduce an objective function based on survival or prediction-error statistics, such as the log-rank test and the concordance error rate. We also describe two alternative cross-validation techniques adapted to the joint task of decision-rule making by recursive peeling and survival estimation. Numerical analyses show the importance of replicated cross-validation and the differences between criteria and techniques in both low and high-dimensional settings. Although several non-parametric survival models exist, none addresses the problem of directly identifying local extrema. We show how SBH efficiently estimates extreme survival/risk subgroups unlike other models. This provides an insight into the behavior of commonly used models and suggests alternatives to be adopted in practice. Finally, our SBH framework was applied to a clinical dataset. In it, we identified subsets of patients characterized by clinical and demographic covariates with a distinct extreme survival outcome, for which tailored medical interventions could be made. An R package \pkg{PRIMsrc} is available on CRAN and GitHub.\\

\end{abstract}

\newpage

%=========================================================================================
\section{Introduction}\label{intro}
%=========================================================================================
\vspace{-0.1in}
\subsubsection*{Non-Parametric Methods for Bump Hunting}
\vspace{-0.1in}
The search for data structures in the form of bumps, modes, components, clusters or classes are important as they often reveal underlying phenomena leading to scientific discoveries. It is a difficult and central problem, applicable to virtually all sort of exact and social sciences with practical applications in various fields such as finance, marketing, physics, astronomy and biology.

It is common to treat the task of finding isolated data structures with the help of a response function as in a regression or classification problem or simply a probabilistic model as in a density estimation problem. Among the non-parametric \emph{unsupervised} methods, this can be done by testing modality \cite{Hartigan_1992, Rozal_1994, Polonik_1995, Burman_2009}, using nonparametric mixture models (see e.g. \cite{Bohning_2003} for a review), pattern recognition or clustering. However, beside the limitations or problems encountered by these methods in higher dimensional settings, model fitting e.g. of finite mixture models is challenged by the estimation of the \emph{true} number of components \cite{Dazard_2010a}. A similar situation exists for clustering procedures where the \emph{true} number of clusters is unknown. Moreover, unsupervised methods may also fail to capture true data structures simply by ignoring a response if available \cite{Dazard_2010a}. Although non-parametric \emph{supervised} approaches such as, for instance, decision trees and their ensemble versions \cite{Breiman_1984, Breiman_2001}, do not have this drawback, these classification and regression procedures may also perform poorly \cite{Dazard_2010a} since they are designed to work when the true number of classes is fixed or assumed in advance.

Exploratory supervised bump hunting procedures are among the few non-parametric methods that have been proposed to address this problem. These methods seek bump supports (possibly disjoint) of the input space of multi variables where a target function (e.g. a regression or density function) is on average larger (or lower) than its average over the entire input space. They cover tasks such as: (i) Mode(s) Hunting, (ii) Local/Global Extremum(a) Finding, (iii) Subgroup(s) Identification, (iv) Outlier(s) Detection. One known as the Patient Rule Induction Method (PRIM) was initially introduced by Friedman \& Fisher \cite{Friedman_1999} and later formalized by Polonik \cite{Polonik_2010}. Essentially, the method is a recursive peeling algorithm that explores the input space to find rectangular solution regions where the response is expected to be larger on average. Some interesting features common and distinct to decision trees such as Classification and Regression Trees (CART) \cite{Breiman_1984} help describe PRIM. As a rule-induction method like CART, PRIM generates simple decision rules describing the solution region of interest. Further, like CART, PRIM is a non-parametric procedure, algorithmic in nature (backwards fitting recursive algorithm), which makes few statistical assumptions about the data. Although PRIM does not explicitly state a model as CART, one can be formulated \cite{Wu_2003, Hastie_2009}. Both algorithms/models have the possibility to recover complex interactions between input variables. Basic difference between the two methods lies in their approach and goal (reviewed in section \ref{peeling_rule}).

To date, only a few extensions of the original PRIM work have been done: This includes a Bayesian model-assisted formulation of PRIM \cite{Wu_2003}, a boosted version of PRIM based on Adaboost \cite{Wang_2004}, an extension of PRIM to censored responses \cite{LeBlanc_2002, LeBlanc_2005} and to discrete variables \cite{Il-Gyo_2008}. Although PRIM is intrinsically multivariate, it was uncertain from the original work how the algorithm would perform in ultra high-dimension where collinearity \cite{Friedman_1999, Fan_2008} and sparsity abound. So, recently, an interesting body of work studied when and why the Principal Component space can be used effectively to optimize the response-predictor relationship in bump hunting. This was first addressed in \cite{Dazard_2010a}, where the computational details of such an approach were laid out for high-dimensional settings. Further, focusing on the properties of PRIM, authors demonstrated using basic geometrical arguments how the PC rotation of the predictor space alone can generate ``improved'' bump estimates \cite{Diaz_2015a, Diaz_2015b}. These developments have important implications for general supervised learning theory and practical applications.  In fact, \cite{Dazard_2010a} first used a sparse PC rotation for improving bump hunting in the context of high dimensional genomic predictors and later showed how this approach can be used to find additional heterogeneity in terms of survival outcomes for colon cancer patients (\cite{Dazard_2012a}).

\vspace{-0.1in}
\subsubsection*{Model Development and Validation in Discovery-Based Research}
\vspace{-0.1in}
The primary problem encountered in discovery-based research has been non-reproducible results. For instance, early biomarker discovery studies using modern high-throughput datasets with large number of features have often been characterized by false or exaggerated claims and eventually disappointment when original results could not be reproduced in an independent study \cite{Simon_2003, Michiels_2005, Dobbin_2005, Ein-Dor_2005, Shi_2006, Dupuy_2007, Subramanian_2010, Leek_2010, Haibe-Kains_2013}. Sadly, these results have been published even in high-profile journals and considered to provide definitive conclusions for both clinical care and biology. The problems of model reliability and reproducibility have usually been characterized by issues of severe model over-fitting, biased model parameter estimates and under-estimated errors. This has been attributed to a lack of proper rules to assess the analytical validity of studies simply because they were either under-developed or not routinely/correctly applied \cite{Ntzani_2003, Ransohoff_2004}. This problem first received the attention of statisticians (see for instance reviews on guidelines and checklists \cite{Baker_2002, Dupuy_2007, McShane_2013b}) as well as editors and US regulators lately \cite{McShane_2013a}.

Meanwhile, considerable development work has been done in the fields of feature selection, predictive model building and model validation to resolve the aforementioned issues. Recent developments include strategies such as variable/feature selection, dimension reduction, coefficient shrinkage and regularization. The challenge is obviously more acute in the context of high-dimensional data where the number of variables greatly exceeds the number of observations (so-called $p \gg n$ paradigm), since usually only a small number of variables truly enter in the model, while the large majority of them just contribute to noise. This noisy situation is even more complicated by the multicollinearity and spurious correlation between variables as well as the endogeneity between variables and model residual errors (see e.g. \cite{Fan_2014} for a recent review).

A common situation where model reliability and reproducibility arise is when, for instance, model performance estimates are calculated from the same data that was used for model building, eventually resulting in initially promising results, but often non-reproducible \cite{Ambroise_2002, Simon_2003, Hastie_2009}. These so-called ``resubstitution estimates'' are severely (optimistically) biased. Another problematic situation is when not all the steps of model building (such as pre-selection, creation of the prediction rule and parameter tuning) are internal to the cross-validation procedure, thereby creating a selection bias \cite{Ambroise_2002, Varma_2006, Hastie_2009}. In addition, findings might not be reproducible even when proper independent sample and validation procedures are used. Problems may arise simply because cross-validated estimates are well-known to have large variance, a situation that is obviously more prevalent when few independent observations or small sample size $n$ are used \cite{Efron_1983, Markatou_2005, Dobbin_2007}.

\vspace{-0.1in}
\subsubsection*{Predictive Survival/Risk Modeling by Rule-Induction Methods}
\vspace{-0.1in}
One important application of survival/risk modeling is to identify and segregate samples for predictive diagnostic and/or prognosis purposes. Direct applications include the stratification of patients by diagnostic and/or prognostic groups and/or responsiveness to treatment. Therefore, survival modeling is usually performed to predict/classify patients into risk or responder groups (not to predict exact survival time) from which one usually derives survival/risk functions estimates (e.g. by Kaplan--Meier estimates). However, for the reasons mentioned above, Kaplan--Meier estimates for the risk groups computed on the same set of data used to develop the survival model may be very biased \cite{Varma_2006, Molinaro_2005}.

In the context of a time-to-event outcome, regression survival trees have proven to be useful.  Several developments have been made for fitting decision trees to non-informative censored survival times \cite{Gordon_1985, Ciampi_1986, Segal_1988, Davis_1989, LeBlanc_1992, LeBlanc_1993, Ahn_1994}. Although regression survival trees are powerful techniques to understand for instance patient outcome and for forming multiple prognostic groups, often times interest focuses only on estimating \emph{extreme} survival/risk groups. In this respect, survival bump hunting aims not at estimating the survival/risk probability function over the entire variable space, but at searching regions where this probability is larger (or smaller) than its average over the entire space.

Also, one possible drawback of decision trees is that the data splits at an exponential rate as the space undergo partitioning (typically by binary splits) as opposed to a more patient rate in decision boxes (typically by controlled quantile). In this sense, bump hunting by recursive peeling may be a more efficient way of learning from the data. With the exception of the work of LeBlanc {\it et al.} on Adaptive Risk Group Refinement \cite{LeBlanc_2005}, it has not been studied whether decision boxes, obtained from box-structured recursive peelings, would yield better estimates for constructing prognostic groups than their tree-structured counterparts.

Although resampling methods are often useful in assessing the prediction accuracy of classifier models, they are not directly applicable for predictive survival modeling applications. Simon et {\it al.} have reviewed the literature of such applications and identified serious deficiencies in the validation of survival risk models \cite{Dupuy_2007, Subramanian_2010, Simon_2011}. They noted for instance that in order to utilize the cross-validation approach developed for classification problems, some studies have dichotomized their survival or disease-free survival data \textellipsis. The problem on how to cross-validate the estimation of survival distributions (e.g. by Kaplan--Meier curves) is not obvious \cite{Simon_2011}. In addition, beside Subramanian and Simon's initial study on the usefulness of resampling methods for assessing survival prediction models in high-dimensional data \cite{Subramanian_2011}, no comparative study has been done for rule-induction methods and specifically recursive peeling methods such as our ``Patient Recursive Survival Peeling'' method (see section \ref{survestimation}).

\vspace{-0.1in}
\subsubsection*{Contribution and Scope}
\vspace{-0.1in}
Our survival/risk bump hunting model is built upon the regular bump hunting framework, which we extended to accommodate a possibly censored time-to-event type of response. To build our survival/risk bump hunting model, we first describe our ``Patient Recursive Survival Peeling'' (PRSP) method, a non-parametric recursive peeling procedure, derived from a rule-induction method, namely the Patient Rule Induction Method (PRIM), which we have extended to allow for survival/risk response, possibly censored. In the process, we describe what appropriate survival estimator and statistic may be used as a peeling criteria to fit our survival/risk bump hunting model.

One of the critiques made in the original PRIM work was the lack of validation procedure and measures of significance of solution regions. So, our objective was also to develop a validation procedure for the purpose of model tuning by means of an optimization criterion of model parameters tuning and a resampling technique amenable to the joint task of decision rule making by recursive peeling (i.e. decision-box) and survival estimation. Specifically, we describe here two alternative, possibly repeated, $K$-fold cross-validation techniques adapted to the task, namely the ``Replicated Combined CV'' (RCCV) and ``Replicated Averaged CV'' (RACV). Moreover, we show how to use survival end-points/prediction statistics for the specific goal of model peeling length optimization by cross-validation.

Results support the claim that optimal survival bump hunting models may be reached using appropriate combination of criterion and technique under certain situations, for which we provide guidelines. Finally, we show empirical results from a real dataset application and from simulated data in low- and high-dimension, illustrating the efficiency of our cross-validation and peeling strategies and the adequacy of our survival bump hunting framework in comparison to other available non-parametric survival models.

We do not describe nor discus the specific treatment of dimension-reduction or variable selection in high-dimensional settings for the only reason that the focus of this study is on cross-validation and peeling strategies. Even though the issue of model unreliability is known to be more severe when there is a large number of variables to choose from \cite{Simon_2003}, it is known to persist even in low-dimensional setting \cite{Subramanian_2013}. So, we posit that the framework described here is relevant and applicable to both low and high-dimensional situations. Nevertheless, the method does include cross-validation procedures to control model size (\# covariates) in addition to model complexity (\# peeling steps). It has been tested in multiple ($> 20$) low and high-dimensional situations where $n \le p$ and even $n \ll p$ (see abstract of application article \cite{Dazard_2016a} and our example datasets in our R package \pkg{PRIMsrc} \cite{Dazard_2015a}) and we show empirical analyses in high-dimensional simulated datasets where $n < p$.

%=========================================================================================
\section{Survival Bump Hunting for Exploratory Survival Analysis} \label{sbh}
%=========================================================================================
%===============================================================
\subsection{Bump Hunting Model}
%===============================================================
\vspace{-0.10in}
\subsubsection{Notation - Goal} \label{goal}
\vspace{-0.10in}
The formal setup of bump hunting is as follows \cite[see also][]{Friedman_1999, Polonik_2010}. Let us consider a supervised problem with a univariate output (response) random variable, denoted ${\bf y} \in \mathbb R$. Further, let us consider a $p$-dimensional random vector ${\bf X} \in \mathbb R^p$ of support $S$, also called input space, in an Euclidean space. Let us denote the $p$ input variables by ${\bf X} = \left[{\bf x}_{j}\right]_{j=1}^{p}$, of joint probability density function $p({\bf X})$ and by $f({\bf x}) = E({\bf y} | {\bf X} = {\bf x})$ the target function to be optimized (e.g. any regression function or e.g. the p.m.f or p.d.f $f_{\bf X}({\bf x})$).

\begin{figure}[!hbt]
  \begin{minipage}{0.42\textwidth}
  Briefly, the goal in bump hunting is to find a sub-space or region $(R \subseteq S)$ of the input space within which the average value $\bar{f}_{R}$ of $f({\bf x})$ is expected to be significantly larger (or smaller) than its average value $\bar{f}_{S}$ over the entire input space $S$ (Figure \ref{Figure01}). In addition, one wishes that the corresponding support (mass) of $R$, say ${\beta_{R}}$, be not too small, that is, greater than a minimal support threshold, say $0 < \beta_{0} < 1$.
  \end{minipage}
  \hfill
  \begin{minipage}{0.55\textwidth}
  \centering\includegraphics[scale=1]{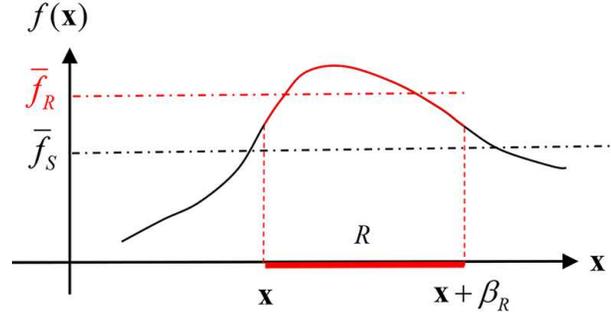}
  \vskip 0pt
  \caption{\sl \footnotesize Schematic representation of bump hunting in the unidimensional case ($p=1$), where the target function $f({\bf x})$ is a regression function of ${\bf x}$ and the estimated region $R$ is a contiguous interval (red segment) corresponding to larger values of $f({\bf x})$ on average. The support $\beta_{R}$ of $R$ and the average values $\bar{f}_{R}$ and $\bar{f}_{S}$ are shown.}
  \label{Figure01}
 \end{minipage}
\end{figure}

Formally, in the continuous case of ${\bf X}$:
\begin{align}
    &\bar{f}_{R} = \frac {\int_{{\bf x} \in R} f({\bf x}) p({\bf x})d{\bf x}} {\int_{{\bf x} \in R} p({\bf x})d{\bf x}} \gg \bar{f}_{S} \label{functionR}\\
    &\beta_{R} = \int_{{\bf x} \in R} p({\bf x})d{\bf x} \gg {\beta_{0}} \label{betaR}
\end{align}

In supervised problems with an output variable (response) ${\bf y}$, one would seek to characterize the conditional expectation $E({\bf y} | {\bf X} = {\bf x})$ and infer the properties of the unknown joint probability density function $p({\bf X})$, whereas in the case of unsupervised learning, one would have to directly infer the properties of $p({\bf X})$, e.g. from some density estimate, without the help of a response.

Let $S_{j}$ be the support of the $j$th variable ${\bf x}_{j}$, such that the input space can be written as the (Cartesian) outer product space $S = \bigtimes_{j=1}^{p} S_{j}$. Let $s_{j} \subseteq S_{j}$ denotes the unknown subset of values of variable ${\bf x}_{j}$ corresponding to the unknown support of the solution region $R$. Let $J \subseteq \{1, \ldots, p\}$ be the subset of indices of selected variables in the process. The goal in bump hunting amounts to finding the value-subsets $\{s_{j}\}_{j \in J}$ of the corresponding variables $\{{\bf x}_{j}\}_{j \in J}$ such that
\begin{equation}
    R = \left\{\bigcap_{j \in J}\left({\bf x}_{j} \in s_{j} \right) \colon (\bar{f}_{R} \gg \bar{f}_{S})(\beta_{R} \gg {\beta_{0}}) \right\}
\label{R}
\end{equation}

\vspace{-0.10in}
\subsubsection{Estimates} \label{estimates}
\vspace{-0.10in}
Since the underlying distribution is not known, the estimates of $\bar{f}_{R}$ and $\beta_{R}$ must be used. Assume a supervised setting, where the outcome response variable is ${\bf y} = (y_{1} \ldots y_{n})^{T}$ and the explanatory/input variables are ${\bf X} = ({\bf x}_{1} \ldots {\bf x}_{n})^{T}$, where each observation is the $p$-dimensional vector of covariates ${\bf x}_{i} = \left[x_{i,1} \ldots x_{i,p}\right]^{T}$, for $i \in \{1, \ldots, n\}$. Plug-in estimates of the average value $\bar{f}_{R}$ of the target function $f({\bf x})$ and of the support ${\beta}_{R}$ (eq. \ref{betaR}) of the region $R$ are respectively derived as:
\begin{align}
    &\hat{\bar{f}}_{R} = \frac{1}{n\hat{\beta}_{R}}\sum_{{\bf x}_{i}\in \hat{R}} y_{i} = \frac{1}{n\hat{\beta}_{R}}\sum_{i=1}^{n} y_{i} I({\bf x}_{i}\in \hat{R})\\
    &\hat{\beta}_{R} = \frac{1}{n}\sum_{{\bf x}_{i} \in \hat{R}} I({\bf x}_{i}\in \hat{R}) = \frac{1}{n}\sum_{i=1}^{n} I({\bf x}_{i}\in \hat{R})
\end{align}

\pagebreak
\vspace{-0.10in}
\subsubsection{Remarks}
\vspace{-0.10in}
\spacingset{0}
\begin{enumerate}
    \item The goal amounts to comparing the conditional expectation of the response over the solution region $R$: $\bar{f}_{R} = E[f({\bf x})|{\bf x} \in R]$ with the unconditional one $\bar{f}_{S} = E[f({\bf x})]$.
    \vspace{-0.1in}
    \item Larger target function average $\bar{f}_{R}$ is associated with smaller support $\beta_{R}$ of the region $R$ (Figure \ref{Figure01}). So, in practice, there is a trade-off between maximizing $\bar{f}_{R}$ and maximizing $\beta_{R}$.
    \vspace{-0.1in}
    \item If the target function to be optimized is for instance the p.m.f or p.d.f $f_{\bf X}({\bf x})$, then $\Pr({\bf x} \in R)$ is the probability mass/density of a local maximum and the task is equivalent to a mode(s) hunting.
    \vspace{-0.1in}
    \item In the case of real-valued inputs, the entire input space is the $p$-dimensional outer product space $S \subseteq \mathbb{R}^{p}$; the support $S_{j}$ of each individual input variable (and of each corresponding value-subset ${s}_{j}$) is the usual interval of the form $S_{j} = \left[t_{j}^{-}, t_{j}^{+}\right]\subset \mathbb{R}$ for $j \in J$; the solution region $R$ has the shape of a $|J|$-dimensional hyper-rectangle in $\mathbb{R}^{|J|}$, called a \emph{box}, denoted $B$, which can be written as the outer product of $|J|$ intervals of the form $B = \bigtimes_{j \in J} [t_{j}^{-},t_{j}^{+}]$.
    \vspace{-0.1in}
    \item In general, region $R$ could be any smooth shape (e.g. a convex hull) possibly disjoint. Describing or modeling such region would be difficult in high dimension and especially when the number of variables is larger than the number of observations ($p \gg n$  paradigm). In general, there is a trade-off between the goodness of fit and the interpretability of the inferences that we want to make. Here, we focus on interpretable models based on rectangular boxes in the input space of variables. Typically these rectangular boxes are aligned to the coordinate axes, but an immediate extension is to use linear combination rules of variables, i.e. a rotated space of input variables, such a the principal components space. We have showed that this strategy may provide a more favorable space to learn from the data (see for instance \cite{Dazard_2010a, Dazard_2012a, Diaz_2015a, Diaz_2015b}).
    \end{enumerate}
\spacingset{1}

\vspace{-0.10in}
\subsubsection{Estimation by the Patient Rule Induction Method (PRIM)} \label{estimation}
\vspace{-0.10in}
The Patient Rule Induction Method (PRIM) is used to get the region estimate $\hat{R}$ with corresponding support estimate $\hat{\beta}_{R}$ and conditional output response mean estimate $\hat{\bar{f}}_{R}$. Essentially, the method is one of recursive peeling/pasting algorithm (a discrete version of the steepest ascent method) that explores the input space solution region, where the response is expected to be larger on average. The method generates a sequence of boxes that collectively cover the region estimate $\hat{R}$. The way the space is covered and the box induction is done as well as how the patience and stopping rules are controlled is detailed in the original article of Friedman \& Fisher \cite{Friedman_1999}, later formalized by Polonik \& Wang \cite{Polonik_2010}.

{\it Covering - Coverage Stopping Rule.} A sequence of boxes $\{B_{m}\}_{m=1}^{M}$ is generated from the data $\{{\bf x}_{i}, y_{i} \}_{i=1}^{n}$ to collectively cover the solution region $R$. Starting from an initial box $B_{1}$ that covers all the data, the box sequence construction algorithm is recursively applied to subsets of the data as follows. At the $m$th iteration $(m > 1)$, a box $B_{m}$ is induced (by the top-down peeling algorithm - see next) using the data remaining after removal of all the observations contained in the previous boxes: $\{({y}_{i},{\bf x}_{i}) \colon {\bf x}_{i} \notin \bigcup_{r=1}^{m-1}B_{r}\}$. At the $M$th iteration of the covering loop, the box sequence $\{B_{m}\}_{m=1}^{M}$ stops either (i) when the estimated individual box support ${\hat{\beta}}_{M}$ becomes too small, say less than an arbitrary threshold $0 < \beta_{0} < 1$ expressed as a fraction of the entire data: ${\hat{\beta}}_{M} < \beta_{0}$, where, or (ii) when the estimated box output mean ${\hat{\bar{y}}}_{M}$ becomes too small, say ${\hat{\bar{y}}}_{M} < \bar{y}$ where $\bar{y} = \frac{1}{n} \sum_{i=1}^{n} y_{i}$ is the global mean, where
\begin{align*}
&{\hat{\beta}}_{M} = \frac{1}{n} \sum_{i=1}^{n} I\left( {\bf x}_{i} \in B_{M} \ \& \ {\bf x}_{i} \notin \bigcup_{m=1}^{M-1}B_{m} \right)\\
&{\hat{\bar{y}}}_{M} = \frac{1}{n\hat{\beta}_{M}} \sum_{i=1}^{n} {y}_{i} I\left( {\bf x}_{i} \in B_{M} \ \& \ {\bf x}_{i} \notin \bigcup_{m=1}^{M-1}B_{m} \right)
\end{align*}

{\it Box Induction.} To induce the box $B_{m}$ at the $m$th iteration $(m > 1)$, the top-down peeling algorithm generates a subsequence of nested boxes $\{B_{m,l}\}_{l=1}^{L}$ starting from an initial box $B_{m,1}$ that covers all the data remaining at the $m$th iteration of the covering loop. How $L$ is estimated is the subject of section \ref{cross}. At the $l$th iteration, a sub-box is peeled off (see next) from within the current box $B_{m,l}$ to produce the next smaller box $B_{m,l+1}$. The particular sub-box $b_{m,l}^{*}$ is chosen to yield the largest box output mean value ${\bar{y}}_{m,l+1}$ within the next box $B_{m,l+1}$, such that:
\begin{align*}
&{\bar{y}}_{m,l+1} = \frac{1}{n\hat{\beta}_{m,l+1}} \sum_{i=1}^{n} y_{i}I \left({\bf x}_{i} \in B_{m,l+1} \ \& \ {\bf x}_{i} \notin \bigcup_{b=1}^{l}B_{m,b}\right)\\
& B_{m,l+1} = B_{m,l} \setminus b_{m,l}^{*} \quad \text{, where} \\
& b_{m,l}^{*} = \text{argmax}_{b_{m,l} \in C(b_{m,l})} \left[{\bar{y}}_{m,l+1} \colon {\bf x}_{i} \in (B_{m,l} \setminus b_{m,l})\right]
\end{align*}
where $C(b_{m,l})$ represents the class of potential sub-boxes $b_{m,l}$ eligible for removal at step or generation $(m,l)$ and $`\setminus`$ represents the set minus operator. The current box $B_{m,l}$ is then updated: $B_{m,l+1} = B_{m,l+1} \setminus b_{m,l}^{*}$ and the peeling procedure is looped until some stopping rule is met (see next). Because a top-down peeling is a greedy search algorithm, it may cause overfitting, so a bottom-up pasting is applied to the minimal candidate box to repeatedly expand along any edge until the expansion fails to increase the output response average within the box.

{\it Patience - Induction Stopping Rule.} There are two important meta-parameters that control the box induction algorithm: (i) the peeling fraction $0 < \alpha_{0} < 1$ that controls the degree of patience, and (ii) the minimal box support threshold $0 < \beta_{0} < 1$, expressed as a fraction of the whole data that is used in the stopping criterion (see next). Only a quantile $\alpha_{0}$ of the data that is in the box $B_{m,l}$ is peeled off at the $l$th iteration of the peeling loop as follows. Each eligible sub-box $b_{m,l}$ is defined by a single input variable ${\bf x}_{j}$. For real valued variables, there are two eligible sub-boxes $b_{j,m,l}^{-} \in C(b_{m,l})$ and $b_{j,m,l}^{+} \in C(b_{m,l})$, which respectively border the lower and upper boundaries of the box $B_{m,l}$ on the $j$th input variable ${\bf x}_{j}$:
\begin{equation*}
\begin{cases}
b_{j,m,l}^{-} = \{ {\bf x}\colon{\bf x}_{j} < {\bf x}_{j}^{(\alpha_{0})} \} \\
b_{j,m,l}^{+} = \{ {\bf x}\colon{\bf x}_{j} > {\bf x}_{j}^{(1-\alpha_{0})} \}
\end{cases}
\end{equation*}

where ${\bf x}_{j}^{(\alpha_{0})}$ and ${\bf x}_{j}^{(1-\alpha_{0})}$ are respectively the $\alpha_{0}$th and $(1-\alpha_{0})$th quantiles of the ${\bf x}_{j}$ values. At the $L$th iteration of the peeling loop, the box sequence $\{B_{m,l}\}_{l=1}^{L}$ stops when the estimated individual support ${\hat{\beta}}_{m,L}$ of the last box $B_{m,L}$ becomes too small, say ${\hat{\beta}}_{m,L} < \beta_{0}$, where $\beta_{0}$ is an arbitrary minimal box support threshold:
\begin{equation*}
\begin{cases}
{\hat{\beta}}_{m,1} = \frac{1}{n} \sum_{i=1}^{n} I({\bf x}_{i} \in B_{m,1}) & \text{for} \quad L = 1 \\
{\hat{\beta}}_{m,L} = \frac{1}{n} \sum_{i=1}^{n} I({\bf x}_{i} \in B_{m,L} \quad \& \quad {\bf x}_{i} \notin \bigcup_{l=1}^{L-1}B_{m,l}) & \text{for} \quad L > 1
\end{cases}
\end{equation*}
\vspace{0.1in}

Note that, with our notation, the last box $B_{m,L}$ of the subsequence is also the next box of the outer box sequence $\{B_{m}\}_{m=1}^{M}$. So, $B_{m,L} = B_{m+1}$, and similarly ${\hat{\bar{y}}}_{m,L} = {\hat{\bar{y}}}_{m+1}$ and ${\hat{\beta}}_{m,L} = {\hat{\beta}}_{m+1}$.

{\it Decision Rules.}
It is desirable that the solution region $R$ be described in an interpretable form by logical statements involving the value-subset of each selected input variable. The above algorithm results in simple decision rules of the input space, where each box $B_{m}$, $m = 1, \dots, M$, is described by the outer product of the value-subsets $s_{j,m}$ of each individual input variable ${\bf x}_{j}$, for $j \in J$. The idea is to describe the solution region $R$ by a disjunctive rule of $M$ conjunctive subrules of the form $\mathcal{R} = \bigcup_{m=1}^{M} \mathcal{R}_{m}$, where $\mathcal{R}_{m} = \left\{{\bf x} \in B_{m}\right\} = \bigcap_{j \in J}({\bf x}_{j} \in s_{j,m})$. In the case of real-valued input variables, each subrule becomes $\mathcal{R}_{m} = \bigcap_{j \in J}({\bf x}_{j} \in [t_{j,m}^{-},t_{j,m}^{+}])$ and the solution region $R$ is fully described by the disjunctive rule:
\begin{equation*}
\hat{\mathcal{R}} = \bigcup_{m=1}^{M} \hat{\mathcal{R}}_{m} = \bigcup_{m=1}^{M} \left\{\bigcap_{j \in J}({\bf x}_{j} \in [t_{j,m}^{-},t_{j,m}^{+}])\right\}
\end{equation*}

%===============================================================
\subsection{Survival Bump Hunting by Recursive Peeling}
%===============================================================
Assume a supervised problem, where the function of interest is a univariate survival/risk response variable (possibly censored) in a multivariate setting of real-valued (continuous or discrete) input variables/covariates ${\bf X}=\left[{\bf x}_{j}\right]_{j=1}^{p}$. The goal is to characterize an extreme-survival-response support in the predictor space and identify the corresponding box-defined group of samples using a recursive peeling method derived from the Patient Rule Induction Method (PRIM).

\vspace{-0.10in}
\subsubsection{Survival-Specific Peeling Rule} \label{peeling_rule}
\vspace{-0.10in}
As mentioned in the introduction, rule-induction methods such as decision tree-based methods have proven to be useful to estimate relative risk in groups in the context of a time-to-event outcome. Several methods have been proposed for fitting trees to non-informative censored survival times \cite{Gordon_1985, Ciampi_1986, Segal_1988, Davis_1989, LeBlanc_1992, LeBlanc_1993, Ahn_1994}.

Basic differences between decision-tree and decision-box methods lie in their approach and goal. Instead of recursively partitioning the space using specific partitioning and stopping criteria, one proceeds by recursively peeling the space to produce box-shaped regions designed to approximate the solution region $R$, using specific peeling and stopping criteria (see details in \ref{estimation}). In decision-trees, a recursive partitioning method attempts to model the target function over the entire data space by generating partitions in which the response averages will be as different as possible, while in decision-boxes, a recursive peeling method generates a box-shaped region in which the response average will be as extreme as possible. So, in contrast to survival decision-trees models, survival bump hunting is not aimed at estimating the survival/risk probability function over the entire covariate space, but at finding regions where this probability is larger than its average over the entire space. Numerical analysis below (\ref{simulation}) show comparisons of relative risk estimates obtained from decision-boxes versus those obtained from decision-trees. Other interesting differences lie in the weaknesses and strengths of the outputs and their applications, which we left for discussion (\ref{discussion}).

In this section, we describe the use of several candidate survival-specific peeling criteria and discuss their merits or strengths. Most of these criteria are borrowed from the survival splitting rules used to grow regression survival trees \cite{Segal_1988, Therneau_1990, LeBlanc_1992, LeBlanc_1993, Ahn_1994} or from their ensemble versions \cite{Ishwaran_2008}. Here, survival-specific peeling criteria are to be used to decide which covariate will be selected to give the best peel between two boxes from two consecutive generations (parent-child descendance) of the box induction/peeling loop in a recursive peeling algorithm (see next section \ref{survestimation}).

To account for censoring, we simply supervise by proxy for extreme time-to-event outcome, turning the censored outcome $y$ into an uncensored ``surrogate'' outcome $z$. Using previous notation (section \ref{estimation}), a peeling at step $(m,l)$ of the box induction/peeling sequence produces a partition of the survival data from the parent box $B_{m,l-1}$ into two partitions, for a given set of covariates: the child box $B_{m,l}$ and its complement. The focus is on selecting a sub-box $b_{m,l}$ at step $(m,l)$ of the box induction/peeling sequence that is to be peeled off from the parent box $B_{m,l-1}$ along one of its faces (i.e. direction of peeling := axis of dimension $j$) to induce the next child box $B_{m,l}$ and its complement. This is done by maximizing the ``surrogate'' outcome rate of increase between two consecutive generations of boxes $B_{m,l-1}$ and $B_{m,l}$ of the box induction/peeling sequence. Denote by $z(m,l)$ the box ``surrogate'' outcome at step or generation $(m,l)$ of the box induction/peeling sequence (Algorithm \ref{algo}). The rate of increase in $z(m,l)$ at step or generation $(m,l)$ between two consecutive generations of boxes $B_{m,l-1}$ and $B_{m,l}$ is defined as:
\begin{equation}
    r(m,l) = \frac{z(m,l) - z(m,l-1)}{\beta_{m,l-1} - \beta_{m,l}}
\label{surrogate}
\end{equation}

Finally, the particular sub-box $b_{m,l}^{*}$ that is chosen to yield the largest box increase rate $r(m,l)$ between box $B_{m,l-1}$ and the next one $B_{m,l}$ is such that
\begin{align}
    \nonumber
    &B_{m,l} = B_{m,l-1} \setminus b_{m,l}^{*} \quad \text{, where} \\
    &b_{m,l}^{*} = \stackrel[b_{m,l} \in C(b_{m,l})]{}{\text{argmax}} \left[r(m,l)\right],
\end{align}
where $C(b_{m,l})$ represents the class of potential sub-boxes $b_{m,l}$ eligible for removal at step or generation $(m,l)$.
\vspace{-0.10in}

\subsubsection{Survival Notation and Definitions} \label{notation}
\vspace{-0.10in}
Let's denote the two child boxes described above by $\{B_{g,m,l}\}_{g=1}^{2}$, where, by convention, let's decide that subscript $g=1$ stands for the ``in-box'' $B_{m,l}$ and $g=2$ for its complement or ``out-of-box''. Dropping further step subscripts $(m,l)$ for simplicity, assume that there are $n$ individuals in parent box $B_{m,l-1}$ and $n_{g}$ in a given child box $B_{g,m,l}$ for fixed $g \in \{1, 2\}$ such that $n=\sum_{g=1}^{2} n_{g}$. Also, we let $\gamma_{i}(g)=I\left({\bf x}_{i} \in B_{g,m,l}\right)$ be the indicator function of individual subject $i$ within a given child box $B_{g,m,l}$ at step $(m,l)$ for fixed $g \in \{1, 2\}$.

The response variable being subject to censoring, we use the general random censoring model. We focus on a univariate right-censored survival outcome under the assumptions of independent observations, non-competitive risks and random (type-I or -II) non-informative censoring. Denote the \emph{true} survival time (or lifetime/failure time) by the random variable $T$ and the \emph{observed} censoring time by the random variable $C$, then the \emph{observed} survival time is the random variable $Y = \min(T,C)$. Also, under our assumptions, $C$ is assumed to be independent of $T$ conditionally on covariates ${\bf X}$. Let the \emph{observed} event indicator random variable be $\mathit{\Delta} = I(T \le C)$.

For each observation $i \in \{1, \ldots, n\}$ in parent box $B_{m,l-1}$, the true survival time, observed censoring time, observed survival time and observed indicator event are the realizations denoted by $T_{i}$, $C_{i}$, $Y_{i} = \min(T_{i},C_{i})$ and $\delta_{i} = I(T_{i} \le C_{i})$, respectively. Also, denote by $t_{(1)} < t_{(2)} < \ldots < t_{(N)}$ for $N \le n$ the \emph{distinct ordered} event times of death (not counting censoring times) in parent box $B_{m,l-1}$. Note that intervals between events $t_{(h)}$ for $h \in \{1, \ldots, N\}$ are not necessarily uniform. Finally, the observed data in parent box $B_{m,l-1}$ consists of $\left(Y_{i}, \delta_{i}, {\bf x}_{i}\right)$, where ${\bf x}_{i} = \left[x_{i,1} \ldots x_{i,p}\right]^{T}$, for $i \in \{1, \ldots, n\}$.

Let $\delta_{i,g} = \gamma_{i}(g)I(T_{i} \le C_{i})$ be the observed indicator event of time point $T_{i}$ for each individual $i \in \{1, \ldots, n_g\}$ in a given child box $B_{g,m,l}$ for $g \in \{1, 2\}$. Also, let $d_{h,g}$ and $n_{h,g}$ be respectively the number of events (deaths) and individuals at risk at time $t_{(h)}$ for $h \in \{1, \ldots, N_g\}$ in a given child box $B_{g,m,l}$ for $g \in \{1, 2\}$, such that $N=\sum_{g=1}^{2} N_{g}$. For simplicity, let's use the same subscript $h \in \{1, \ldots, N_g\}$ and $i \in \{1, \ldots, n_g\}$ from parent and child boxes for indexing events and individuals, respectively. Note that $n_{h,g}$ is the number of individuals in child box $B_{g,m,l}$ who either have not yet had an event (or been right-censored) just until time $t_{(h)}$ or who had an event at time $t_{(h)}$. Formally, if considering all individuals in child box $B_{g,m,l}$ only, $n_{h,g} = \sum_{i=1}^{n_g} I\left(Y_{i} \ge t_{(h)}\right)$ or, if considering all individuals in parent box $B_{g,m,l-1}$, $n_{h,g} = \sum_{i=1}^{n} \gamma_{i}(g)I\left(Y_{i} \ge t_{(h)}\right)$. Likewise, one can write $d_{h,g} = \sum_{i=1}^{n_g} \delta_{i,g}I\left(Y_{i} \ge t_{(h)}\right)$, or $d_{h,g} = \sum_{i=1}^{n} \delta_{i}\gamma_{i}(g)I\left(Y_{i} \ge t_{(h)}\right)$. Also, denote $d_{h}=\sum_{g=1}^{2} d_{h,g}$ and $n_{h}=\sum_{g=1}^{2} n_{h,g}$.

Let $S(t)=\Pr(T \ge t)$ be the survival probability that an individual from the population of interest will have a lifetime $T$ free of the event until time $t$. As usual, denote by $\Lambda(t)=-\log(S(t))$ the corresponding cumulative hazard function and by $\lambda(t)=\frac{d\Lambda(t)}{dt}$ the hazard rate function. To come up with decision-box survival-specific peeling criteria (see next section \ref{criterion}), the following non/semi-parametric estimators can be used with respect to the box-defined subgroups: the Nelson--Aalen estimator, denoted by $\hat{H}_{g,m,l}(t)$, to estimate the cumulative hazard function; and the hazard rate function estimator derived from a Cox Proportional Hazards (CPH) regression model. By definition, these estimators are given as follows for individuals in a given child box $B_{g,m,l}$, for fixed $g \in \{1, 2\}$, at step $(m,l)$:
\begin{align}
	&\hat{H}_{g,m,l}(t)  = \sum_{h:t_{(h)} \le t} \frac{d_{h,g}}{n_{h,g}}            \qquad \nonumber
\end{align}

As usual, the hazard rate function may be estimated by regressing the subject-specific hazard rate on the covariates in a CPH regression model, assuming proportional hazards \cite{Cox_1972}. With the above notation,
\begin{align}
    \hat{\lambda}_{i,g,m,l}(t|{\bf x}_{i}) &= \lambda_{0}(t)\exp\left[\eta_{g,m,l}({\bf x}_{i})\right] \nonumber\\
                                           &= \lambda_{0}(t)\exp\left[\eta_{g,m,l}\gamma_{i}(g)\right] \nonumber
\end{align}
where the regression function $\eta_{g,m,l}({\bf x}_{i}) = \boldsymbol{\eta}_{g,m,l}^T {\bf x}_{i} = \sum_{j=1}^{p}\eta_{j,g,m,l}x_{i,j}$ with $p$-dimensional vectors of regression coefficients $\boldsymbol{\eta}_{g,m,l} = \left[\eta_{1,g,m,l} \ldots \eta_{p,g,m,l}\right]^{T}$ and covariate ${\bf x}_{i} = \left[x_{i,1} \ldots x_{i,p}\right]^{T}$ reduce respectively to a scalar $\boldsymbol{\eta}_{g,m,l} = \eta_{1,g,m,l} = \eta_{g,m,l}$ times a simple box indicator variable ${\bf x}_{i} = x_{i,1} = I\left({\bf x}_{i} \in B_{g,m,l}\right)= \gamma_{i}(g)$.

\vspace{-0.10in}
\subsubsection{Non-Parametric Survival Peeling Criteria} \label{criterion}
\vspace{-0.10in}
The choice of \emph{uncensored} surrogate outcome $z(m,l)$ in equation \ref{surrogate}, that is, which estimator to choose as a box peeling criterion at a peeling step $(m,l)$, is central to the PRSP algorithm (see Algorithm \ref{algo}). Currently, our Survival Bump Hunting implementation in our R package \pkg{PRIMsrc} \cite{Dazard_2015a} offers three statistics derived from the above non/semi-parametric estimators: (i) the Log-Rank Test statistic, (ii) the Nelson--Aalen Summary statistic and (iii) the CPH-derived Log Hazard Ratio statistic (assuming proportional hazards).

\spacingset{0}
\begin{itemize}
    \vspace{-0.1in}
    \item The (two-sample) log-rank test can be used at a peeling step $(m,l)$ to compare estimates of the hazard functions from each child box-defined subgroups (``in-box'' and its ``out-of-box'' complement). We recently proposed to use it as a survival-specific box peeling criterion \cite{Dazard_2014}. Using the (two-sample) log-rank test statistic is actually a natural candidate for survival decision-box, having been a well-established concept for splitting trees in survival decision-trees \cite{Segal_1988, Therneau_1990, LeBlanc_1992, LeBlanc_1993, Ahn_1994} and for being robust in non-proportional hazard settings \cite{LeBlanc_1993}. The approximate log-rank test introduced by LeBlanc and Crowley can be used instead to greatly reduce computations \cite{LeBlanc_1993}. Formally, one can derive the Log-Rank Test (LRT) statistic, denoted $\hat{\chi}_{LRT}(m,l)$, for the individuals in a given child box $B_{g,m,l}$, for fixed $g \in \{1, 2\}$ (e.g. $g=1$), at step $(m,l)$ as follows:
        \begin{align}
            \hat{\chi}_{LRT}(m,l) = \frac{\sum_{h=1}^{N} \left(d_{h,1} - n_{h,1}\frac{d_{h}}{n_{h}}\right)}
                                         {\sqrt{\sum_{h=1}^{N} n_{h,1}\frac{d_{h}}{n_{h}} \left(1 - \frac{n_{h,1}}{n_{h}}\right) \left(\frac{n_{h}-d_{h}}{n_{h}-1}\right)}}
            \label{LRTcriterion}
        \end{align}
    \vspace{-0.1in}
    \item If the Nelson--Aalen estimator is used, one can derive an overall summary statistic across all observed time points $Y_{i}$ for $i \in \{1, \ldots, n_g\}$ of the individuals in a given child box $B_{g,m,l}$, for fixed $g \in \{1, 2\}$ (e.g. $g=1$), at step $(m,l)$. This is done by adding the Nelson--Aalen estimators over all these time points to obtain a so-called Cumulative Hazard Summary (CHS), denoted $\hat{\Lambda}_{CHS}(m,l)$:
        \begin{align}
            \hat{\Lambda}_{CHS}(m,l) &= \sum_{i=1}^{n_1}\hat{H}_{1,m,l}(Y_{i})
                                        \quad \nonumber\\
                                     &= \sum_{i=1}^{n_1}\left(\sum_{h:t_{h} \le Y_{i}} \frac{d_{h,1}}{n_{h,1}}\right)
                                        \quad \nonumber\\
                                     &= \sum_{i=1}^{n_1}\left(\sum_{h:t_{h} \le Y_{i}} \frac{\sum_{i=1}^{n_1}\delta_{i,1}I\left(Y_{i} \ge t_{(h)}\right)}{\sum_{i=1}^{n_1} I\left(Y_{i} \ge t_{(h)}\right)}\right)
                                        \quad \nonumber\\
                                     &= \sum_{i=1}^{n_1}\left(\frac{\sum_{i=1}^{n_1} \delta_{i,1}}{n_1}\right)
                                        \quad \nonumber\\
                                     &= \sum_{i=1}^{n_1}\delta_{i,1}
                                        \quad
            \label{CHScriterion}
        \end{align}
    \vspace{-0.1in}
    \item Alternatively, the use of a Hazard Ratio or Relative Risk was originally proposed by LeBlanc et al \cite{LeBlanc_2002, LeBlanc_2005}. If the Cox-PH hazard rate estimate is used, then one derives the Log-Hazard Ratio (LHR) statistic, denoted $\hat{\lambda}_{LHR}(m,l)$, for the individuals in both child boxes $B_{g,m,l}$, for $g \in \{1, 2\}$, at step $(m,l)$:
        \begin{align}
            \hat{\lambda}_{LHR}(m,l)
            &= \log \left\{\frac{\lambda_{0}(t)\exp\left[\eta_{1,m,l}\gamma_{i}(1)\right]}
                                {\lambda_{0}(t)\exp\left[\eta_{2,m,l}\gamma_{i}(2)\right]}\right\}
                          \qquad \quad \nonumber\\
            &= \log \left\{\frac{\exp(\eta_{1,m,l})}{\exp(0)}\right\}
                          \qquad \quad \nonumber\\
            &= \eta_{1,m,l}
                          \qquad \quad
            \label{LHRcriterion}
        \end{align}
        where $\gamma_{i}(1) = 1$ and $\gamma_{i}(2) = 0$ by convention.
\end{itemize}
\spacingset{1}

Finally, all the above three peeling criteria statistics can be used to maximize the differences in survival outcomes between two consecutive boxes $\hat{B}_{m,l-1}$ and $\hat{B}_{m,l}$ of the box induction/peeling sequence. This leads to the derivation of the corresponding box rate of increase estimate $\hat{r}(m,l)$, at step $(m,l)$, according to equation \ref{surrogate}:
\begin{align}
    \hat{r}_{LRT}(m,l) &= \frac{\hat{\chi}_{LRT}(m,l) - \hat{\chi}_{LRT}(m,l-1)} {\hat{\beta}_{m,l-1} - \hat{\beta}_{m,l}}\\
    \hat{r}_{CHS}(m,l) &= \frac{\hat{\Lambda}_{CHS}(m,l) - \hat{\Lambda}_{CHS}(m,l-1)} {\hat{\beta}_{m,l-1} - \hat{\beta}_{m,l}} = \frac{\sum_{i=1}^{n_1,m,l}\delta_{i,1,m,l} - \sum_{i=1}^{n_1,m,l}\delta_{i,1,m,l-1}}{\hat{\beta}_{m,l-1} - \hat{\beta}_{m,l}}\\
    \hat{r}_{LHR}(m,l) &= \frac{\hat{\lambda}_{LHR}(m,l) - \hat{\lambda}_{LHR}(m,l-1)} {\hat{\beta}_{m,l-1} - \hat{\beta}_{m,l}} = \frac{\eta_{1,m,l} - \eta_{1,m,l-1}}{\hat{\beta}_{m,l-1} - \hat{\beta}_{m,l}}
\end{align}

\vspace{-0.10in}
\subsubsection{Comments} \label{comments}
\vspace{-0.10in}
An alternative estimator is to consider the conditional probability $P_{g,m,l}(t | {\bf x}_{i}) = \text{Pr}(T_{i} \le t | {\bf x}_{i} \in B_{g,m,l})$, which amounts to computing the Nelson-Aalen estimator $\hat{H}_{g,m,l}(t)$ conditioning on the data in a given child box $B_{g,m,l}$. Although this probability is interpretable and estimable, it is by definition a function of an observed event (death) at time $t$ (in a given child box $B_{g,m,l}$). So, one would need to fix a meaningful survival time $t$. One could think, for instance, of the box median survival time (as is commonly done) or the box maximal event time. In addition, this would induce a likely loss of ``power'' in contrast to an estimator based on the global survival distribution. As a result, for a given choice of $t$, this probability may be easy to estimate in some boxes but not estimable in other boxes after sufficient peeling.

Note that the Nelson--Aalen estimator is known to imply conservation of events \cite{Naftel_1985}, that is in this case, the total number of deaths is conserved in each child box $B_{g,m,l}$. In fact, that is what the Cumulative Hazard Summary (CHS) statistic amounts to (see eq: \ref{CHScriterion}).

A modified summary statistic derived from the Nelson--Aalen is also possible by normalizing $\hat{\Lambda}_{CHS}(m,l)$ to the total box sample size $n_g$. This would have the advantage of hedging against large versus small box bias. In our experience, this could be important in the case of discrete covariates.

In addition to the above assumption on the censoring mechanism, the CPH-derived Log-Hazard Ratio or Relative Risk statistic to be used in equation \ref{surrogate} assumes proportional hazards, which may not be realistic. For this reason, this survival peeling criterion is referred to the reader as not preferred and left as a means of comparison to potentially better alternative survival peeling criteria described above (section \ref{criterion}).

\vspace{-0.10in}
\subsubsection{Alternative Survival Peeling Criteria} \label{alternative}
\vspace{-0.10in}
Further discussion of the use of the above estimators is found in our comparisons of numerical results (section \ref{building}) and in the discussion (\ref{discussion}). Additionally, we mention below a few more alternative survival peeling criteria, although none of these is preferred nor implemented in our R package.

\spacingset{0}
\begin{enumerate}
    \vspace{-0.1in}
    \item It is common to estimate the hazard rate of the simplest parametric survival model (exponential survival model) by the parametric Maximum Likelihood Estimator (MLE). Using above notation and dropping further step subscripts $(m,l)$ for simplicity, if we let $T_{i} \sim \text{Exp}(\lambda)$ for $i \in \{1, \ldots, n_g\}$ then the parametric MLE is: $\hat{\lambda}_{g}(t)=\frac{\sum_{i=1}^{n_g}\delta_{i}\gamma_{i}(g)}{\sum_{i=1}^{n_g}Y_{i}\gamma_{i}(g)}$. The use of this simple parametric estimator of hazard rate for a box was originally proposed by LeBlanc et al. \cite{LeBlanc_2002, LeBlanc_2005}. Since it is always estimable, it could be used to maximize the box rate of increase $r(m,l)$ (eq. \ref{surrogate}) at each step $(m,l)$. It also does come with likely the least variance. However, the underlying assumption of constant hazard rate over the duration of time makes it potentially unrealistic and therefore not preferred.
    \vspace{-0.1in}
    \item The Log-Rank Score Test statistic for splitting in trees (see \cite{Hothorn_2003}) or their ensemble versions \cite{Ishwaran_2008} is another potential criterion available. Note that if there are no tied event times, the Log-Rank Test and the Log-Rank Score Test statistics are identical and, unless there are a large number of tied times, will give very similar results.
\end{enumerate}
\spacingset{1}

Although some may argue that residuals are counter intuitive to use as a peeling criterion, others have used them. For instance, the use of Martingale residuals is strongly recommended by Kehl et al. \cite{Kehl_2006}. Their claim is that they perform better than the deviance residuals, which are a transformation of the Martingale residuals correcting for long tails of the residual distribution. However, others have found that the deviance residuals lead to better rule induction results for bump hunting (Steve Horvath et al.'s personal communication and \cite{Liu_2004}). Therneau et al \cite{Therneau_1990} have also found that using the deviance residuals as a splitting criterion in regression trees leads to better results than the Martingale residuals.

\spacingset{0}
\begin{enumerate}
    \vspace{-0.1in}
    \item Martingale Residuals $M_{i} = \delta_{i} - \hat{\Lambda}(Y_{i},{\bf x}_{i}) = \delta_{i} - \hat{\Lambda}_{0}(Y_{i})\exp(\boldsymbol{\eta}^T {\bf x}_{i})$, for the $i$th observation, result from fitting an intercept-only Cox regression to the censored survival times. The idea is to use these as new (uncensored) outcomes in the model instead of time \cite{Therneau_1990}, where $\delta_{i}$ is the event indicator and $\hat{\Lambda}_{0}(Y_{i})$ is a non-parametric estimate of the baseline cumulative hazard function for the entire sample.
    \vspace{-0.2in}
    \item Deviance Residuals $D_{i}=\text{sign}(\hat{M}_{i})\sqrt{2\left[\delta_{i}\log\left( \frac{\delta_{i}}{\hat{\Lambda}_{0}(Y_{i})}\right) - \hat{M}_{i}\right]}$, for the $i$th observation, have a less symmetric distribution than Martingale residuals \cite{Therneau_1990}. LeBlanc and Crowley \cite{LeBlanc_1992} also demonstrated that (i) using deviance residuals in regression trees is similar to the survival tree methods presented by Segal \cite{Segal_1988} and Ciampi et al. \cite{Ciampi_1986}, and that (ii) using deviance residuals is more efficient than using Martingale residuals with regression trees.
\end{enumerate}
\spacingset{1}

\vspace{-0.10in}
\subsubsection{Box End-Point Statistics} \label{end-points}
\begin{figure}[!hbt]
  \centering\includegraphics[width=6in]{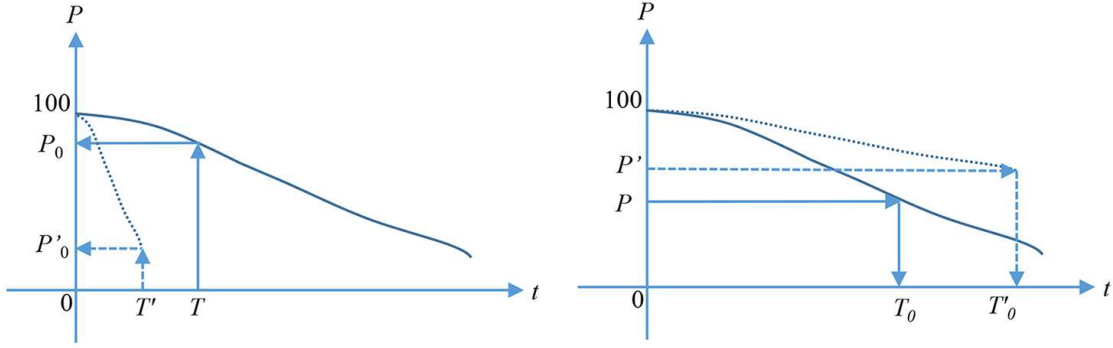}
  \caption{\sl \footnotesize Survival end points statistics used in Survival Bump Hunting at each step $(m,l)$ of the box generation. Left: Event-Free Probability $P_{0}$ and Minimal Event-Free Probability (\emph{MEFP}) $P_{0}^{\prime}$. Right: Event-Free Time $T_{0}$ and Maximal Event-Free Time (\emph{MEFT}) $T_{0}^{\prime}$. Subscripts $(m,l)$ are dropped for simplification but understood.}
  \label{Figure02}
  \vskip -10pt
\end{figure}

\vspace{-0.10in}
Below is a summary of box end-point statistics of interest one can derive in our Survival Bump Hunting method. Each is defined for each step $(m,l)$ and all are implemented in our R package \pkg{PRIMsrc} \cite{Dazard_2015a}:

\spacingset{0}
\begin{enumerate}
    \vspace{-0.1in}
    \item Log Hazards Ratios (\emph{LHR}), denoted $\lambda(m,l)$ between the highest-risk group/box and lower-risk groups/boxes of the same generation.
    \vspace{-0.1in}
    \item Log-Rank Test statistic (\emph{LRT}), denoted $\chi(m,l)$ between the highest-risk group/box and lower-risk groups/boxes of the same generation.
    \vspace{-0.1in}
    \item Concordance Error Rate (\emph{CER}), denoted $\theta(m,l)$ in the highest-risk group/box, that is a prediction performance metric taking censoring into account. For each step $(m,l)$, $\theta(m,l) = 1 - C(m,l)$, where $C$ is Harrel\textquoteright s Concordance Index for censored data \cite{Harrell_1982}, a rank correlation U-statistic, to estimate the probability of concordance between predicted and observed survival times.
    \vspace{-0.1in}
    \item Event-Free Probability (EFP), denoted $P_{0}(m,l)$ or probability of non-event until a certain time $T(m,l)$ in the highest-risk group/box (Figure \ref{Figure02} left). For instance, the Probability of Event-Free Survival (PEFS) or the Survival Rate (SR) are frequently used. However, $P_{0}(m,l)$ may not always be reached for a specified time $T(m,l)$. In this case, we determine the limit end-point $P_{0}^{\prime}(m,l)$ or Minimal Event-Free Probability (\emph{MEFP}) and corresponding maximal time $T^{\prime}(m,l)$, which are always observable (see Figure \ref{Figure02} left).
    \vspace{-0.1in}
    \item Event-Free Time (EFT), denoted $T_{0}(m,l)$ or time to reach a certain end-point probability $P(m,l)$ in the highest-risk group/box (Figure \ref{Figure02} right). For instance, the Median Survival (MS) is frequently used to indicate the period of time where 50\% of subjects have reached survival. However, $T_{0}(m,l)$ may not always be reached for a certain probability $P(m,l)$. In this case, we determine the limit end-point $T_{0}^{\prime}(m,l)$ or Maximal Event-Free Time (\emph{MEFT}) and corresponding minimal probability $P^{\prime}(m,l)$, which are always observable (see Figure \ref{Figure02} right).
    \vspace{-0.1in}
    \item Box characteristics:
    \vspace{-0.1in}
        \begin{itemize}
            \item $2p$ box edges $\left[t_{j}^{-}(m,l), t_{j}^{+}(m,l)\right]_{j=1}^{p}$,
            \vspace{-0.1in}
            \item box support (mass) $\beta(m,l)$
            \item box membership indicator $\boldsymbol{\gamma}(m,l)$
            \vspace{-0.1in}
        \end{itemize}
    \vspace{-0.1in}
    \item Traces of Covariate Usage $VU(m,l)$ and Covariate Importance $VI(m,l)$
    \vspace{-0.1in}
    \item Kaplan--Meir curves of survival probability values with log-rank test permutation $p$-values $p(m,l)$
\end{enumerate}
\spacingset{1}

\vspace{-0.20in}
\subsubsection{Estimation by Patient Recursive Survival Peeling} \label{survestimation}
\vspace{-0.10in}
The strategy employed here is a recursive peeling algorithm for survival bump hunting. Our ``Patient Recursive Survival Peeling'' method proceeds similarly ato which it is done in PRIM except for the box induction peeling/pasting criteria and the induction stopping rule (see section \ref{estimation}):
\spacingset{0}
\begin{algorithm}[!ht]
    \caption{Patient Recursive Survival Peeling (annotated below w.l.o.g for a maximization problem).}
    \label{algo}
    \begin{itemize}
        \item Start with the training data $\mathcal{L}_{(1)}$ and a maximal box $\hat{B}_{1}$ containing it
        \vspace{-0.1in}
        \item For $m \in \{1, \dots, M\}$:
        \begin{algorithmic}[1]
            \State Generate a box $\hat{B}_{m}$ using the remaining training data $\mathcal{L}_{(m)}$
            \State For $l \in \{1, \dots, L\}$:
            \begin{itemize}
                \item Top-down peeling: Generate a box $\hat{B}_{m,l}$ by conducting a stepwise covariate selection/usage: shrink the box by compressing one face (peeling), so as to peel off a quantile $\alpha_{0}$ of observations of a covariate ${\bf x}_{j}$ for $j \in \{1,\ldots,p\}$. Choose the direction of peeling $j$ that yields the largest box increase rate $\hat{r}(m,l)$ of the statistic used as peeling criterion between box $\hat{B}_{m,l-1}$ and $B_{m,l}$ in the next generation: Log-Rank Test $\hat{\chi}_{LRT}(m,l)$, Cumulative Hazard Summary $\hat{\Lambda}_{CHS}(m,l)$, Log Hazards Ratio $\hat{\lambda}_{LHR}(m,l)$. The current box $\hat{B}_{m,l-1}$ is then updated: $\hat{B}_{m,l} = \hat{B}_{m,l-1} \setminus \hat{b}_{m,l}^{*}$, where $\hat{b}_{m,l}^{*} = \stackrel[\hat{b}_{m,l} \in C(b_{m,l})]{}{\text{argmax}} \left[\hat{r}(m,l)\right]$
                \vspace{-0.05in}
                \item Bottom-up pasting: Expand the box along any face (pasting) as long as the resulting box increase rate $\hat{r}(m,l) > 0$
                \vspace{-0.05in}
                \item Stop the peeling loop until a minimal box support $\hat\beta_{m,L}$ of $\hat{B}_{m,L}$ is such that it reached a minimal box support $0 \le \beta_{0} \le 1$, expressed as a fraction of the data: $\hat{\beta}_{m,L} \le \beta_{0}$
                \vspace{-0.05in}
                \item $l \leftarrow l + 1$
            \end{itemize}
            \State Step \#2 give a sequence of nested boxes $\{\hat{B}_{m,l}\}_{l=1}^{L}$, where $L$ is the estimated number of peeling/pasting steps with different numbers of observations in each box. Call the next box $\hat{B}_{m+1}=\hat{B}_{m,L}$. Remove the data in box $\hat{B}_{m}$ from the training data: $\mathcal{L}_{(m+1)} = \mathcal{L}_{(m)} \setminus \hat{B}_{m}$
            \State Stop the covering loop when running out of data or when a minimal number of observations remains within the last box $\hat{B}_{M}$, say $\hat{\beta}_{M} \le \beta_{0}$
            \State $m \leftarrow m + 1$
        \end{algorithmic}
        \vspace{-0.1in}
        \item Steps \#1 \text{--} \#5 produce a sequence of (not necessarily nested) boxes $\{\hat{B}_{m}\}_{m=1}^{M}$, where $M$ is the estimated total number of boxes covering $\mathcal{L}_{(1)}$
        \vspace{-0.1in}
        \item Collect the decision rules of all boxes $\{\hat{B}_{m}\}_{m=1}^{M}$ into a simple final decision rule $\hat{\mathcal{R}}$ of the solution region $\hat{R}$ of the form: $\hat{\mathcal{R}} = \bigcup_{m=1}^{M} \hat{\mathcal{R}}_{m}$, where $\hat{\mathcal{R}}_{m} = \bigcap_{j \in J}({\bf x}_{j} \in [t_{j,m}^{-},t_{j,m}^{+}])$ giving a full description of the estimated bumps in the entire input space
    \end{itemize}
\end{algorithm}
\spacingset{1}

%=========================================================================================
\section{Cross-Validation for Recursive Peeling Methods and a Survival/Risk Outcome} \label{sbhcv}
%=========================================================================================
%===============================================================================
\subsection{Split-Sample-Validation} \label{validation}
%===============================================================================
\vspace{-0.10in}
\subsubsection{Setup} \label{setup}
\vspace{-0.10in}
We previously tested the possibility of finding survival bumps in a small dataset, namely the Veteran's Administration lung cancer trial data from Kalbfleisch and Prentice \cite{Kalbfleisch_2002}. We could unravel interesting subgroups of patients with a poor survival time that could be characterized by a set of descriptive rules on the predictors including treatment intervention. Typically, this was indicative that an alternative intervention therapy could be required for these non-responders. While this approach showed promising results, it remained naive in that possible issues of bias and overfitting were not kept in check by model validation.

Assessment of model performance (e.g. prediction accuracy) requires the use of separation of the whole data $\mathcal{L}$ between a ``training set'' $\mathcal{L}^{\smallsetminus t}$ used to build a model and an independent ``testing set'' $\mathcal{L}^{t}$ used to assess model performance. To do so, the Split-Sample Validation technique (a.k.a Full Validation) is possible. Using this approach, a model is entirely developed on the training set $\mathcal{L}^{\smallsetminus t}$. Then, samples in the independent testing set $\mathcal{L}^{t}$ are used to determine the error rates. The samples in the testing set are never to be used for any aspect of model development such as variable selection and calibration and can therefore be used to check model performance \cite{Varma_2006, Molinaro_2005}.

Here, cross-validation of box estimates should include all steps of the box generation sequence $\{B_{m}\}_{m=1}^{M}$ i.e. for the (outer) coverage loop of our ``Patient Recursive Survival Peeling'' method (Algorithm \ref{algo}), each step of which involves a peeling sequence $\{B_{m,l}\}_{l=1}^{L}$ of the (inner) box peeling/induction loop. However, for simplicity, cross-validation designs of box estimates and resulting decision rule $\hat{\mathcal{R}}_{m}$ are shown below for \emph{fixed} $m \in \{1, \dots, M\}$ of the complete box sequence $\{\{\hat{B}_{m,l}\}_{l=1}^{L}\}_{m=1}^{M}$, so that subscript $m$ is further dropped. Without loss of generality, fix $m = 1$ (first coverage box).

\vspace{-0.10in}
\subsubsection{Estimated Box Quantities of Interest} \label{splitestimates}
\vspace{-0.10in}
Using previous notation, if we let $\hat{B}_{l}$ be the $l$th trained box and $\hat{\beta}_{l}$ be its estimated box support for $l \in \{1, \ldots, L\}$ of a box peeling sequence $\{\hat{B}_{l}\}_{l=1}^{L}$, then the test-set mean estimate of a box quantity of interest $q$ for the $l$th peeling step is indexed by the $l$th test box support $\hat{\beta}_{l}^{t}$ as follows:
\begin{equation}
    q(\hat{\beta}_{l}^{t}) = \frac{1}{n^{t}\hat{\beta}_{l}^{t}} \sum_{i=1}^{n^{t}} \hat{q}_{i}^{t} I\left( {\bf x}_{i}^{t} \in \hat{B}_{l} \right)
    \label{spliteq}
\end{equation}

where $q(\cdot)$ is the functional corresponding to the quantity $q$, $\hat{\beta}_{l}^{t} = \frac{1}{n^{t}} \sum_{i=1}^{n^{t}} I\left( {\bf x}_{i}^{t} \in \hat{B}_{l}^{t} \right)$ and $\hat{q}_{i}^{t}, {\bf x}_{i}^{t}, n^{t}$ are test-set quantities. Useful test-set quantities for the highest-risk box are box end-point statistics mentioned in section \ref{end-points}.

%=================================================================================
\subsection{$K$-fold Cross-Validation} \label{cross}
%=================================================================================
\vspace{-0.10in}
\subsubsection{Resampling Design - Notation} \label{resampling}
\vspace{-0.10in}
Although using a fully independent test set for evaluating a predictive bump hunting model is always advisable, the sample size $n$ in discovery-based studies is often too small to effectively split the data into training and testing sets and provide accurate estimates \cite{Baker_2002, Simon_2003, Dobbin_2007}. In such cases, resampling techniques such as $K$-fold Cross-Validation (CV) are required \cite{Ambroise_2002, Molinaro_2005}.

In resampling based on full $K$-fold cross-validation, the whole data $\mathcal{L}$ is randomly partitioned into $K$ approximately equal parts of test samples or test-sets $(\mathcal{L}_{1}, \ldots, \mathcal{L}_{k}, \dots, \mathcal{L}_{K})$. For each test-set $\mathcal{L}_{k}$, for $k \in \{1, \dots, K\}$, a training set $\mathcal{L}_{(k)}$ is formed from the union of the remaining $K-1$ subsets: $\mathcal{L}_{(k)} = \mathcal{L} \smallsetminus \mathcal{L}_{k}$. The process is repeated $K$ times, so that $K$ test-sets $\mathcal{L}_{k}$ are formed of about equal size and $K$ corresponding training subsets $\mathcal{L}_{(k)}$, for $k \in \{1, \ldots, K\}$. Typically, $K \in \{3,\ldots,10\}$. The training samples are approximately of size $\approx n(K-1)/K$ and the test samples are of size $n^{t} \approx n/K$.

\vspace{-0.10in}
\subsubsection{Cross-Validation Techniques} \label{techniques}
\vspace{-0.10in}
Recently, we described a cross-validation technique for recursive peeling methods in a survival/risk setting \cite{Dazard_2014}. The subject of this section is to give a more in-depth development of this strategy and compare it to standard cross-validation techniques.

There are issues when dealing with $K$-fold CV: first, how to cross-validate a simple peeling trajectory $\{\hat{B}_{l}\}_{l=1}^{L}$ and related statistics is not straightforward; second, how to cross-validate survival curve estimates and related statistics is also not intuitive (see also \cite{Simon_2011}); third, the data splitting in the cross-validation step should balance the class distributions of the outcome (i.e. here the censoring rate) within the cross-validation splits, which we call ``stratified random splitting by conservation of events''. So, regular $K$-fold cross-validation is not directly applicable to the joint task of box decision rules making by recursive peeling and survival estimation. One must design a specific cross-validation technique(s) of survival bump hunting that is amenable to this joint task.

Hence, we propose two techniques by which $K$-fold cross-validation estimates can be computed:
\spacingset{0}
\begin{itemize}
    \vspace{-0.1in}
    \item \emph{Averaging Technique}: Estimations are first computed for each ``in-box'' test subset samples, then averaged over the cross-validation loops of random splitting to give the ``Averaged Cross-Validation'' estimates (see details in section below \ref{averaged}).
    \vspace{-0.1in}
    \item \emph{Combining Technique}: All ``in-box'' test subset samples are first collected from all the cross-validation loops of random splitting to build a \emph{combined} test ``in-box'' and corresponding \emph{combined} test ``in-box'' samples to compute \emph{once} the final ``Combined Cross-Validation'' estimates (see details in section below \ref{combined}).
\end{itemize}
\spacingset{1}

Note that, unlike in the averaging technique, cross-validated combined estimates are computed on test samples of size $n$ instead of $n^{t} \approx n/K$, which could be an advantage in the case of tiny sample size $n$. In our numerical analyses, both strategies were compared with each other and with the situation of no cross-validation (see result section \ref{building}).

Finally, to account for the high variability of cross-validated estimates \cite{Efron_1983, Markatou_2005, Dobbin_2007}, we iterate each cross-validation procedure several times over some replicates $B$ (typically, $B \ge 10$) to average the estimates and reduce their variance. This so-called ``Replicated Cross-Validation'' approach is further detailed in section below (\ref{replication}). Also, aside the Split-Sample-Validation, mentioned in the previous section (\ref{validation}), other resampling techniques are available, which we left for discussion (section \ref{discussion}).

\vspace{-0.10in}
\subsubsection{Model Peeling Length Optimization Criterion} \label{optimization}
\vspace{-0.10in}
In model tuning, a trade-off between under-fitting and over-fitting can be achieved by optimizing an empirical function or objective criterion that takes censoring into account using cross-validation. The ``optimization criterion'' that we derive below is adapted to the task of of fitting a survival bump hunting model by recursive peeling with a survival outcome. Specifically, we tune a peeling model by optimizing its complexity, that is, the final length or number of peeling steps $L$ of the peeling sequence. The reason is that, for a given set of variables/covariates, the final length $L$ of a peeling model only depends on the peeling meta-parameters $\alpha_{0}$ (assumed fixed here) and $\beta_{0}$ (see section \ref{estimation}). In fact, an upper bound on the length of all possible peeling trajectories is given by $L_{\alpha_{0},\beta_{0}}=\left\lceil \frac{\log(\beta_{0})}{\log(1-\alpha_{0})} \right\rceil$ (see \cite{Friedman_1999} for details). So, a cheaper cross-validation can be achieved on $L$ only rather than on $\alpha_{0}$ and $\beta_{0}$ simultaneously.

Assuming $m$ fixed (see step \#2 of Algorithm \ref{algo}) and dropping subscript $m$ for simplification, the process of model building is repeated $K$ times for $k \in \{1, \dots, K\}$ as follows. First, let $l(k)$ denote the $l$th peeling step in the $k$th trajectory for $l \in \{1, \ldots, L(k)\}$ and $k \in \{1, \dots, K\}$, where $L(k)$ denotes the final length of a trained peeling model. Note that $L(k) \le L_{\alpha_{0},\beta_{0}}$, for all $k \in \{1, \dots, K\}$, but, in general this inequality is strict for large enough sample sizes. Let $\hat{B}_{l(k)}$ be the trained box of support $\hat{\beta}_{l(k)}$ of the box peeling sequence $\{\hat{B}_{l(k)}\}_{l=1}^{L}$ that is constructed from training set $\mathcal{L}_{(k)}$, leaving out the test-set $\mathcal{L}_{k}$ during all aspects of model building including covariate selection.

Second, once a resulting trained decision rule, abbreviated $\mathcal{R}_{k}$, and box definition estimates are generated from each training set $\mathcal{L}_{(k)}$, cross-validated estimates of box end-points statistics (described in \ref{end-points}) are computed using the left-out test-set $\mathcal{L}_{k}$. Three of these are the Log Hazard Ratio (\emph{LHR}), Log-Rank Test (\emph{LRT}) and the cross-validated estimate of prediction performance, namely the Concordance Error Rate (\emph{CER}) that is obtained by calculating the test-set error rate using the left-out test-set $\mathcal{L}_{k}$.

In the subsequent sections, we denote by superscript $cv$ any cross-validated estimate on the test-set $\mathcal{L}_{k}$. Since peeling lengths $L(k)$ are not necessarily equal for all $k \in \{1, \dots, K\}$, we use the following cross-validated maximum peeling length $\hat{L}_{m}^{cv}$ over the $K$ trajectories:
\begin{equation}
    \hat{L}_{m}^{cv} = \stackrel[k \in \{1, \dots, K\}]{}{\min}\left[L(k)\right]
    \label{Lm}
\end{equation}

After $K$ rounds of training and testing are complete and (averaged or combined) test profiles of \emph{LHR}, \emph{LRT} or \emph{CER} estimates are determined for each step $l \in \{1, \ldots, \hat{L}_{m}^{cv}\}$, model tuning is done by determining the optimal peeling length $\hat{L}^{cv}$ of the peeling sequence. To that end, one uses the maximization of the (averaged or combined) test profiles of \emph{LHR} and \emph{LRT} or the minimization of the (averaged or combined) test profile of \emph{CER} as criterion. Formally:
\begin{equation}
    \hat{L}^{cv} = \stackrel[l \in \{1, \dots, \hat{L}_{m}^{cv}\}]{}{\text{argmax}} \left[\hat{\lambda}^{cv}(l)\right]
    \quad \text{or} \quad
    \hat{L}^{cv} = \stackrel[l \in \{1, \dots, \hat{L}_{m}^{cv}\}]{}{\text{argmax}} \left[\hat{\chi}^{cv}(l)\right]
    \quad \text{or} \quad
    \hat{L}^{cv} = \stackrel[l \in \{1, \dots, \hat{L}_{m}^{cv}\}]{}{\text{argmin}} \left[\hat{\theta}^{cv}(l)\right],
    \label{optimizationeq}
\end{equation}
where $\hat{\lambda}^{cv}(l)$ is the cross-validated \emph{LHR)} in the high-risk box at step $l$, $\hat{\chi}^{cv}(l)$ is the cross-validated \emph{LRT} between the high vs. low-risk box at step $l$ and $\hat{\theta}^{cv}(l)$ is the cross-validated \emph{CER} between high-risk box predicted and observed survival times at step $l$.

Depending on the desired degree of conservativeness, the usual one-standard-error rule \cite{Hastie_2009} may be applied in combination with the profiles minimizer or maximizer to get smaller estimates corresponding to one standard-error below the maximum of \emph{LHR} and \emph{LRT} or standard-error above the minimum of \emph{CER}. In the subsequent sections, we denote by superscript $cv$ any cross-validated estimate on the test-set $\mathcal{L}_{k}$.

%=============================================================================================
\subsection{$K$-fold Averaged Cross-Validation} \label{averaged}
%=============================================================================================
In $K$-fold Averaged Cross-Validation, the \emph{averaged} cross-validated estimate of a box quantity $q$ at the $l(k)$th step of the box peeling sequence is based on the test samples falling within the trained box $\hat{B}_{l(k)}$. The \emph{averaged} cross-validated estimate of $q$ at step $l$ is simply computed by averaging the estimates obtained from all test boxes computed over all $K$ cross-validation loops. Specifically, each test-set $\mathcal{L}_{k}$ is used to

\begin{figure}[!hbt]
    \begin{minipage}{0.50\textwidth}
        \vskip -10pt
        \indent estimate the $l(k)$th test box membership indicator $\hat{\boldsymbol{\gamma}}_{l(k)}^{t}$ from the model grown on the training set $\mathcal{L}_{(k)}$. The corresponding test box support $\hat{\beta}_{l(k)}^{t}$ is directly derived from $\hat{\boldsymbol{\gamma}}_{l(k)}^{t}$ by computing the fraction of test data falling within the trained box $\hat{B}_{l(k)}$. The $l(k)$th estimate of the box quantity $q$ is indexed by the corresponding test box support $\hat{\beta}_{l(k)}^{t}$. For each training set $\mathcal{L}_{(k)}$, a trajectory curve $q(x)$ of a box quantity $q$ is defined as a piecewise constant curve, evaluated at the $l(k)$th test box support $\hat{\beta}_{l(k)}^{t}$, so that each trajectory curve is: $q(x)=q(\hat{\beta}_{l(k)}^{t})$ for $\hat{\beta}_{l(k)+1}^{t} \le x \le \hat{\beta}_{l(k)}^{t}$ (Figure \ref{Figure03}), where $q(\hat{\beta}_{l(k)}^{t})$ is derived as in equation \ref{spliteq}. The \emph{averaged} CV trajectory curve $\hat{q}^{cv}(x)$ of length $\hat{L}_{m}^{cv}$ is simply the average of the $K$ trajectory curves over the $K$ cross-validation loops: $\hat{q}^{cv}(x) = \frac{1}{K}\sum_{k=1}^{K} q(\hat{\beta}_{l(k)}^{t})$ for $\hat{\beta}_{l(k)+1}^{t} \le x \le \hat{\beta}_{l(k)}^{t}$.
    \end{minipage}
  \hfill
    \begin{minipage}{0.45\textwidth}
        \vskip -10pt
        \centering\includegraphics[scale=1]{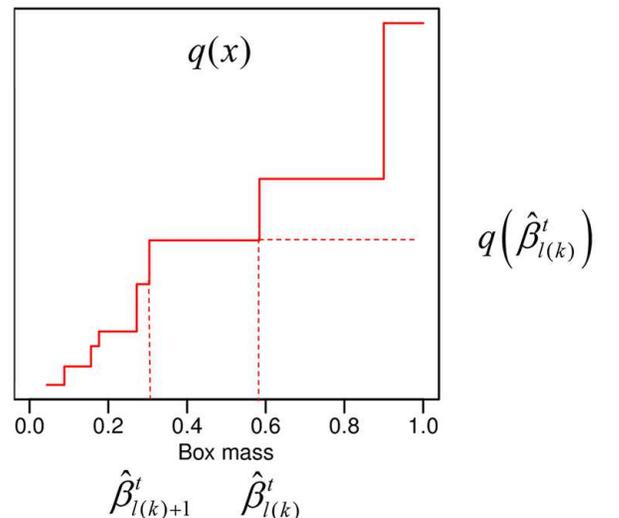}
        \vskip -20pt
        \caption{\sl \footnotesize Example of decreasing trajectory curve $q(x)$ of a box quantity $q$. Notice the piecewise constant curve of $q(x)$ for $\hat{\beta}_{l(k)+1}^{t} \le x \le \hat{\beta}_{l(k)}^{t}$. By convention, we define $\hat{\beta}_{0}^{t}=1$, $\hat{\beta}_{L(k)}^{t}=\beta_{0}$ and $\hat{\beta}_{L(k)+1}^{t}=0$.}
        \label{Figure03}
    \end{minipage}
\end{figure}

Formally, we show below how things are computed from an initial set of $K$ trained peeling trajectories:
\vspace{0.1in}

\noindent
\begin{minipage}{0.15\textwidth}
\parbox{1\textwidth}{For the $k$th training set $\mathcal{L}_{(k)}$:}

\vspace{-0.00in}
\begin{displaymath}
\parbox{1\textwidth}{$k$th training trajectory with box definitions} \ \rightarrow \!
\begin{cases}
\\ \\
\end{cases}
\end{displaymath}

\vspace{-0.20in}
\begin{displaymath}
\parbox{1\textwidth}{$k$th test box membership indicators} \ \rightarrow \!
\begin{cases}
\\ \\ \\
\end{cases}
\end{displaymath}

\vspace{-0.20in}
\begin{displaymath}
\parbox{1\textwidth}{$k$th test box supports} \ \rightarrow \!
\begin{cases}
\\ \\
\end{cases}
\end{displaymath}

\vspace{-0.20in}
\begin{displaymath}
\parbox{1\textwidth}{$k$th test box estimated quantities} \ \rightarrow \!
\begin{cases}
\\ \\
\end{cases}
\end{displaymath}

\vspace{-0.20in}
\begin{displaymath}
\parbox{1\textwidth}{Averaged CV test box quantities over $K$ trajectories} \ \rightarrow \!
\begin{cases}
\\ \\
\end{cases}
\end{displaymath}

\end{minipage}
\hfill
\noindent
\begin{minipage}{0.85\textwidth}
\centering{\framebox{\parbox{1in}{\centering First peeling step in the $k$th trajectory}} \ \qquad \qquad
{\centering direction of peeling} \ \qquad \qquad
\framebox{\parbox{1in}{\centering Last peeling step in the $k$th trajectory}}}\\
\vspace{-0.20in}
\centering{\raisebox{1in}{$\rightarrow$}}\\

\vspace{-0.90in}
\centering{
           $1(k)$ \qquad\qquad\qquad\qquad\qquad
           $l(k)$ \qquad\qquad\qquad\qquad\qquad
           $L(k)$}\\
\centering{\framebox{\parbox{0.35in}{\centering $\hat{B}_{1(k)}$}}
           $\qquad\qquad \cdots \qquad\quad \ $
           \framebox{\parbox{0.35in}{\centering $\hat{B}_{l(k)}$}}
           $\qquad\qquad \ \cdots \qquad\quad \ $
           \framebox{\parbox{0.35in}{\centering $\hat{B}_{L(k)}$}}}\\

\vspace{-0.00in}
\centering{\begin{displaymath}
                \hat{\boldsymbol{\gamma}}_{1(k)}^{t \ T} \!=\! \left[\hat{\gamma}_{i,1(k)}^{t}\right]_{i=1}^{n^t}
                \qquad \cdots \quad
                \hat{\boldsymbol{\gamma}}_{l(k)}^{t \ T} \!=\! \left[\hat{\gamma}_{i,l(k)}^{t}\right]_{i=1}^{n^t}
                \qquad \cdots \quad
                \hat{\boldsymbol{\gamma}}_{L(k))}^{t \ T} \!=\! \left[\hat{\gamma}_{i,L(k)}^{t}\right]_{i=1}^{n^t}
            \end{displaymath}}
\vspace{-0.20in}
\centering{\begin{displaymath}
                              \!=\! \left[I[{\bf x}_{i}^{t} \in \hat{B}_{1(k)}]\right]_{i=1}^{n^{t}}
                \qquad \quad \!=\! \left[I[{\bf x}_{i}^{t} \in \hat{B}_{l(k)}]\right]_{i=1}^{n^{t}}
                \qquad \quad \!=\! \left[I[{\bf x}_{i}^{t} \in \hat{B}_{L(k)}]\right]_{i=1}^{n^{t}}
            \end{displaymath}}

\vspace{-0.40in}
\centering{\begin{displaymath}
                \hat{\beta}_{1(k)}^{t} \!=\! \frac{1}{n^t}\smashoperator[r]{\sum_{i=1}^{n^{t}}} \hat{\gamma}_{i,1(k)}^{t}
                \quad \cdots \quad
                \hat{\beta}_{l(k)}^{t} \!=\! \frac{1}{n^t}\smashoperator[r]{\sum_{i=1}^{n^{t}}} \hat{\gamma}_{i,l(k)}^{t}
                \quad \cdots \quad
                \hat{\beta}_{L(k)}^{t} \!=\! \frac{1}{n^t}\smashoperator[r]{\sum_{i=1}^{n^{t}}} \hat{\gamma}_{i,L(k)}^{t}
            \end{displaymath}}

\vspace{-0.00in}
\centering{$            q \left(\hat{\beta}_{1(k)}^{t}\right)$
           $\quad \qquad \cdots \quad \quad$
           $\quad \quad q \left(\hat{\beta}_{l(k)}^{t}\right)$
           $\quad \qquad \cdots \quad \quad$
           $\quad \quad q \left(\hat{\beta}_{L(k)}^{t}\right)$}\\

\vspace{0.10in}
\centering{\begin{displaymath}
                \hat{q}^{cv}(1) \!=\! \frac{1}{K}\smashoperator[r]{\sum_{k=1}^{K}} q \left(\hat{\beta}_{1(k)}^{t}\right)
                \ \cdots \
                \hat{q}^{cv}(l) \!=\! \frac{1}{K}\smashoperator[r]{\sum_{k=1}^{K}} q \left(\hat{\beta}_{l(k)}^{t}\right)
                \ \cdots \
                \hat{q}^{cv}(\hat{L}_{m}^{cv}) \!=\! \frac{1}{K}\smashoperator[r]{\sum_{k=1}^{K}} q \left(\hat{\beta}_{\hat{L}_{m}^{cv}}^{t}\right)
            \end{displaymath}}
\end{minipage}
\newline
\newline

\indent From the $K$ test trajectories, one derives first the ``Averaged CV'' optimal peeling length of the peeling trajectory, according to the optimization criterion for model selection as in equation \ref{optimizationeq}:
\begin{equation*}
    \hat{L}^{cv} = \stackrel[l \in \{1, \dots, \hat{L}_{m}^{cv}\}]{}{\text{argmax}} \left[\hat{\lambda}^{cv}(l)\right] \quad \text{or} \quad
    \hat{L}^{cv} = \stackrel[l \in \{1, \dots, \hat{L}_{m}^{cv}\}]{}{\text{argmax}} \left[\hat{\chi}^{cv}(l)\right] \quad \text{or} \quad
    \hat{L}^{cv} = \stackrel[l \in \{1, \dots, \hat{L}_{m}^{cv}\}]{}{\text{argmin}} \left[\hat{\theta}^{cv}(l)\right]
\end{equation*}

Then, one derives ``Averaged CV'' estimates for each step $l \in \{1, \ldots, \hat{L}^{cv}\}$ as follows:
\spacingset{0}
\begin{itemize}
    \item The ``Averaged CV'' box definition, which can be written as the outer product of $|J|$ intervals as in equation \ref{R}, is formed by taking the rectangular box where each of the $2|J|$ edge is averaged over the $K$ cross-validation loops:
        \begin{equation*}
            \hat{B}^{cv}(l) = \smashoperator[r]{\bigtimes_{j \in J}} [\hat{t}_{j,l}^{-},\hat{t}_{j,l}^{+}]
            \quad \text{where for each} \ j \in J \text{,} \quad
            \begin{cases}
                \hat{t}_{j,l}^{-} = \stackrel[k \in \{1,\dots,K\}]{}{\text{ave}} \ [\hat{t}_{j,l(k)}^{-}]\\
                \hat{t}_{j,l}^{+} = \stackrel[k \in \{1,\dots,K\}]{}{\text{ave}} \ [\hat{t}_{j,l(k)}^{+}]
            \end{cases}
        \end{equation*}
        \vspace{-0.0in}
    \item The ``Averaged CV'' box membership indicator is formed by counting the data within the ``Averaged CV'' box:
        \begin{equation*}
            \hat{\boldsymbol{\gamma}}^{cv \ T}(l) = \left[I[{\bf x}_{i} \in \hat{B}^{cv}(l)]\right]_{i=1}^{n}
        \end{equation*}
        \vspace{-0.2in}
    \item The ``Averaged CV'' box support is computed as the fraction of data within the ``Averaged CV'' box:
        \begin{equation*}
            \hat{\beta}^{cv}(l) = \frac{1}{n}\sum_{i=1}^{n} I[{\bf x}_{i} \in \hat{B}^{cv}(l)]
        \end{equation*}
        \vspace{-0.2in}
    \item The ``Averaged CV'' box end-point quantity $q$ is taken as the averaged CV trajectory curve evaluated at the $l(k)$th test box support $\hat{\beta}_{l(k)}^{t}$:
        \begin{align*}
            &\hat{q}^{cv}(l) = \frac{1}{K}\sum_{k=1}^{K} q \left( \hat{\beta}_{l(k)}^{t} \right) \ \text{, where} \\ \nonumber
            &q \left( \hat{\beta}_{l(k)}^{t} \right) = \frac{1}{n^{t}\hat{\beta}_{l(k)}^{t}} \sum_{i=1}^{n^{t}} \hat{q}_{i}^{t} I\left( {\bf x}_{i}^{t} \in \hat{B}_{l(k)} \right) \ \text{as in equation } \ref{spliteq}.
        \end{align*}
        The latter is done for the ``Averaged CV'' box estimates of: (i) The Log Hazard Ratio (\emph{LHR}) in the high-risk box: $\hat{\lambda}^{cv}(l) = \frac{1}{K}\sum_{k=1}^{K} \lambda \left( \hat{\beta}_{l(k)}^{t} \right)$; (ii) The Log-Rank Test (\emph{LRT}) between the high vs. low-risk box: $\hat{\chi}^{cv}(l) = \frac{1}{K}\sum_{k=1}^{K} \chi \left( \hat{\beta}_{l(k)}^{t} \right)$; (iii) The Concordance Error Rate (\emph{CER}) between high-risk box predicted and observed survival times: $\hat{\theta}^{cv}(l) = \frac{1}{K}\sum_{k=1}^{K} \theta \left( \hat{\beta}_{l(k)}^{t} \right)$; (iv) The Minimal Event-Free Probability (\emph{MEFP}): $\widehat{P_{0}^{\prime}}^{cv}(l) = \frac{1}{K}\sum_{k=1}^{K} P_{0}^{\prime} \left( \hat{\beta}_{l(k)}^{t} \right)$; (v) The Minimal Event-Free Time (\emph{MEFT}): $\widehat{T_{0}^{\prime}}^{cv}(l) = \frac{1}{K}\sum_{k=1}^{K} T_{0}^{\prime} \left( \hat{\beta}_{l(k)}^{t} \right)$.
\end{itemize}
\spacingset{1}

%=============================================================================================
\subsection{$K$-fold Combined Cross-Validation} \label{combined}
%=============================================================================================
In $K$-fold Combined Cross-Validation, for each loop, samples from the training set are used to train a peeling model of a certain length, then samples from the test-set are used to determine the ``in-box'' test-set samples falling into the trained box. Eventually, all ``in-box'' test samples are combined together and all ``out-of-box'' test samples are combined together as well. So, in $K$-fold Combined CV, estimate of a box quantity $q$ is computed \emph{once} on the collective test-set ``in-box'' samples, formed over the $K$ cross-validation loops. This allows the estimation of box quantities and box survival distribution curves for both ``in-box'' and ``out-of-box'' samples.

Specifically, each test-set $\mathcal{L}_{k}$ is used to estimate the test box membership indicator $\tilde{\boldsymbol{\gamma}}_{l(k)}^{t}$ from the model grown on the $k$th training set $\mathcal{L}_{(k)}$. The $l$th \emph{combined} cross-validated test box membership indicator $\tilde{\boldsymbol{\gamma}}^{cv}(l)$ is formed \emph{once} by taking the vector concatenation of all the cross-validated test box membership indicators $\{\tilde{\boldsymbol{\gamma}}_{l(k)}^{t}\}_{k=1}^{K}$ over the $K$ cross-validation loops. The corresponding $l$th combined cross-validated test box support $\tilde{\beta}^{cv}(l)$ is then directly derived from $\tilde{\boldsymbol{\gamma}}^{cv}(l)$.

The \emph{combined} cross-validated estimate of a box quantity $q$ at the $l$th step of the peeling trajectory is then computed \emph{once} from the combined cross-validated test box membership indicator $\tilde{\boldsymbol{\gamma}}(l)^{cv}$ and indexed by the corresponding test box support $\tilde{\beta}(l)^{cv}$. Here, the \emph{combined} cross-validated trajectory curve $\tilde{q}_{k}(x)$ is defined as the piecewise constant curve of length $\tilde{L}_{m}^{cv}$, evaluated at the $l$th combined cross-validated test box membership indicator $\tilde{\boldsymbol{\gamma}}^{cv}(l)$.

Formally, we show below how things are computed from an initial set of $K$ trained peeling trajectories (where $\bigparallel$ denotes the concatenation operator).
\newline

\noindent
\begin{minipage}{0.15\textwidth}
\parbox{1\textwidth}{For the $k$th training set $\mathcal{L}_{(k)}$:}

\begin{displaymath}
\parbox{1\textwidth}{$k$th training trajectory with box definitions} \ \ \rightarrow \!
\begin{cases}
\\ \\
\end{cases}
\end{displaymath}

\vspace{-0.00in}
\begin{displaymath}
\parbox{1\textwidth}{$k$th test box membership indicators} \ \ \rightarrow \!
\begin{cases}
\\ \\
\end{cases}
\end{displaymath}

\vspace{-0.10in}
\begin{displaymath}
\parbox{1\textwidth}{Combined CV test box membership indicators over $K$ trajectories} \ \ \rightarrow \!
\begin{cases}
\\ \\ \\ \\
\end{cases}
\end{displaymath}

\vspace{-0.20in}
\begin{displaymath}
\parbox{1\textwidth}{Combined CV test box supports over $K$ trajectories} \ \ \rightarrow \!
\begin{cases}
\\ \\
\end{cases}
\end{displaymath}

\vspace{-0.20in}
\begin{displaymath}
\parbox{1\textwidth}{Combined CV test box quantities over $K$ trajectories} \ \ \rightarrow \!
\begin{cases}
\\ \\
\end{cases}
\end{displaymath}

\end{minipage}
\hfill
\noindent
\begin{minipage}{0.80\textwidth}
\centering{\framebox{\parbox{1in}{\centering First peeling step in the $k$th trajectory}} \ \qquad \qquad
{\centering direction of peeling} \ \qquad \qquad
\framebox{\parbox{1in}{\centering Last peeling step in the $k$th trajectory}}}\\
\vspace{-0.20in}
\centering{\raisebox{1in}{$\rightarrow$}}\\

\vspace{-0.90in}
\centering{
           $1(k)$ \qquad\qquad\qquad\qquad\qquad
           $l(k)$ \qquad\qquad\qquad\qquad\qquad
           $L(k)$}\\
\centering{\framebox{\parbox{0.35in}{\centering $\tilde{B}_{1(k)}$}}
           $\qquad\qquad \cdots \qquad\quad \ $
           \framebox{\parbox{0.35in}{\centering $\tilde{B}_{l(k)}$}}
           $\qquad\qquad \ \cdots \qquad\quad \ $
           \framebox{\parbox{0.35in}{\centering $\tilde{B}_{L(k)}$}}}\\

\vspace{0.20in}
\centering{\begin{displaymath}
                \tilde{\boldsymbol{\gamma}}_{1(k)}^{t \ T} \!=\! \left[I[{\bf x}_{i}^{t} \in \tilde{B}_{1(k)}]\right]_{i=1}^{n^{t}}
                \ \cdots \
                \tilde{\boldsymbol{\gamma}}_{l(k)}^{t \ T} \!=\! \left[I[{\bf x}_{i}^{t} \in \tilde{B}_{l(k)}]\right]_{i=1}^{n^{t}}
                \ \cdots \
                \tilde{\boldsymbol{\gamma}}_{L(k)}^{t \ T} \!=\! \left[I[{\bf x}_{i}^{t} \in \tilde{B}_{L(k)}]\right]_{i=1}^{n^{t}}
            \end{displaymath}}

\vspace{-0.20in}
\centering{\begin{displaymath}
                \tilde{\boldsymbol{\gamma}}^{cv}(1) \!=\! \left[\tilde{\gamma}_{i}^{cv}(1)\right]_{i=1}^{n}
                \quad \cdots \quad
                \tilde{\boldsymbol{\gamma}}^{cv}(l) \!=\! \left[\tilde{\gamma}_{i}^{cv}(l)\right]_{i=1}^{n}
                \quad \cdots \quad
                \tilde{\boldsymbol{\gamma}}^{cv}(\tilde{L}_{m}^{cv}) \!=\! \left[\tilde{\gamma}_{i}^{cv}(\tilde{L}_{m}^{cv})\right]_{i=1}^{n}
            \end{displaymath}}
\vspace{-0.20in}
\centering{\begin{displaymath}
                                            \!=\! \smashoperator[r]{\bigparallel_{k=1}^{K}} \tilde{\boldsymbol{\gamma}}_{1(k)}^{t}
                \qquad \qquad \qquad \qquad \!=\! \smashoperator[r]{\bigparallel_{k=1}^{K}} \tilde{\boldsymbol{\gamma}}_{l(k)}^{t}
                \qquad \qquad \qquad \qquad \!=\! \smashoperator[r]{\bigparallel_{k=1}^{K}} \tilde{\boldsymbol{\gamma}}_{\tilde{L}_{m}^{cv}}^{t}
            \end{displaymath}}

\vspace{-0.10in}
\centering{\begin{displaymath}
                \tilde{\beta}^{cv}(1) \!=\! \frac{1}{n}\smashoperator[r]{\sum_{i=1}^{n}} \tilde{\gamma}_{i}^{cv}(1)
                \ \cdots \
                \tilde{\beta}^{cv}(l) \!=\! \frac{1}{n}\smashoperator[r]{\sum_{i=1}^{n}} \tilde{\gamma}_{i}^{cv}(l)
                \ \cdots \
                \tilde{\beta}^{cv}(\tilde{L}_{m}^{cv}) \!=\! \frac{1}{n}\smashoperator[r]{\sum_{i=1}^{n}} \tilde{\gamma}_{i}^{cv}(\tilde{L}_{m}^{cv})
            \end{displaymath}}

\vspace{-0.10in}
\centering{\begin{displaymath}
                \tilde{q}^{cv}(1) \!=\! q \left[ \tilde{\boldsymbol{\beta}}^{cv}(1) \right]
                \quad \cdots \quad
                \tilde{q}^{cv}(l) \!=\! q \left[ \tilde{\boldsymbol{\beta}}^{cv}(l) \right]
                \quad \cdots \quad
                \tilde{q}^{cv}(\tilde{L}_{m}^{cv}) \!=\! q \left[ \tilde{\boldsymbol{\beta}}^{cv}(\tilde{L}_{m}^{cv}) \right]
            \end{displaymath}}
\end{minipage}
\newline
\newline

\indent From the $K$ test trajectories, one derives first the ``Combined CV'' optimal peeling length of the peeling trajectory, according to the optimization criterion as in equation \ref{optimizationeq}:
\begin{equation*}
    \tilde{L}^{cv} = \stackrel[l \in \{1, \dots, \tilde{L}_{m}^{cv}\}]{}{\text{argmax}} \left[\tilde{\lambda}^{cv}(l)\right] \quad \text{or} \quad
    \tilde{L}^{cv} = \stackrel[l \in \{1, \dots, \tilde{L}_{m}^{cv}\}]{}{\text{argmax}} \left[\tilde{\chi}^{cv}(l)\right] \quad \text{or} \quad
    \tilde{L}^{cv} = \stackrel[l \in \{1, \dots, \tilde{L}_{m}^{cv}\}]{}{\text{argmin}} \left[\tilde{\theta}^{cv}(l)\right]
\end{equation*}

Likewise, from the $K$ test trajectories, one derives ``Combined CV'' estimates for each step $l \in \{1, \ldots, \tilde{L}^{cv}\}$ as follows:
\spacingset{0}
\begin{itemize}
   \item The ``Combined CV'' box membership indicator (Boolean $n$-vector) is formed by vector-concatenation of all the test box membership indicators over the $K$ cross-validation loops:
        \begin{equation*}
            \tilde{\boldsymbol{\gamma}}^{cv \ T}(l) = \left[\tilde{\gamma}_{i}^{cv}(l)\right]_{i=1}^{n} = \bigparallel_{k=1}^{K} \tilde{\boldsymbol{\gamma}}_{l(k)}^{t} = \bigparallel_{k=1}^{K} \left[I[{\bf x}_{i}^{t} \in \tilde{B}_{l(k)}]\right]_{i=1}^{n^{t}}
        \end{equation*}
        \vspace{-0.2in}
   \item The ``Combined CV'' box definition, which can be written as the outer product of $|J|$ intervals as in equation \ref{R}, is formed by taking the rectangular box ($2|J|$ edges) circumscribing all the ``in-box'' test samples over the $K$ cross-validation loops:
        \begin{equation*}
            \tilde{B}^{cv}(l) = \smashoperator[r]{\bigtimes_{j \in J}} [\tilde{t}_{j,l}^{-},\tilde{t}_{j,l}^{+}]
            \quad \text{where for each} \ j \in J \text{,} \quad
            \begin{cases}
                \tilde{t}_{j,l}^{-} = \smashoperator[r]{\min_{k \in \{1,\dots,K\}}} \quad [x_{i,j}^{t}, i \in \{1,\dots,n^{t}\} \colon x_{i,j}^{t} \in \tilde{B}_{l(k)}]\\
                \tilde{t}_{j,l}^{+} = \smashoperator[r]{\max_{k \in \{1,\dots,K\}}} \quad [x_{i,j}^{t}, i \in \{1,\dots,n^{t}\} \colon x_{i,j}^{t} \in \tilde{B}_{l(k)}]
            \end{cases}
        \end{equation*}
        \vspace{-0.1in}
   \item The ``Combined CV'' box support is computed as the fraction of data within the ``Combined CV'' box:
        \begin{equation*}
            \tilde{\beta}^{cv}(l) = \frac{1}{n}\smashoperator[r]{\sum_{i=1}^{n}} \tilde{\gamma}_{i}^{cv}(l)
        \end{equation*}
        \vspace{-0.2in}
   \item The ``Combined CV'' box end-point quantity $q$ is taken as the result of the functional $q(\cdot)$ evaluated at the $l$th ``Combined CV'' test box support $\tilde{\beta}^{cv}(l)$:
        \begin{equation*}
            \tilde{q}^{cv}(l) = q \left[ \tilde{\boldsymbol{\beta}}^{cv}(l) \right]
        \end{equation*}
        The latter is done for the ``Combined CV'' box estimates of: (i) The Log Hazard Ratio (\emph{LHR}) in the high-risk box: $\tilde{\lambda}^{cv}(l) = \lambda \left[\tilde{\boldsymbol{\beta}}^{cv}(l)\right]$; (ii) The Log-Rank Test (\emph{LRT}) between the high vs. low-risk box: $\tilde{\chi}^{cv}(l) = \chi \left[\tilde{\boldsymbol{\beta}}^{cv}(l)\right]$; (ii) The Concordance Error Rate (\emph{CER}) between high-risk box predicted and observed survival times: $\tilde{\theta}^{cv}(l) = \theta \left[\tilde{\boldsymbol{\beta}}^{cv}(l)\right]$; (iv) The Minimal Event-Free Probability (\emph{MEFP}): $\widetilde{P_{0}^{\prime}}^{cv}(l) = P_{0}^{\prime} \left[\tilde{\boldsymbol{\beta}}^{cv}(l)\right]$; (v) The Minimal Event-Free Time (\emph{MEFT}): $\widetilde{T_{0}^{\prime}}^{cv}(l) = T_{0}^{\prime} \left[\tilde{\boldsymbol{\beta}}^{cv}(l)\right]$.
\end{itemize}
\spacingset{1}

%============================================================================================
\subsection{$K$-fold Cross-Validation of \emph{P}-Values} \label{cvpval}
%============================================================================================
The log-rank test statistic (e.g. $\chi_{1}^{2}$ for a two group comparison) is a classical measure to evaluate the statistical significance of separation between survival curves. However, the null distribution of the log-rank test is not valid for cross-validated curves because the observations used to cross-validate the curves are not independent anymore.

For each step $l \in \{1, \ldots, \tilde{L}^{rcv}\}$, we generate the null distribution of the cross-validated log-rank statistic $\tilde{\chi}^{cv(a)}(l)$ for $a \in \{1,\dots,A\}$ by randomly permuting the correspondence of survival times and censoring indicators of the data and by computing the corresponding cross-validated survival curves and cross-validated log-rank statistic for that permutation. By repeating $A$ times the entire $K$-fold cross-validation process for many random permutations (typically $A = 1000$), one generates a null distribution of the permuted log-rank statistics (annotated below w.l.o.g. for the case of ``Combined CV''):
\begin{equation*}
   \{\tilde{\chi}^{cv(a)}(l)\}_{a=1}^{A}
\end{equation*}

The proportion of replicates with log-rank statistic greater than or equal to the observed statistic $\tilde{\chi}^{cv}(l)$ for the un-permuted data is the statistical significance level for the test. Log-rank test permutation $p$-values are then calculated for each step $l \in \{1, \ldots, \tilde{L}^{rcv}\}$ as:
\begin{equation}
    \tilde{p}^{cv}(l) = \frac{1}{A} \smashoperator[r]{\sum_{a=1}^{A}} I\left[\tilde{\chi}^{cv(a)}(l) \geq \tilde{\chi}^{cv}(l)\right]
\end{equation}

These $p$-values may be discrete: the precision depends on the number $A$ of random permutations and the lower bound $1/A$ may be reached in practise.

%=================================================================================================
\subsection{Replicated $K$-fold Cross-Validation} \label{replication}
%=================================================================================================
Typically, $K$-fold cross-validation is repeated $B = 10 - 100$ times and resulting replicated cross-validated estimates are somehow ``averaged'' over the replicates. We denote these by the superscript $rcv$ and each replicate by the superscript $cv(b)$, for $b \in \{1, \dots, B\}$. This is done for either cross-validation technique (shown below w.l.o.g. for the case of ``Combined CV'').

Formally, one first derives the ``Replicated CV'' maximal peeling length of the peeling model from the cross-validation replicates, denoted $\bar{L}_{m}^{rcv}$. To do so, one uses the cross-validated maximum peeling length $\hat{L}_{m}^{cv(b)}$ of the peeling trajectory, defined in equation \ref{Lm}, for $b \in \{1, \dots, B\}$. Formally, the ``Replicated CV'' maximal peeling length  $\bar{L}_{m}^{rcv}$ is calculated as the ceiling-mean of the cross-validated quantities $\hat{L}_{m}^{cv(b)}$:
\begin{equation}
    \bar{L}_{m}^{rcv} = \left\lceil \frac{1}{B} \smashoperator[r]{\sum_{b=1}^{B}} \hat{L}_{m}^{cv(b)} \right\rceil
    \label{rep.Lm}
\end{equation}

Next, depending on the optimization criterion used, one gets the ``Replicated CV'' optimal length of the peeling trajectory:
\begin{equation}
    \bar{L}^{rcv} = \stackrel[l \in \{1, \dots, \bar{L}_{m}^{rcv}\}]{}{\text{argmax}} \left[\bar{\lambda}^{rcv}(l)\right] \quad \text{or} \quad
    \bar{L}^{rcv} = \stackrel[l \in \{1, \dots, \bar{L}_{m}^{rcv}\}]{}{\text{argmax}} \left[\bar{\chi}^{rcv}(l)\right] \quad \text{or} \quad
    \bar{L}^{rcv} = \stackrel[l \in \{1, \dots, \bar{L}_{m}^{rcv}\}]{}{\text{argmin}} \left[\bar{\theta}^{rcv}(l)\right]
    \label{rep.optimizationeq}
\end{equation}
where each optimization criterion: (i) The Log Hazard Ratio (\emph{LHR}) in the high-risk box: $\bar{\lambda}^{rcv}(l)$, (ii) The Log-Rank Test (\emph{LRT}) between the high vs. low-risk box: $\bar{\chi}^{rcv}(l)$, and (iii) The Concordance Error Rate (\emph{CER}) between high-risk box predicted and observed survival times: $\widebar{\theta}^{rcv}(l)$ is taken as the average estimate over the $B$ replicates for each step $l \in \{1, \ldots, \bar{L}_{m}^{rcv}\}$ as follows:
\begin{equation}
    \bar{\lambda}^{rcv}(l) = \frac{1}{B} \smashoperator[r]{\sum_{b=1}^{B}} \tilde{\lambda}^{cv(b)}(l)
     \quad \text{or} \quad
    \bar{\chi}^{rcv}(l) = \frac{1}{B} \smashoperator[r]{\sum_{b=1}^{B}} \tilde{\chi}^{cv(b)}(l)
     \quad \text{or} \quad
    \bar{\theta}^{rcv}(l) = \frac{1}{B} \smashoperator[r]{\sum_{b=1}^{B}} \tilde{\theta}^{cv(b)}(l)
\end{equation}

Using the above ``Replicated CV'' optimal length of the peeling trajectory, one finally derives ``Replicated CV'' box end points from the $B$ replicates for each step $l \in \{1, \ldots, \bar{L}^{rcv}\}$ as follows:
\spacingset{0}
\begin{itemize}
    \item The ``Replicated CV'' box definition ($2|J|$ edges) is taken as the average-box over the $B$ replicates:
        \vspace{-0.1in}
        \begin{equation}
            \bar{B}^{rcv}(l) = \stackrel[b \in \{1, \dots, B\}]{}{\text{ave}}\left[\tilde{B}^{cv(b)}(l)\right]
        \end{equation}
        where $\text{ave}(\cdot)$ denotes the averaging function by edge or dimension $j$ for $j \in J$.
    \item The ``Replicated CV'' box membership indicator (Boolean $n$-vector) is taken as the average-box membership indicator, observed to be nearly equal to the point-wise majority vote over the $B$ replicates:
        \vspace{-0.2in}
        \begin{align}
            \bar{\boldsymbol{\gamma}}^{rcv}(l) = \left[ I[ {\bf x}_{i} \in \bar{B}^{rcv}(l) ] \right]_{i=1}^{n}
                                               \approx \left[ I\left( \smashoperator[r]{\sum_{b=1}^{B}} \tilde{\gamma}_{i}^{cv(b)}(l) \geq \left\lceil\frac{B}{2}\right\rceil \right) \right]_{i=1}^{n}
        \end{align}
   \item The ```Replicated CV'' box support is taken as the average estimate over the $B$ replicates:
        \begin{equation}
            \bar{\beta}^{rcv}(l) = \frac{1}{B}\smashoperator[r]{\sum_{b=1}^{B}} \tilde{\beta}^{cv(b)}(l)
        \end{equation}
        \vspace{-0.2in}
   \item Other ``Replicated CV'' box end-point quantities $q$ estimates, taken as the average estimate over the $B$ replicates:
        \begin{equation}
            \bar{q}^{rcv}(l) = \frac{1}{B} \smashoperator[r]{\sum_{b=1}^{B}} \tilde{q}^{cv(b)}(l)
        \end{equation}
        This is done for: (i) The Minimal Event-Free Probability (\emph{MEFP}): $\widebar{P_{0}^{\prime}}^{rcv}(l)$ and (ii) The Minimal Event-Free Time (\emph{MEFT}): $\widebar{T_{0}^{\prime}}^{rcv}(l)$.
\end{itemize}
\spacingset{1}

%=========================================================================================
\section{Numerical Analyses} \label{simulation}
%=========================================================================================
%===================================
\subsection{Simulation Design} \label{design}
%===================================
The $p$-dimensional covariates ${\bf x}_{i} = \left[x_{i,1} \ldots x_{i,p}\right]^{T}$, for $i \in \{1, \ldots, n\}$, were identically and independently drawn from either: a (i) $p$-multivariate normal distribution with mean vector $\boldsymbol{\mu}$ and variance-covariance matrix $\boldsymbol{\Sigma}$: ${\bf x}_{i} \sim N_{p}(\boldsymbol{\mu},\boldsymbol{\Sigma})$; or (ii) from a $p$-multivariate uniform distribution on the interval $[a,b]$: ${\bf x}_{i} \sim U_{p}(a,b)$.

Simulations were carried out according to the assumptions stated in section \ref{notation}. Simulated realizations of true survival times $T_{i}$ were identically and independently drawn from an exponential distribution with rate parameter $\lambda$ (and mean $\frac{1}{\lambda}$): $T_{i} \sim \text{Exp}(\lambda)$. Simulated realizations $C_{i}$ of true censoring times were identically and independently sampled from a uniform distribution: $C_{i} \sim U(0,v)$ with $v > 0$, so that approximately $100 \times \pi (\%)$ of the simulated realizations of observed survival times $Y_{i} = \min(T_{i},C_{i})$ were censored, where $\pi \in \{0.3,0.5,0.7\}$. Finally, the simulated realizations of observed event (non-censoring) random variable indicator were $\delta_{i} = I(T_{i} \le C_{i})$.

To simulate survival models with various types of relationship between survival times (or hazards) and covariates (i.e. variable informativeness) including saturated regression models and noise models, the individual hazard rate parameter $\lambda_{i}$ was simulated as an exponential regression function of individual covariate ${\bf x}_{i}$: $\lambda_{i}(t|{\bf x}_{i}) = \lambda_{0}(t)\exp\left[\eta({\bf x}_{i})\right]$, where the regression function $\eta({\bf x}_{i})$ can take different forms depending on the simulated survival model. In summary, our simulation was done using the following parameters (see documentation in our R package for more details \cite{Dazard_2015a}):

\spacingset{0}
\begin{itemize}
    \vspace{-0.05in}
    \item ${\bf x}_{i} \sim U_{p}(0,1)$ with $n = 250$ and $p = 3$, or ${\bf x}_{i} \sim N_{p}([0 \ 0 \ 0]^{T},\sigma^{2}\bf{I})$ with $n = 100$ and $p = 1000$.
    \vspace{-0.1in}
    \item by characterization of the first coverage box ${B}_{1}$ (i.e. for $m = 1$), using constrained/directed peeling, without pasting and with meta-parameter values $(\alpha_{0},\beta_{0}) \in \{(0.10,0.05)\}$.
    \vspace{-0.1in}
    \item with censoring rate $\pi$ = 0.5 and five concurrent simulated survival models, where $n > p$ and $n \ll p$, representing low- and high-dimensional situations, and where the regression function is as follows:
        \begin{itemize}
            \item In simulated models \#1-4, $\eta({\bf x}_{i}) = \boldsymbol{\eta}^T {\bf x}_{i}$ with regression parameters $\boldsymbol{\eta} = [\eta_{1} \ldots 0_{j} \ldots \eta_{p}]^T$, for $j \in \varnothing \cup \{1, \ldots, p\}$:
                \vspace{-0.10in}
                \begin{align}
                    \nonumber
                    \begin{cases}
                        \text{Low-dim. Saturated Model \#1:}    &n=250, p=3    \ \qquad \boldsymbol{\eta} = [12 \ -15 \  -5]^T\\
                        \text{Low-dim. Un-saturated Model \#2:} &n=250, p=3    \ \qquad \boldsymbol{\eta} = [12 \ -15 \ \ 0]^T\\
                        \text{Low-dim. Noise Model \#3:}        &n=250, p=3    \ \qquad \boldsymbol{\eta} = [ 0 \ \ 0 \ \ 0]^T\\
                        \text{High-dim. Un-saturated Model \#4:}&n=100, p=1000   \quad  \boldsymbol{\eta} = [\eta_1 \ldots \eta_{100} \ 0 \ldots \ 0]^T
                    \end{cases}
                \end{align}
                \vspace{-0.10in}
            \item In simulated model \#1b, a Low-dim. Saturated Model with $n=250$ and $p=3$, where
                \vspace{-0.10in}
                \begin{align}
                    \nonumber
                    \eta({\bf x}_{i}) =
                    \begin{cases}
                        \boldsymbol{\eta}^T {\bf x}_{i} \quad \text{with regression parameters} \ \boldsymbol{\eta} = [12 \ -15 \  -5]^T
                        &\quad \text{for} \quad {\bf x}_{i} \in R\\
                        {u}_{i} \sim U(0,1)
                        &\quad \text{for} \quad {\bf x}_{i} \notin R
                    \end{cases}
                \end{align}
                \qquad \ \ where $R = [0.7,1] \times [0,0.2] \times [0,0.4]$ is an arbitrary box in $\mathbb R^3$.
        \end{itemize}
    \vspace{-0.1in}
    \item using $K = 5$-fold cross-validation, $A = 1024$ for the permutation \emph{p}-values and $B = 128$ independent replications.
\end{itemize}
\spacingset{1}

%===================================
\subsection{Summary of Outputs} \label{outputs}
%===================================
We explain below how the main diagnostic and descriptive output plots are used.

\vspace{-0.10in}
\subsubsection{Cross-Validated Tuning Profiles}
\begin{figure}[!hbt]
    \centering\includegraphics[width=6in]{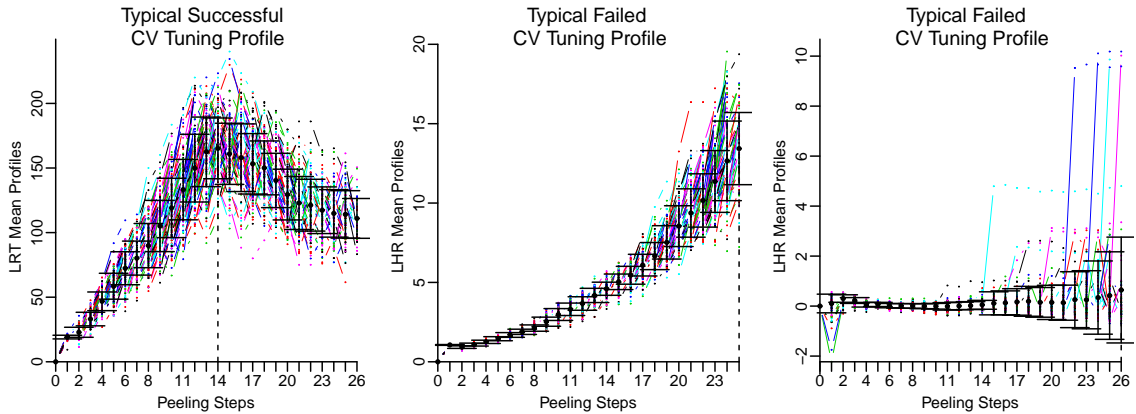}
    \caption{\sl \footnotesize Illustrations of typical successful (left) and failed (center and right) cross-validated tuning profiles of box end-point statistics. Left: Successful peeling stops with a ``Replicated CV'' optimal peeling length $\bar{L}^{rcv}$ (see eq. \ref{rep.optimizationeq}) reached within the $[1,\bar{L}_{m}^{rcv}]$ boundaries of possible peeling lengths; Center: Failure to reach a maximum before running out of data; Right: Failure to reach a reliable maximum because of a flat profile. The ``Replicated CV'' optimal peeling length $\bar{L}^{rcv}$ of the peeling trajectory is shown in each plot with the vertical black dashed line. Each colored profile corresponds to one of the replications ($B = 128$). The cross-validated mean profile of the statistic used in the optimization criterion is shown by the dotted black line with standard error of the sample mean.}
    \label{Figure04}
\end{figure}

\vspace{-0.10in}
Cross-validated tuning profiles plot values of the box end-points statistics (section \ref{end-points}) Log Hazard Ratio (\emph{LHR}), Log-Rank Test (\emph{LRT}) or Concordance Error Rate (\emph{CER}), depending on the optimization criterion chosen, as a function of peeling length or peeling steps of the peeling trajectory (model complexity). A peeling step includes step \#0 corresponding to the situation where the starting box covers the entire test-set data $\mathcal{L}_{k}$ before peeling (Algorithm \ref{algo}). These statistics are used internally or interactively to get the ``Replicated CV'' optimal length of the peeling trajectory: $\bar{L}^{rcv}$ (section \ref{replication}). In order to successfully determine the profiles minimizer or maximizer (section \ref{optimization}), the cross-validated tuning profile should be approximately non-monotone up to sampling variability (Figure \ref{Figure04}). In addition, one expects an inflation of variance of cross-validated point estimates towards the right-end of the cross-validated tuning profile corresponding to an increase in overfitting and model uncertainty for more complex models (Figure \ref{Figure04}, Supporting\_Figures \ref{SuppFigure01}, \ref{SuppFigure02}, \ref{SuppFigure03}).

The choice of the optimization criteria for controlling the peeling length is crucial. Two typical situations of failure of cross-validation can happen from the cross-validated tuning profiles of the box end-point statistics: either an extremum cannot be reached in the profile before the peeling sequence runs out of data, or the profile is essentially flat due to noise or the absence of any effect in the data (Figure \ref{Figure01}). In the former case, this results in excessive peeling steps and cross-validated values of optimal peeling lengths $\bar{L}^{rcv}$ (eq. \ref{rep.optimizationeq}) at, or near, the maximal peeling lengths $\bar{L}_{m}^{rcv}$ (eq. \ref{rep.Lm}). In the latter case, this results in un-reliable optimal peeling lengths $\bar{L}^{rcv}$ that can take any value between the $[1,\bar{L}_{m}^{rcv}]$ boundaries. In both cases, this leads to likely under-fitted or over-fitted models (see asterisk-annotated models in Table \ref{Table01}).

\vspace{-0.10in}
\subsubsection{Peeling Trajectories}
\vspace{-0.10in}
Cross-validated peeling trajectories are estimated by step functions of the covariates box cuts as a function of box support/mass (Figures \ref{Figure05}, \ref{Figure07}). They are read from right to left as they track the top-down peeling of the box induction process (peeling loop) of our ``Patient Recursive Survival Peeling'' method (Algorithm \ref{algo}). These peeling trajectories are, up to sampling variability:
\spacingset{0}
\begin{itemize}
    \vspace{-0.1in}
    \item Monotone (increasing or decreasing) functions for each input covariate ${\bf x}_{j}$, for $j \in \{1,\dots,p\}$.
    \vspace{-0.1in}
    \item Non-monotone (increasing then decreasing) functions for \emph{LHR} $\bar{\lambda}^{rcv}(l)$.
    \vspace{-0.1in}
    \item Non-monotone (increasing then decreasing) functions for \emph{LRT} $\bar{\chi}^{rcv}(l)$.
    \vspace{-0.1in}
    \item Non-monotone (decreasing then increasing) functions of \emph{CER} $\bar{\theta}^{rcv}(l)$.
    \vspace{-0.1in}
    \item Monotone decreasing functions for \emph{MEFP} $\widebar{P_{0}^{\prime}}^{rcv}(l)$.
    \vspace{-0.1in}
    \item Monotone decreasing functions for \emph{MEFT} $\widebar{T_{0}^{\prime}}^{rcv}(l)$.
\end{itemize}
\spacingset{1}

\vspace{-0.10in}
\subsubsection{Trace Curves}
\vspace{-0.10in}
Cross-validated trace curves of covariate importance and covariate usage are estimated by piece-wise linear and step functions, respectively, as a function of box support/mass (Figures \ref{Figure06}, \ref{Figure08}). Similarly to peeling trajectories, they are read from right to left. Trace curves of covariate importance show on a single plot: (i) the amplitude of used covariates, (ii) the order (prioritization) with which these covariates are used, and (iii) the extent of the number of peeling steps by which each covariate is used. Covariate traces are reminiscent of the concept of variable selection from the fields of decision tree and regularization, that is:
\spacingset{0}
\begin{itemize}
    \item In ``Variable Importance'', a prediction-based statistics borrowed from the existing theory of decision trees \cite{Breiman_1984} and their ensemble version \cite{Breiman_2001}.
    \vspace{-0.1in}
    \item In ``Selective Shrinkage'' of variable coefficients/parameters from the existing theory of regularization and variable selection (e.g. LARS \cite{Efron_2004b}, Lasso \cite{Tibshirani_1996}, Elastic Net \cite{Zou_2005a} and Spike \& Slab \cite{Ishwaran_2005b}).
\end{itemize}
\spacingset{1}

\vspace{-0.10in}
\subsubsection{Survival Curves}
\vspace{-0.10in}
Each subplot of Figure \ref{Figure09}, \ref{Figure11} and \ref{Figure13} corresponds to a peeling step of our Patient Recursive Survival Peeling method for a tested model and cross-validation technique (including none). Each subplot shows cross-validated Kaplan--Meir estimates of the survival functions, as a function of survival time, of both ``in-box'' (red) and ``out-of-box'' (black) samples, corresponding respectively to the high-risk and low-risk groups. Each subplot also displays the corresponding step number along with cross-validated Log Hazard Ratio (\emph{LHR}), Log-Rank Test (\emph{LRT}) and log-rank permutation $p$-value $\tilde{p}^{cv}(l)$ of survival distribution separation (see section \ref{cvpval}). A single survival curve always exists at Step \#0 corresponding to the situation where the starting box covers the entire test-set data $\mathcal{L}_{k}$ before peeling (Algorithm \ref{algo}). As the peeling progresses, the survival curves of ``in-box'' and ``out-of-box'' samples further separate until the peeling stops.

%===================================
\subsection{Effect of Model Tuning} \label{building}
%===================================
\subsubsection{Effect of the Optimization Criteria} \label{optcrit}
\vspace{-0.10in}
We first compare the effect of the three optimization criteria (Log Hazard Ratio \emph{LHR}, Log-Rank Test \emph{LRT} or Concordance Error Rate \emph{CER} - section \ref{optimization}) used for model tuning and selection. Evaluations are reported on (i) the ``Replicated CV'' optimal peeling lengths $\bar{L}^{rcv}$ (eq. \ref{rep.optimizationeq}) obtained from the cross-validated tuning profiles of the box end-point statistics (Log Hazard Ratio \emph{LHR}, Log-Rank Test \emph{LRT} or Concordance Error Rate \emph{CER}), and (ii) on the cross-validated numbers of used covariates by the PRSP algorithm (see Algorithm \ref{algo}) out of the total number of pre-selected ones. Results are presented in Table \ref{Table01}, Supporting\_Table \ref{SuppTable01} and Supporting\_Figures \ref{SuppFigure01}, \ref{SuppFigure02}, \ref{SuppFigure03} for the three peeling criteria used (Log Hazard Ratio \emph{LHR}, Log-Rank Test \emph{LRT} or Cumulative Hazard Summary \emph{CHS} - section \ref{criterion}) as well as for our two cross-validation techniques (``Replicated Averaged CV'' (RACV) vs. ``Replicated Combined CV'' (RCCV)), whether in low- or high-dimensional simulated survival regression models \#1, \#2, \#3 and \#4.

\begin{table}[!hbt]
    \caption{\sl \footnotesize Effect of peeling and optimization criteria as well as cross-validation techniques on the cross-validated tuning profiles of the box end-point statistics (Log Hazard Ratio LHR, Log-Rank Test LRT or Concordance Error Rate CER) and the resulting ``Replicated CV'' optimal peeling length $\bar{L}^{rcv}$ (see eq. \ref{rep.optimizationeq}). Cross-validated optimal peeling lengths $\bar{L}^{rcv}$  are reported for the combined effects of: (i) the three peeling criteria (by rows: Log Hazard Ratio (LHR), Log-Rank Test (LRT) or Cumulative Hazard Summary (CHS)), (ii) the three optimization criteria (by columns: Log Hazard Ratio (LHR), Log-Rank Test (LRT) or Concordance Error Rate (CER)), (iii) the two cross-validation techniques (by columns: ``Replicated Averaged CV'' or RACV and ``Replicated Combined CV'' or RCCV), and (iv) the four tested simulation models (by rows: Model \#1, \#2, \#3 or \#4). Asterisks denote situations where $\bar{L}^{rcv}$ reaches either a (quasi-)minimal or (quasi-)maximal value of optimal peeling lengths, corresponding to failed cross-validations / likely under-fitted or over-fitted models.}
    \begin{center}
        \footnotesize
        \begin{tabular}{llrr@{}rrr@{}rrr}
            \hline
            \hline
            \ Model \#1                                                                                            \\
            \hline
            \hline
            \                & &\multicolumn{8}{c}{Optimization Criterion}                                         \\
            \cline{3-10}
            \                & &\multicolumn{2}{c}{$LHR$} & &\multicolumn{2}{c}{$LRT$} & &\multicolumn{2}{c}{$CER$}\\
            \cline{3-4} \cline{6-7} \cline{9-10}
            \                &            &RACV &RCCV                & &RACV &RCCV                & &RACV &RCCV    \\
            \hline
            \Lower{Peeling}  &$LHR$       &26*  &25*                 & &20   &20                  & &10   &11      \\
            \Lower{Criterion}&$LRT$       &26*  &25*                 & &17   &20                  & &10   &10      \\
            \                &$CHS$       &26*  &26*                 & &14   &14                  & &09   &09      \\
            \hline
            \\[-1pt]
            \hline
            \hline
            \ Model \#2                                                                                            \\
            \hline
            \hline
            \                & &\multicolumn{8}{c}{Optimization Criterion}                                         \\
            \cline{3-10}
            \                & &\multicolumn{2}{c}{$LHR$} & &\multicolumn{2}{c}{$LRT$} & &\multicolumn{2}{c}{$CER$}\\
            \cline{3-4} \cline{6-7} \cline{9-10}
            \                &            &RACV &RCCV                & &RACV &RCCV                & &RACV &RCCV    \\
            \hline
            \Lower{Peeling}  &$LHR$       &26*  &25*                 & &17   &17                  & &12   &12      \\
            \Lower{Criterion}&$LRT$       &26*  &24*                 & &10   &11                  & &10   &10      \\
            \                &$CHS$       &26*  &26*                 & &14   &14                  & &10   &10      \\
            \hline
            \\[-1pt]
            \hline
            \hline
            \ Model \#3                                                                                            \\
            \hline
            \hline
            \                & &\multicolumn{8}{c}{Optimization Criterion}                                         \\
            \cline{3-10}
            \                & &\multicolumn{2}{c}{$LHR$} & &\multicolumn{2}{c}{$LRT$} & &\multicolumn{2}{c}{$CER$}\\
            \cline{3-4} \cline{6-7} \cline{9-10}
            \                &            &RACV &RCCV                & &RACV &RCCV                & &RACV &RCCV    \\
            \hline
            \Lower{Peeling}  &$LHR$       &23*  &01                  & &23*  &01                  & &05   &02      \\
            \Lower{Criterion}&$LRT$       &26*  &02                  & &26*  &02                  & &03   &02      \\
            \                &$CHS$       &24*  &01                  & &23*  &02                  & &04   &02      \\
            \hline
            \\[-1pt]
            \hline
            \hline
            \ Model \#4                                                                                            \\
            \hline
            \hline
            \                & &\multicolumn{8}{c}{Optimization Criterion}                                         \\
            \cline{3-10}
            \                & &\multicolumn{2}{c}{$LHR$} & &\multicolumn{2}{c}{$LRT$} & &\multicolumn{2}{c}{$CER$}\\
            \cline{3-4} \cline{6-7} \cline{9-10}
            \                &            &RACV &RCCV                & &RACV &RCCV                & &RACV &RCCV    \\
            \hline
            \Lower{Peeling}  &$LHR$       &17*  &09                  & &16*   &06*                & &05   &05      \\
            \Lower{Criterion}&$LRT$       &16*  &01*                 & &16*   &08*                & &06   &08      \\
            \                &$CHS$       &16*  &01*                 & &16*   &08*                & &05   &05      \\
            \hline
        \end{tabular}
    \end{center}
    \label{Table01}
\end{table}

From Table \ref{Table01} and Supporting\_Figures \ref{SuppFigure01}, \ref{SuppFigure02}, \ref{SuppFigure03}, it results that both Log-Rank Test (\emph{LRT}) and Concordance Error Rate (\emph{CER}) optimization criteria give satisfactory results in low-dimensional simulated models (\#1 and \#2), other than the simulated noise model (\#3), where the peeling length is optimally pruned ($\bar{L}^{rcv}=9-20$). This is in sharp contrast to the Log Hazard Ratio (\emph{LHR}) optimization criterion that fails to control the peeling length, regardless of the cross-validation technique or the peeling criterion used ($\bar{L}^{rcv}=26-27$).

The situation differs in high-dimensional simulated models: the Concordance Error Rate (\emph{CER}) appears to be the only optimization criterion that reliably controls the peeling length in simulated model \#4, regardless of the peeling criterion or cross-validation technique used (Table \ref{Table01}, Supporting\_Figures \ref{SuppFigure01}, \ref{SuppFigure02}, \ref{SuppFigure03}).

Finally, note that the Concordance Error Rate \emph{CER}) optimization criterion tends to yield slightly more conservative results than the Log-Rank Test (\emph{LRT}) in terms of peeling length (Table \ref{Table01}, Supporting\_Figures \ref{SuppFigure01}, \ref{SuppFigure02}, \ref{SuppFigure03}) and number of used covariates (Supporting\_Table \ref{SuppTable01}). Also, note that the Concordance Error Rate \emph{CER}) has systematically less variance than the other Log-Rank Test (\emph{LRT}) and Log Hazard Ratio (\emph{LHR}) optimization criteria (Table \ref{Table01}, Supporting\_Figures \ref{SuppFigure01}, \ref{SuppFigure02}, \ref{SuppFigure03}).

For the above reasons, we recommend using the Concordance Error Rate \emph{CER} as optimization criterion in every situation or the Log-Rank Test \emph{LRT} in low-dimensional situation only.

\vspace{-0.10in}
\subsubsection{Effect of the Peeling Criteria} \label{peelcrit}
\vspace{-0.10in}
Overall, in all simulation models tested, the effect of the peeling criteria, used for model fitting of the survival bump hunting model, is relatively marginal compared to that of the optimization criterion and/or cross-validation technique used for model tuning.

However, for any combination of the optimization criterion and cross-validation technique used, both Log-Rank Test \emph{LRT} and Cumulative Hazard Summary \emph{CHS} peeling criteria tend to induce slightly shorter profiles and use less covariates than the Log Hazard Ratio \emph{LHR} criterion  (Table \ref{Table01}, Supporting\_Table \ref{SuppTable01}, Supporting\_Figures \ref{SuppFigure01}, \ref{SuppFigure02}, \ref{SuppFigure03}). Moreover, the Cumulative Hazard Summary \emph{CHS} peeling criterion tends to be slightly more conservative than the Log-Rank Test \emph{LRT} (and the Log Hazard Ratio \emph{LHR}) in terms of peeling length and number of used covariates, especially in high-dimensional models (Table \ref{Table01}, Supporting\_Table \ref{SuppTable01}, Supporting\_Figures \ref{SuppFigure01}, \ref{SuppFigure02}, \ref{SuppFigure03}).

So, we recommend using primarily the Cumulative Hazard Summary \emph{CHS} and secondly the Log-Rank Test \emph{LRT} as peeling criterion (in low or high-dimensional data) to reduce the risk of over-fitting (or conversely the Log Hazard Ratio \emph{LHR} to avoid excessive conservativeness).

\vspace{-0.10in}
\subsubsection{Effect of the Cross-Validation Technique} \label{cvtech}
\vspace{-0.10in}
The difference of results between cross-validation techniques (including none) is striking in simulated noise model \#3. Here, only the ``Replicated Combined CV'' (RCCV) cross-validation technique gives satisfactory results, regardless of the optimization or peeling criterion used. As expected in this situation, the model is extensively pruned ($\bar{L}^{rcv}=1-2$) using RCCV. The same consistency is not observed for ``Replicated Averaged CV'' (RACV) that fails to properly control the peeling length when the other Log-Rank Test (\emph{LRT}) or Log Hazard Ratio (\emph{LHR}) optimization criteria are used ($\bar{L}^{rcv} \approx \bar{L}_{m}^{rcv}$) (Table \ref{Table01}, Supporting\_Figures \ref{SuppFigure01}, \ref{SuppFigure02}, \ref{SuppFigure03}). Differences between cross-validation techniques are also significant in high-dimensional simulated models (Table \ref{Table01}, Supporting\_Figures \ref{SuppFigure01}, \ref{SuppFigure02}, \ref{SuppFigure03}).

\begin{figure}[!hbt]
    \centering\includegraphics[width=4in]{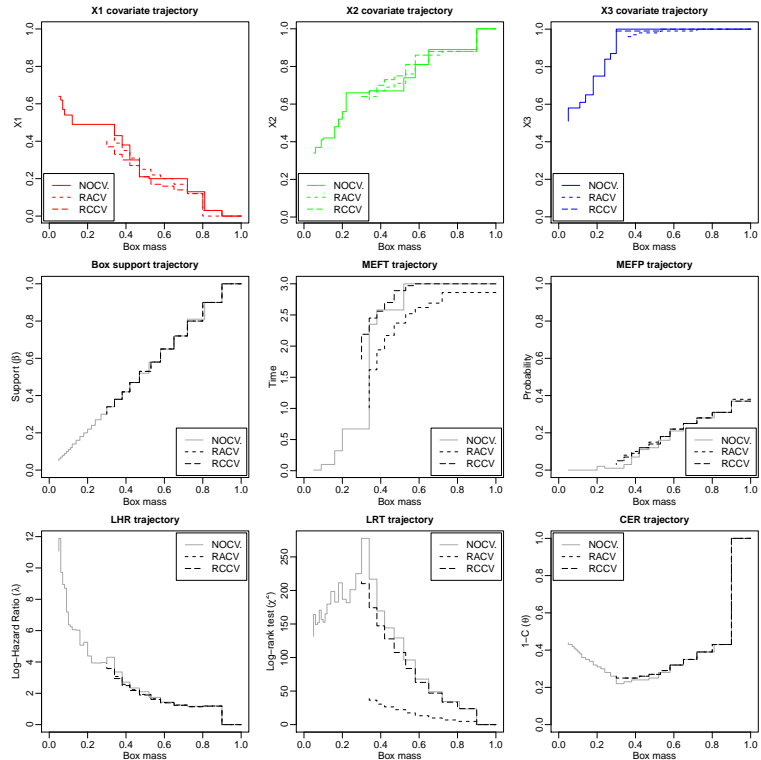}
    \caption{\sl \footnotesize Comparison of cross-validated peeling trajectories between situations when either cross-validation technique ``Replicated Combined CV'' (RCCV) or ``Replicated Averaged CV'' (RACV) and no cross-validation (NOCV) was done. Results are for simulated model \#2 and the LRT statistic used in both peeling and optimization criteria. Compare the trajectory lengths between either cross-validation technique and in the absence of either one. Notice also the flat trajectory profile of covariate ${\bf x}_{3}$ in the presence of either cross-validation technique (RACV or RCCV) as opposed to the situation where no cross-validation (NOCV) was done.}
    \label{Figure05}
\end{figure}

\begin{figure}[!hbt]
    \centering\includegraphics[width=3in]{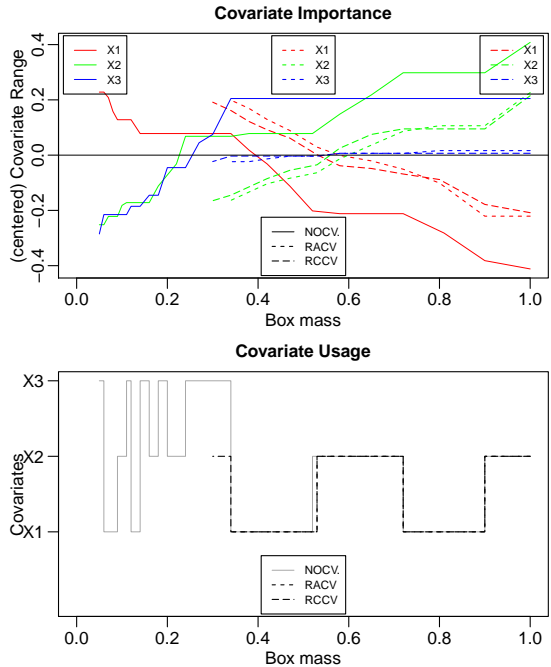}
    \caption{\sl \footnotesize Comparison of cross-validated trace plots of covariate importance $\widebar{VI}(l)$ (top) and covariate usage $\widebar{VU}(l)$ (bottom) between situations when either cross-validation technique ``Replicated Combined CV'' (RCCV) or ``Replicated Averaged CV'' (RACV) and no cross-validation (NOCV) was done. Results are for simulated model \#2 and the LRT statistic used in both peeling and optimization criteria. Compare the trace lengths between either cross-validation technique and in the absence of either one. Notice also the flat trace of covariate ${\bf x}_{3}$ about 0 in the presence of either cross-validation technique (RACV or RCCV) as opposed to the situation where no cross-validation (NOCV) was done.}
    \label{Figure06}
\end{figure}

Using from now on a given peeling and optimization criterion in low-dimensional simulated models, we compare the performance of our two cross-validation techniques with each other (RACV vs. RCCV) and with the situation of no cross-validation (NOCV). Peeling trajectory and covariate usage/importance results are shown for model \#2 where it is possible to specifically assess effects of cross-validation on a noise/random covariate (${\bf x}_{3}$). Figures \ref{Figure05} and \ref{Figure06} show peeling trajectory profiles and covariate traces for model \#2. Table \ref{Table02} gives the corresponding rules.

Clearly, both cross-validation techniques are effective in terms of (i) smoothing peeling trajectories out and (ii) pruning peeling trajectories off. Compare for instance results of simulation model \#2: $\bar{L}^{rcv} = 26$ without cross-validation (NOCV), $\bar{L}^{rcv} = 10$ with RACV and $\bar{L}^{rcv} = 11$ with RCCV (Figures \ref{Figure05} and \ref{Figure06}, Table \ref{Table02}). In fact, all cross-validated trajectory profiles in Figure \ref{Figure05} and covariate traces in Figure \ref{Figure06} stop at $\bar{\beta}^{rcv}(l=11) \lessapprox 0.30$ for RCCV and $\bar{\beta}^{rcv}(l=10) \lessapprox 0.34$ for RACV as compared to $\bar{\beta}^{rcv}(l=27) \lessapprox 0.05$ in the absence of cross-validation (NOCV).

Notice from the peeling trajectories and trace plots of simulation model \#2 (compared to \#1) how a noise/random covariate (${\bf x}_{3}$) is effectively eliminated from the model after using either cross-validation technique (RCCV or RACV), while it is not in the absence of cross-validation (NOCV). In fact, ${\bf x}_{3}$'s RCCV and RACV peeling trajectories are mostly flat, that is, ${\bf x}_{3}$ is unused in the decision rule (blue dashed curves in Figure \ref{Figure05}). Consistently, ${\bf x}_{3}$'s RCCV and RACV covariate importance trace plots are mostly flat (top of Figure \ref{Figure06}) and ${\bf x}_{3}$'s RCCV and RACV covariate usage trace plots show that ${\bf x}_{3}$ is not used at all (bottom of Figure \ref{Figure06}). Similar conclusions are drawn with respect to simulated model \#3 (compared to \#1).

The non-monotone behavior of the \emph{LRT} and the overly large \emph{LHR} values obtained in the non-cross-validated (NOCV) results of simulated model \#2 ($\bar{\lambda}^{rcv}(l=26) = 11.08$) clearly reflect over-fitting and sub-optimal models. This is evident when comparing to the much more conservative values obtained from the corresponding cross-validated peeling profiles: $\bar{\lambda}^{rcv}(l=11) = 3.90$ for RCCV and $\bar{\lambda}^{rcv}(l=10) = 3.76$ for RACV (Figure \ref{Figure05} and Table \ref{Table02}). This non-monotone behavior of \emph{LRT} peeling profile is precisely what allows us to use it in the optimization criterion. We suggest that this could be due to a greater sensitivity of \emph{LRT} to small sample sizes at deep peeling steps.

\begin{table}[!hbt]
 \caption{\sl \footnotesize Comparison of cross-validated decision rules (upper Table) and box end points statistics of interest (lower Table) between situations when either cross-validation technique ``Replicated Combined CV'' (RCCV) or ``Replicated Averaged CV'' (RACV) and no cross-validation (NOCV) was done.  For conciseness, only the initial and final decision rules ($\bar{L}^{rcv}$th step) are shown.  Values are sample mean estimates with corresponding standard errors in parenthesis (NA in the case of NOCV, where no replication was performed - see manual of R package \pkg{PRIMsrc} for details \cite{Dazard_2015a}). Step \#0 corresponds to the situation where the starting box covers the entire test-set data $\mathcal{L}_{k}$ before peeling. Results are for simulated model \#2 and the LRT statistic used in both peeling and optimization criteria. Notice the non-usage of covariate ${\bf x}_{3}$ in the presence of either cross-validation technique (RACV or RCCV) as opposed to the situation where no cross-validation (NOCV) was done.}
 \footnotesize
 \begin{tabular}{lrccc}
    \hline
    \hline
    \hspace{-0.2in}               &Step $l$  &${\bf x}_{1}$                         &${\bf x}_{2}$                         &${\bf x}_{3}$                       \\
    \hline
                 \hspace{-0.2in}  &$ 0$      &${\bf x}_{1} \ge 0.00 \ (0.00)$       &${\bf x}_{2} \le 1.00 \ (0.00)$       &${\bf x}_{3} \le 1.00 \ (0.00)$     \\
    \Lower{RCCV} \hspace{-0.2in}  &$ 1$      &${\bf x}_{1} \ge 0.03 \ (0.00)$       &${\bf x}_{2} \le 0.88 \ (0.01)$       &${\bf x}_{3} \le 1.00 \ (0.00)$     \\
                 \hspace{-0.2in}  &$\vdots$  &$\vdots$                              &$\vdots$                              &$\vdots$                            \\
                 \hspace{-0.2in}  &$11$      &${\bf x}_{1} \ge 0.40 \ (0.06)$       &${\bf x}_{2} \le 0.62 \ (0.04)$       &${\bf x}_{3} \le 0.97 \ (0.05)$     \\
    \hline
                 \hspace{-0.2in}  &$ 0$      &${\bf x}_{1} \ge 0.00 \ (0.00)$       &${\bf x}_{2} \le 1.00 \ (0.00)$       &${\bf x}_{3} \le 1.00 \ (0.00)$     \\
    \Lower{RACV} \hspace{-0.2in}  &$ 1$      &${\bf x}_{1} \ge 0.00 \ (0.00)$       &${\bf x}_{2} \le 0.88 \ (0.00)$       &${\bf x}_{3} \le 1.00 \ (0.00)$     \\
                 \hspace{-0.2in}  &$\vdots$  &$\vdots$                              &$\vdots$                              &$\vdots$                            \\
                 \hspace{-0.2in}  &$10$      &${\bf x}_{1} \ge 0.42 \ (0.03)$       &${\bf x}_{2} \le 0.61 \ (0.02)$       &${\bf x}_{3} \le 0.96 \ (0.03)$     \\
    \hline
                 \hspace{-0.2in}  &$ 0$      &${\bf x}_{1} \ge 0.00 \ (\text{NA})$  &${\bf x}_{2} \le 1.00 \ (\text{NA})$  &${\bf x}_{3} \le 1.00 \ (\text{NA})$\\
    \Lower{NOCV} \hspace{-0.2in}  &$ 1$      &${\bf x}_{1} \ge 0.03 \ (\text{NA})$  &${\bf x}_{2} \le 0.89 \ (\text{NA})$  &${\bf x}_{3} \le 1.00 \ (\text{NA})$\\
                 \hspace{-0.2in}  &$\vdots$  &$\vdots$                              &$\vdots$                              &$\vdots$                            \\
                 \hspace{-0.2in}  &$26$      &${\bf x}_{1} \ge 0.64 \ (\text{NA})$  &${\bf x}_{2} \le 0.34 \ (\text{NA})$  &${\bf x}_{3} \le 0.51 \ (\text{NA})$\\
    \hline
 \end{tabular}\\
 \begin{tabular}{lrccccccc}
        \hline
        \hline
        \hspace{-0.2in}               &Step $l$  &$n(l)$        &$\bar{\beta}^{rcv}(l)$  &$\widebar{T_{0}^{\prime}}^{rcv}(l)$  &$\widebar{P_{0}^{\prime}}^{rcv}(l)$  &$\bar{\lambda}^{rcv}(l)$  &$\bar{\chi}^{rcv}(l)$  &$\bar{\theta}^{rcv}(l)$\\
        \hline
                     \hspace{-0.2in}  &$ 0$      &250 \ (0.00)  &1.00 \ (0.00)           &3.00 \ (0.00)                        &0.37 \ (0.00)                        & 0.00 \ (0.00)            &  0.00 \  (0.00)     &1.00 \ (0.00)        \\
        \Lower{RCCV} \hspace{-0.2in}  &$ 1$      &225 \ (0.00)  &0.90 \ (0.00)           &3.00 \ (0.00)                        &0.31 \ (0.01)                        & 0.18 \ (0.02)            & 23.86 \  (1.14)     &0.43 \
                     (0.00)        \\
                     \hspace{-0.2in}  &$\vdots$  &$\vdots$      &$\vdots$                &$\vdots$                             &$\vdots$                             &$\vdots$                  &$\vdots$               &$\vdots$               \\
                     \hspace{-0.2in}  &$11$      & 75 \ (2.50)  &0.30 \ (0.01)           &1.79 \ (0.71)                        &0.03 \ (0.02)                        & 3.90 \ (0.46)            &214.61 \ (33.56)     &0.26 \ (0.01)        \\
        \hline
                     \hspace{-0.2in}  &$ 0$      &250 \ (0.00)  &1.00 \ (0.00)           &2.86 \ (0.04)                        &0.38 \ (0.02)                        & 0.00 \ (0.00)            &  0.00 \  (0.00)     &1.00 \ (0.00)        \\
        \Lower{RACV} \hspace{-0.2in}  &$ 1$      &225 \ (2.50)  &0.90 \ (0.01)           &2.86 \ (0.05)                        &0.31 \ (0.02)                        & 1.19 \ (0.02)            &  4.84 \  (0.27)     &0.43 \
                     (0.01)        \\
                     \hspace{-0.2in}  &$\vdots$  &$\vdots$      &$\vdots$                &$\vdots$                             &$\vdots$                             &$\vdots$                  &$\vdots$               &$\vdots$               \\
                     \hspace{-0.2in}  &$10$      & 85 \ (2.50)  &0.34 \ (0.01)           &1.01 \ (0.34)                        &0.05 \ (0.02)                        & 3.76 \ (0.41)            & 43.64 \  (6.01)     &0.25 \ (0.02)        \\
        \hline
                     \hspace{-0.2in}  &$ 0$      &250 \ (NA)    &1.00 \ (NA)             &3.00 \ (NA)                          &0.37 \ (NA)                          & 0.00 \ (NA)              &  0.00 \ (NA)        &1.00 \ (NA)           \\
        \Lower{NOCV} \hspace{-0.2in}  &$ 1$      &225 \ (NA)    &0.90 \ (NA)             &3.00 \ (NA)                          &0.31 \ (NA)                          & 1.19 \ (NA)              & 23.43 \ (NA)        &0.43 \
                     (NA)           \\
                     \hspace{-0.2in}  &$\vdots$  &$\vdots$      &$\vdots$                &$\vdots$                             &$\vdots$                             &$\vdots$                  &$\vdots$               &$\vdots$               \\
                     \hspace{-0.2in}  &$26$      & 12 \ (NA)    &0.05 \ (NA)             &0.01 \ (NA)                          &0.00 \ (NA)                          &11.08 \ (NA)              &131.73 \ (NA)        &0.44 \ (NA)           \\
        \hline
 \end{tabular}
 \label{Table02}
\end{table}

To further compare cross-validation techniques, we generated empirical distributions of various cross-validated estimates of box decision rules and box survival end-points/prediction statistics (section \ref{end-points}) for each technique and end-point as a function of peeling steps. Distributions were obtained by generating $B = 128$ Monte-Carlo simulated datasets according to simulated model \#1, where the \emph{LRT} statistic was used in both peeling and optimization criteria. The replication design accounts for two folds of variability: the one due to random splitting by cross-validation and the one due to sampling from the simulated model. Then, Box Coefficient of Variation (BCV) of decision rules (as defined in \cite{Dazard_2010a}) and coefficient of variation of box survival end-points/prediction statistics were computed and plotted as a function of peeling steps. Here, cross-validated coefficient of variation profiles do not show a consistent advantage of one cross-validation technique over the other considering all end-point analyzed. This remains true for a range of realistic sample sizes $n \in \{50, 100, 200\}$ (Supporting\_Figure \ref{SuppFigure04}).

Overall, our two cross-validation techniques, although not equivalent in design, give similar results on most profiles for the sample size and simulation models tested, confirming that both techniques are appropriate to the task in most situations. However, ``Replicated Averaged CV'' (RACV) appears to be less conservative than ``Replicated Combined CV'' (RCCV), especially in high-dimensional settings, which could be a problem if ones cares about reducing the risk of over-fitting. Also, RACV failed to prune simulated model \#3 (Table \ref{Table01} and Supporting\_Figures \ref{SuppFigure01}, \ref{SuppFigure02}, \ref{SuppFigure03}). This indicates that our RCCV cross-validation technique is more robust in noisy situations, possibly because RCCV uses larger test-set samples of size $n$ to make estimations than RACV, which uses test-set samples of size $n^{t} \approx n/K$ (section \ref{techniques}). For these reasons, we recommend using RCCV preferably to RACV.

\vspace{-0.10in}
\subsubsection{Comparison Between Simulated Survival Models}
\begin{figure}[!hbt]
  \centering\includegraphics[width=4in]{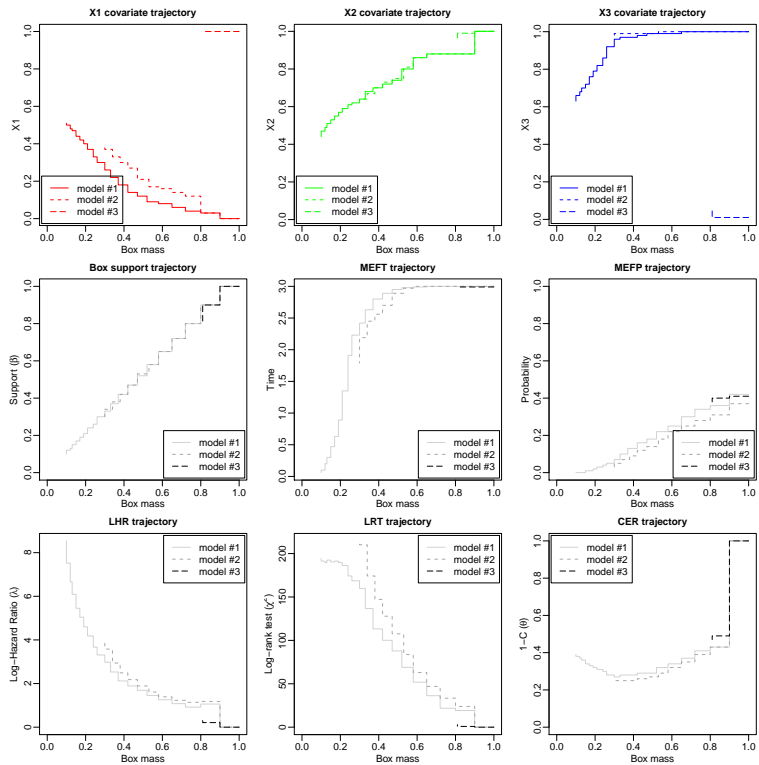}
  \caption{\sl \footnotesize Comparison of replicated combined cross-validated results for the peeling trajectories between simulated models \#1, \#2 and \#3 for the ``Replicated Combined CV'' (RCCV) technique and the LRT statistic used in both peeling and optimization criteria. Notice the usage of all covariates $({\bf x}_{1},{\bf x}_{2},{\bf x}_{3})$ in model \#1 as opposed to the selective usage of covariates $({\bf x}_{1},{\bf x}_{2})$ in model \#2 and the abortive usage of all covariates in noise model \#3.}
  \label{Figure07}
\end{figure}

\begin{figure}[!hbt]
  \centering\includegraphics[width=3in]{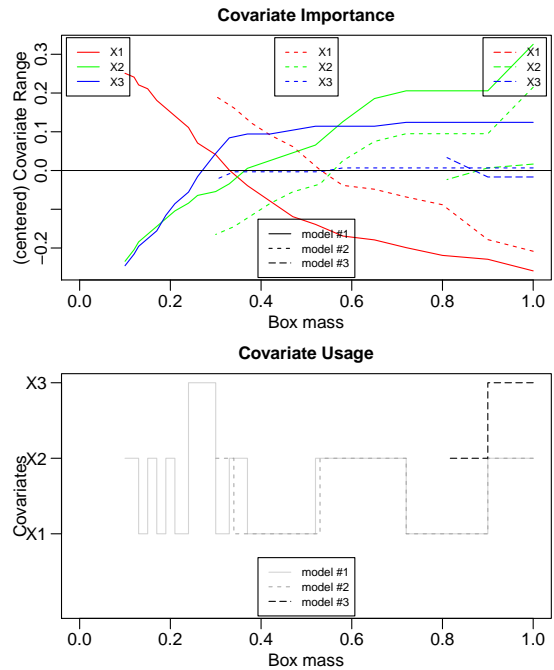}
  \caption{\sl \footnotesize Comparison of replicated combined cross-validated trace plots of covariate importance $\widebar{VI}(l)$ (top) and covariate usage $\widebar{VU}(l)$ (bottom) between simulated models \#1, \#2 and \#3 for the ``Replicated Combined CV'' (RCCV) technique and the LRT statistic used in both peeling and optimization criteria. Notice the usage of all covariates $({\bf x}_{1},{\bf x}_{2},{\bf x}_{3})$ in model \#1 as opposed to the selective usage of covariates $({\bf x}_{1},{\bf x}_{2})$ in model \#2 and the abortive usage of all covariates in noise model \#3.}
  \label{Figure08}
\end{figure}

\begin{figure}[!hbt]
    \centering \rotatebox{-90}{\includegraphics[width=1.94in]{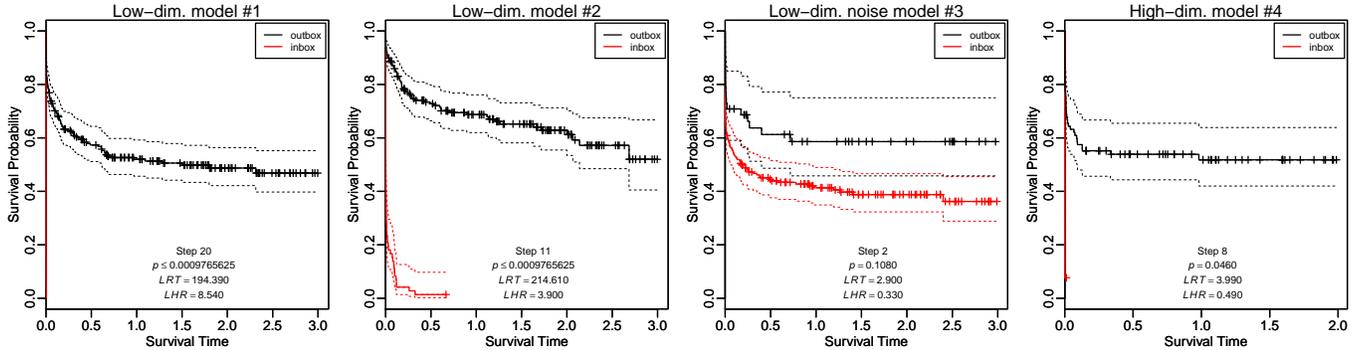}}
    \caption{\sl \footnotesize Comparison of cross-validated Kaplan--Meir survival probability curves of the high-risk (red curve ``in-box'') and low-risk (black curve ``out-of-box'') groups in simulated models \#1, \#2, \#3 and \#4. Results are for the ``Replicated Combined CV'' (RCCV) technique and the CHS statistic used as peeling criterion and CER used as optimization criteria. Left column: model \#1, middle column: model \#2, right column: model \#3. For conciseness, only the last peeling step of the peeling sequence is shown for each model. Cross-validated LRT, LHR and permutation p-values of ``in-box'' samples are shown at the bottom of the plot with the corresponding peeling step for each method. P-values $\hat{p}^{cv}(l) \le 9.7e-5$ correspond to 1/10th of the precision limit (see section \ref{cvpval}). Notice how the survival curves of ``in-box'' and ``out-of-box'' samples separates in models \#1, \#2 and \#4 in contrast to the overlapping situation in noise model \#3 with the corresponding significant and non-significant log-rank permutation $p$-value $\hat{p}^{cv}(l)$ of survival distribution separation.}
    \label{Figure09}
\end{figure}

\vspace{-0.10in}
In line with the above guidelines (sections \ref{peelcrit} and \ref{cvtech}), we used the following criteria and techniques in our numerical analyses for fitting and tuning/selecting our survival bump hunting model: (i) The Log-Rank Test \emph{LRT} both as peeling and optimization criterion in low-dimensional settings; (ii) The Cumulative Hazard Summary \emph{CHS} as peeling criterion and the Concordance Error Rate \emph{CER} as optimization criteria in high-dimensional settings; and our ``Replicated Combined Cross-Validation'' (RCCV) technique. We compared the performance of our Survival Bump Hunting procedure in terms of peeling trajectories (Figures \ref{Figure07}, \ref{Figure08}, Table \ref{Table03}, Supporting\_Figures \ref{SuppFigure05}, \ref{SuppFigure06}, Supporting\_Table \ref{SuppTable02}) and survival distribution curves (Figure \ref{Figure09}) between all our models.

Notice the striking differences in cross-validated peeling trajectories (Figure \ref{Figure07}) and covariate traces (Figure \ref{Figure08}): (i) when all covariates $({\bf x}_{1},{\bf x}_{2},{\bf x}_{3})$ are noise (model \#3), or (ii) when one covariate only (${\bf x}_{3}$) is noise (model \#2) or (iii) when all are informative (model \#1). As expected, \emph{all} peeling trajectories related to model \#3 are much shorter than in the other models, indicating an abortive PRSP procedure with little or no covariate usage during the peeling process (Figure \ref{Figure07}) nor involvement in the decision rule (Table \ref{Table03}). Similarly, one expects little or no usage of covariate ${\bf x}_{3}$ in models \#2 and \#3, as seen in their covariate peeling trajectories (Figure \ref{Figure07}, Table \ref{Table03}).

Consistent observations can be made from the cross-validated covariate importance and covariate usage traces of model \#3. In fact, ${\bf x}_{3}$'s covariate importance trace stops after only $\bar{L}^{rcv} = 2$ steps with box mass $\bar{\beta}^{rcv}(\bar{L}^{rcv}) \approx 0.81$ for model \#3, as compared to $\bar{L}^{rcv} = 20$ steps with $\bar{\beta}^{rcv}(\bar{L}^{rcv}) \approx 0.10$ for model \#1 and $\bar{L}^{rcv} = 11$ steps with $\bar{\beta}^{rcv}(\bar{L}^{rcv}) \approx 0.30$ for model \#2 (Table \ref{Table03} and Figure \ref{Figure08}). Also notice the limited usage of covariate ${\bf x}_{3}$ in the covariate usage traces of models \#2 and \#3 as compared to \#1 and the fact that all cross-validated peeling trajectories in model \#1 extend further than in other models \#2 and \#3 (Figure \ref{Figure07}, \ref{Figure08} and Table \ref{Table03}). This is consistent with our simulation design in that all covariates in regression model \#1 additively contribute to the hazards and to the separation of survival distributions.

\begin{table}[!hbt]
 \caption{\sl \footnotesize Comparison of cross-validated decision rules (upper Table) and box end points statistics of interest (lower Table) between simulated models \#1, \#2 and \#3 for the ``Replicated Combined CV'' (RCCV) technique and the LRT statistic used in both peeling and optimization criteria. For conciseness, only the initial and final decision rules ($\bar{L}^{rcv}$th step) are shown. Step \#0 corresponds to the situation where the starting box covers the entire test-set data $\mathcal{L}_{k}$ before peeling. Values are sample mean estimates with corresponding standard errors in parenthesis. Notice the usage of all covariates $({\bf x}_{1},{\bf x}_{2},{\bf x}_{3})$ in model \#1 as opposed to the selective usage of covariates $({\bf x}_{1},{\bf x}_{2})$ in model \#2 and the abortive usage of all covariates in noise model \#3.}
 \footnotesize
 \begin{tabular}{lrccc}
    \hline
    \hline
    \hspace{-0.2in}                    &Step $l$  &${\bf x}_{1}$                    &${\bf x}_{2}$                    &${\bf x}_{3}$                  \\
    \hline
                      \hspace{-0.2in}  &$ 0$      &${\bf x}_{1} \ge 0.00 \ (0.00)$  &${\bf x}_{2} \le 1.00 \ (0.00)$  &${\bf x}_{3} \le 1.00 \ (0.00)$\\
    \Lower{model \#1} \hspace{-0.2in}  &$ 1$      &${\bf x}_{1} \ge 0.03 \ (0.00)$  &${\bf x}_{2} \le 0.88 \ (0.00)$  &${\bf x}_{3} \le 1.00 \ (0.00)$\\
                      \hspace{-0.2in}  &$\vdots$  &$\vdots$                           &$\vdots$                           &$\vdots$                   \\
                      \hspace{-0.2in}  &$20$      &${\bf x}_{1} \ge 0.51 \ (0.04)$  &${\bf x}_{2} \le 0.44 \ (0.06)$  &${\bf x}_{3} \le 0.63 \ (0.08)$\\
    \hline
                      \hspace{-0.2in}  &$ 0$      &${\bf x}_{1} \ge 0.00 \ (0.00)$  &${\bf x}_{2} \le 1.00 \ (0.00)$  &${\bf x}_{3} \le 1.00 \ (0.00)$\\
    \Lower{model \#2} \hspace{-0.2in}  &$ 1$      &${\bf x}_{1} \ge 0.03 \ (0.00)$  &${\bf x}_{2} \le 0.88 \ (0.00)$  &${\bf x}_{3} \le 1.00 \ (0.00)$\\
                      \hspace{-0.2in}  &$\vdots$  &$\vdots$                           &$\vdots$                           &$\vdots$                   \\
                      \hspace{-0.2in}  &$11$      &${\bf x}_{1} \ge 0.40 \ (0.06)$  &${\bf x}_{2} \le 0.62 \ (0.04)$  &${\bf x}_{3} \le 0.97 \ (0.05)$\\
    \hline
                      \hspace{-0.2in}  &$ 0$      &${\bf x}_{1} \le 1.00 \ (0.00)$  &${\bf x}_{2} \le 1.00 \ (0.00)$  &${\bf x}_{3} \ge 0.01 \ (0.00)$\\
    model \#3         \hspace{-0.2in}  &$ 1$      &${\bf x}_{1} \le 1.00 \ (0.00)$  &${\bf x}_{2} \le 0.99 \ (0.01)$  &${\bf x}_{3} \ge 0.01 \ (0.01)$\\
                      \hspace{-0.2in}  &$ 2$      &${\bf x}_{1} \le 1.00 \ (0.00)$  &${\bf x}_{2} \le 0.96 \ (0.05)$  &${\bf x}_{3} \ge 0.06 \ (0.02)$\\
    \hline
 \end{tabular}\\
 \begin{tabular}{lrccccccc}
        \hline
        \hline
        \hspace{-0.2in}                    &Step $l$  &$n(l)$        &$\bar{\beta}^{rcv}(l)$  &$\widebar{T_{0}^{\prime}}^{rcv}(l)$  &$\widebar{P_{0}^{\prime}}^{rcv}(l)$  &$\bar{\lambda}^{rcv}(l)$ &$\bar{\chi}^{rcv}(l)$  &$\bar{\theta}^{rcv}(l)$\\
        \hline
                          \hspace{-0.2in}  &$ 0$      &250 \ (0.00)  &1.00 \ (0.00)           &3.00 \ (0.00)                        &0.42 \ (0.00)                        & 0.00 \ (0.00)           &  0.00 \  (0.00)     &1.00 \ (0.00)        \\
        \Lower{model \#1} \hspace{-0.2in}  &$ 1$      &225 \ (0.00)  &0.90 \ (0.00)           &3.00 \ (0.00)                        &0.36 \ (0.00)                        & 1.06 \ (0.03)           & 19.21 \  (1.13)     &0.43 \
                          (0.00)        \\
                          \hspace{-0.2in}  &$\vdots$  &$\vdots$      &$\vdots$                &$\vdots$                             &$\vdots$                             &$\vdots$                 &$\vdots$             &$\vdots$               \\
                          \hspace{-0.2in}  &$20$      & 25 \ (2.50)  &0.10 \ (0.01)           &0.06 \ (0.05)                        &0.00 \ (0.00)                        & 8.54 \ (1.29)           &194.39 \ (30.24)     &0.39 \ (0.01)        \\
        \hline
                          \hspace{-0.2in}  &$ 0$      &250 \ (0.00)  &1.00 \ (0.00)           &3.00 \ (0.00)                        &0.37 \ (0.00)                        & 0.00 \ (0.00)           &  0.00 \  (0.00)     &1.00 \ (0.00)        \\
        \Lower{model \#2} \hspace{-0.2in}  &$ 1$      &225 \ (0.00)  &0.90 \ (0.00)           &3.00 \ (0.00)                        &0.31 \ (0.00)                        & 1.18 \ (0.02)           & 23.86 \  (1.14)     &0.43 \
                          (0.00)        \\
                          \hspace{-0.2in}  &$\vdots$  &$\vdots$      &$\vdots$                &$\vdots$                             &$\vdots$                             &$\vdots$                 &$\vdots$             &$\vdots$               \\
                          \hspace{-0.2in}  &$11$      & 75 \ (2.50)  &0.30 \ (0.01)           &1.79 \ (0.71)                        &0.03 \ (0.71)                        & 3.90 \ (0.46)           &214.61 \ (33.56)     &0.26 \ (0.01)        \\
        \hline
                          \hspace{-0.2in}  &$ 0$      &250 \ (0.00)  &1.00 \ (0.00)           &2.99 \ (0.00)                        &0.41 \ (0.00)                        & 0.00 \ (0.00)           &0.00 \ (0.00)        &1.00 \ (0.00)        \\
        model \#3         \hspace{-0.2in}  &$ 1$      &225 \ (2.50)  &0.90 \ (0.01)           &2.99 \ (0.00)                        &0.40 \ (0.01)                        & 0.21 \ (0.14)           &0.86 \ (0.79)        &0.49 \
                          (0.01)        \\
                          \hspace{-0.2in}  &$ 2$      &202 \ (5.00)  &0.81 \ (0.02)           &2.99 \ (0.00)                        &0.38 \ (0.01)                        & 0.33 \ (0.11)           &2.90 \ (1.75)        &0.47 \ (0.01)        \\
        \hline
 \end{tabular}
 \label{Table03}
\end{table}

Finally, Figure \ref{Figure09} shows the cross-validated Kaplan--Meir survival probability curves of the highest-risk group vs. lower-risk group in all low- and high-dimensional simulated models with their corresponding log-rang permutation \emph{p}-values of survival distribution curve separation. The separation is especially evident in results of models \#1, \#2 and \#4 in contrast to the overlap seen in model \#3. The permutation \emph{p}-values are: $\tilde{p}^{cv}(l=20) \le 9.7e-5$, $\tilde{p}^{cv}(l=11) \le 9.7e-5$ and $\tilde{p}^{cv}(l=8) \approx 0.046$ for models \#1, \#2 and \#4 respectively, and $\tilde{p}^{cv}(l=2) \approx 0.1080$ for model \#3 (Figure \ref{Figure09}).

Overall, Figures \ref{Figure07}, \ref{Figure08}, \ref{Figure09}, Table \ref{Table03}, Supporting\_Figures \ref{SuppFigure05}, \ref{SuppFigure06} and Supporting\_Table \ref{SuppTable02} collectively support that our ``Replicated Combined CV'' (RCCV) cross-validation technique, when used with an appropriate combination of \emph{LRT} and \emph{CER} statistics as peeling and/or optimization criteria, is efficient at fitting and tuning a survival bump hunting model, whether in low- or high-dimensional settings. Moreover, we show that our PRSP algorithm (Algorithm \ref{algo}) successfully selects/uses a subset of (or all) the covariates that are informative (i.e. that truly enter into the model) in the box decision rules.

\pagebreak
%===================================
\subsection{Comparisons Against Other Methods} \label{performance}
%===================================
\vspace{-0.10in}
\subsubsection{Design and Choice of Various Non-Parametric Survival Models}
\vspace{-0.10in}
Next, we compared our Survival Bump Hunting (SBH) procedure by our PRSP algorithm (\ref{algo}) to other competitive non-parametric survival models or methods in terms of survival and prediction end-points statistics. In all our performance analyses below, we used \emph{LRT} in both peeling and optimization criteria and RCCV as our cross-validation technique. Comparisons include (i) Survival Bump Hunting by our PRSP method (Algorithm \ref{algo} and \cite{Dazard_2014}), (ii) Regression Survival Trees (RST) by recursive partitioning \cite{Gordon_1985, Ciampi_1986, Segal_1988, Davis_1989, LeBlanc_1992, LeBlanc_1993, Ahn_1994}, (iii) Random Survival Forest (RSF) by ensemble tree-based method \cite{Ishwaran_2008}, (iv) Cox Proportional Hazard Regression (CPHR) \cite{Cox_1972}, (v) Survival Supervised PCA (SSPCA) \cite{Bair_2006a}, (vi) Survival Supervised Clustering (SSC) \cite{Bair_2004}.

The simulated survival model was according to simulated model \#1b, that is, by generating a box-shaped region $R$ of the input covariate space with higher hazards than a uniform background (see \ref{design}). For comparisons, $B = 128$ repeated Monte-Carlo simulated datasets \#1b were used to generate empirical sampling distributions of cross-validation estimates of box statistics, survival end-points and prediction performance metrics (section \ref{end-points}) and make points and confidence intervals inferences. This replication design accounts for random splitting and simulated model sampling variabilities.

For each method, an internal cross-validation was carried out for model fitting/training that was done by optimizing a specific empirical objective function of Goodness of Fit or Prediction Error measure on the corresponding test-set, such as: (i) maximization of the Log-Rank Test statistic or minimization of a Concordance Error Rate (SBH), maximization of the Deviance Residuals statistic (RST), maximization of the Concordance Index (RSF), maximization of the Likelihood Ratio Statistic between the reduced vs. full model (SSPCA), maximization of the Concordance Index (SSC). Then, cross-validation was used again to make estimations and predictions on the combined test-sets as described before (section \ref{techniques}).

Whether the goal is to make estimations or predictions, one wants to classify samples into two survival/risk groups. However, unlike Survival Bump Hunting that inherently generates ``in-box'' and ``out-of-box'' groups, all other methods do not necessarily give directly two survival/risk groups. For comparisons purposes, one needs to come up with a calibrated way across all other methods to output two groups only. One way, shown to work well empirically, is by using the median survival time threshold (\cite{Huang_2006}).

\vspace{-0.10in}
\subsubsection{Comparison of End-Point Estimates}
\vspace{-0.10in}
Specifically, the trained models generate cross-validated fits from which cross-validated estimates of highest-risk/group support and survival end-points statistics (described in section \ref{end-points}) are made using the left-out test-set $\mathcal{L}_{k}$. We report the results for all methods in Figure \ref{Figure10} below. The figure shows the highest-risk/group end-points distributions of RCCV estimates of support and survival end-points statistics computed over $B = 128$ repeated Monte-Carlo simulated models \#1b for all competitive non-parametric survival models under study.

\begin{figure}[!hbt]
    \centering\includegraphics[width=6in]{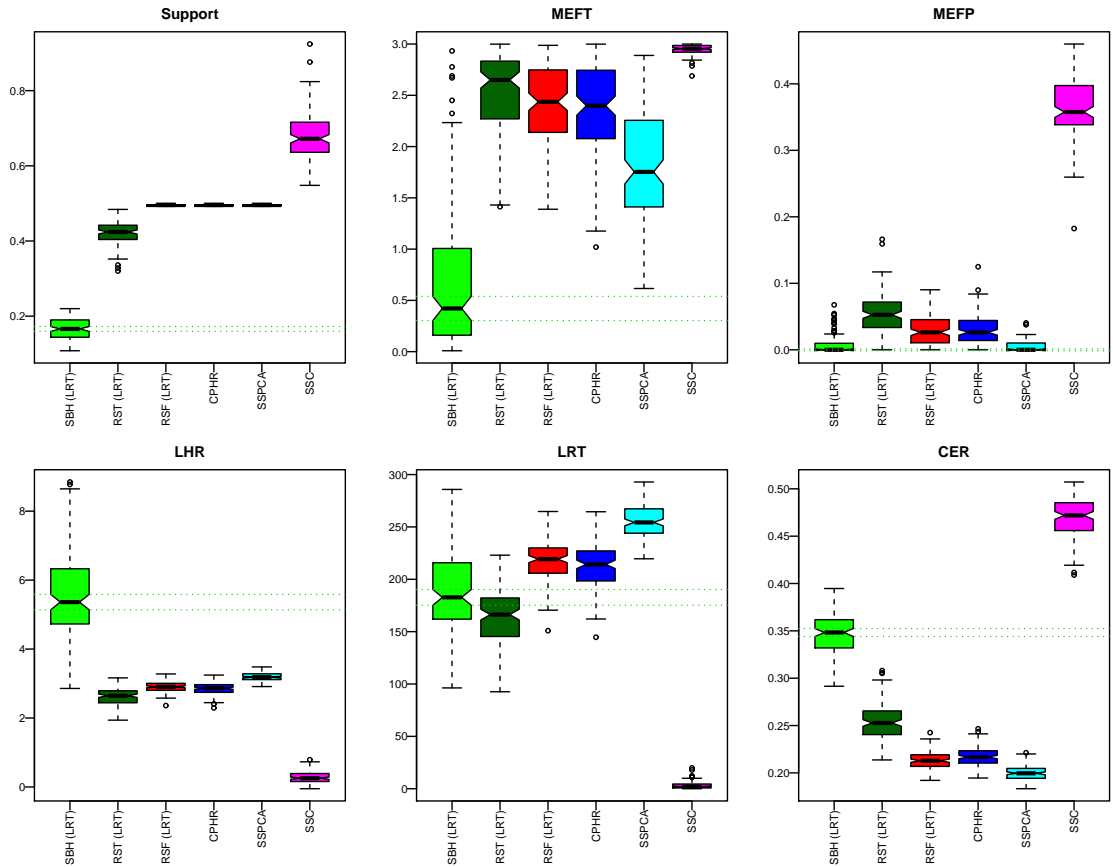}
    \caption{\sl \footnotesize Distributions of RCCV estimates of highest-risk/group end-points, computed over $B = 128$ repeated Monte-Carlo simulated models \#1 and for all competitive non-parametric survival models under study. Comparisons include (i) Survival Bump Hunting (SBH), (ii) Regression Survival Trees (RST), (iii) Random Survival Forest (RSF), (iv) Cox Proportional Hazard Regression (CPHR), (v) Survival Supervised PCA (SSPCA), (vi) Survival Supervised Clustering (SSC). In parenthesis is shown the criterion used for peeling or partitioning as it applies. For each SBH boxplot, the pair of horizontal dotted lines delineates the approximate (95\%) confidence interval of the median. Results are for the ``Replicated Combined CV'' (RCCV) technique and the LRT statistic used in the optimization criteria.}
    \label{Figure10}
\end{figure}

In Figure \ref{Figure11} below, a Kaplan--Meier estimate of RCCV survival probability curve is shown for each group and competitive non-parametric survival models under study from one replicate out of $B=128$. As expected, it shows the extremeness of the survival distribution of the highest-risk box/group found by SBH as compared to all other methods. The box sample sizes in the highest-risk box/group (out of $n=250$ samples) were as follows for each method (and that replicate): SBH: $n_{SBH} = 39$, RST: $n_{RST} = 105$, RSF: $n_{RSF} = 124$, CPHR: $n_{CPHR} = 124$, SSPCA: $n_{SSPCA} = 124$, SSC: $n_{SSC} = 170$.

\begin{figure}[!hbt]
    \centering\includegraphics[width=6in]{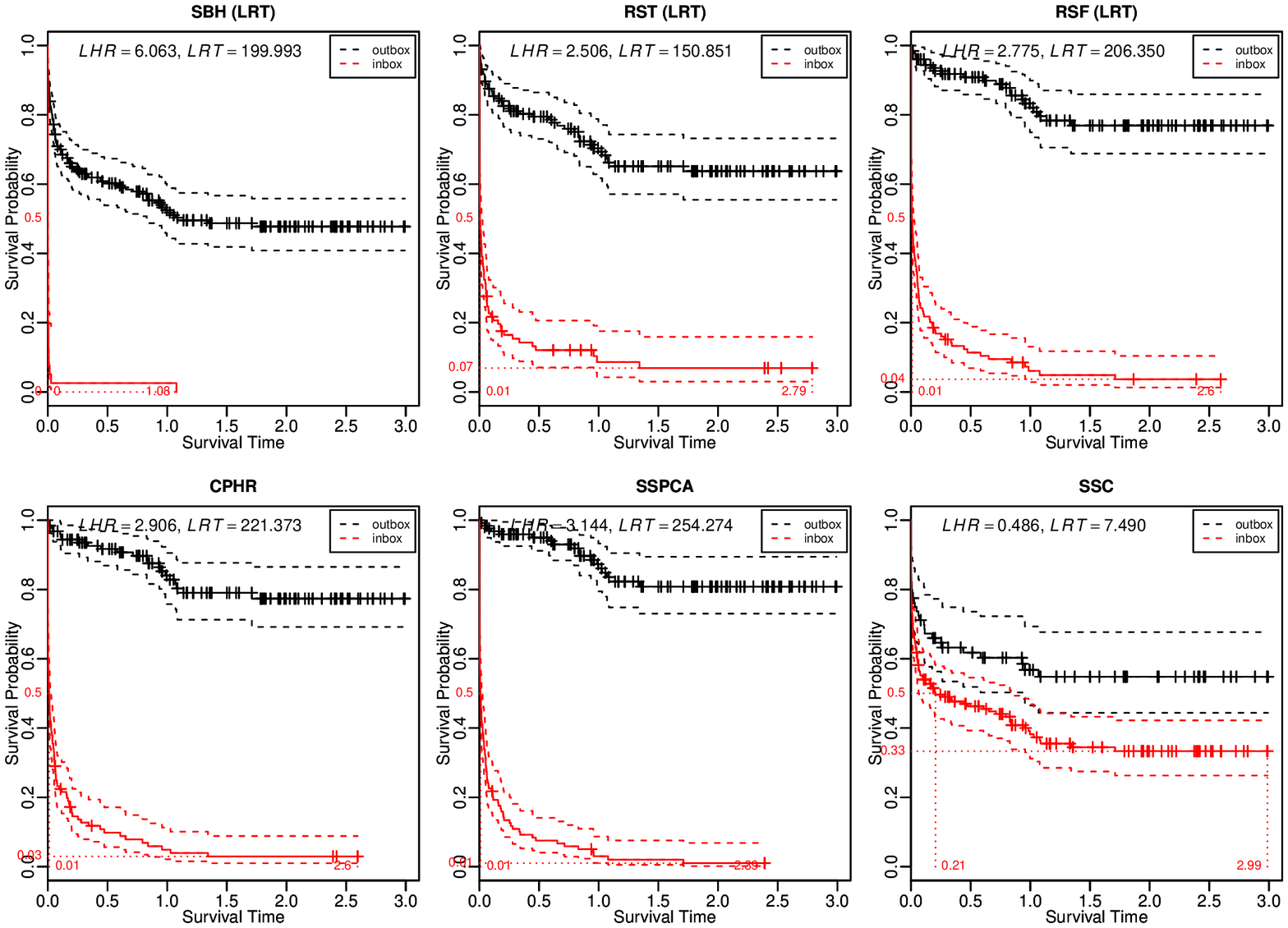}
    \caption{\sl \footnotesize Kaplan--Meier plots of RCCV survival probability curves for all competitive non-parametric survival models under study. Plots are illustrative of one replication out of $B=128$. Comparisons include (i) Survival Bump Hunting (SBH), (ii) Regression Survival Trees (RST), (iii) Random Survival Forest (RSF), (iv) Cox Proportional Hazard Regression (CPHR), (v) Survival Supervised PCA (SSPCA), (vi) Survival Supervised Clustering (SSC). In parenthesis is shown the criterion used for peeling or partitioning as it applies. The ``in-box'' legends (red) corresponds to the highest-risk box/group. Cross-validated LRT, LHR of ``in-box'' samples are shown at the top of the plot for each method (and that replicate). Results are for the ``Replicated Combined CV'' (RCCV) technique and LRT statistic used in the optimization criteria.}
    \label{Figure11}
\end{figure}

Overall, results from Figures \ref{Figure10} and \ref{Figure11} point out that the highest-risk box/group found by SBH is, as expected, smaller in size (support) and more extreme in terms of survival hazards ($LHR$) or risks, with consistent smaller event-free end-point times ($MEFT$) and probabilities ($MEFP$) than any other method/model under study. This was the goal. However, we did not expect the separation of the estimated survival distributions to be necessarily larger. In fact, the distributions of the log-rank test statistics ($LRT$) are not significantly different between most methods (except SSC). Interestingly, the Concordance Error Rates ($CER$) are slightly higher for SBH than most methods (except SSC). So, it's possible that the task of finding extreme survival/risks subgroups comes with some trade-off between achieving high levels of extremeness and high accuracy of survival time prediction.

\vspace{-0.10in}
\subsubsection{Comparative Prediction Performance}
\vspace{-0.10in}
The simulated survival model we drew from was according to model \#1b as follows: samples contained within a box-shaped region $R$ of the input covariate space had increased risk/hazards while samples outside of it had a uniform background risk. The survival times were generated as in section (\ref{design}), using the exponential distribution with random uniform censoring. The regression function for samples within $R$ was as in model \#1.

Due to the rule-induction nature of our ``Patient Recursive Survival Peeling'' method (Algorithm \ref{algo}), the box decision rule can be used as the classification rule. The cross-validated classification error is estimated from the discrepancies between the true and predictive classifications of the independent observations. Specifically, for each loop $k \in \{1, \dots, K\}$ of the cross-validation, we compute a cross-validated estimate of the error by matching the SBH test-set ``in-box'' prediction samples to the true ones in the high-risk box-shaped region $R$ of simulated model \#1b using the left-out test-set $\mathcal{L}_{k}$. The final cross-validated estimate of the Miss-classification Error Rate ($MER$) is given by the average of the cross-validated errors from the $K$ models $\{\tilde{\mathcal{R}}_{k}\}_{k=1}^{k=K}$ generated from each loop of the cross-validation. This is repeated $B$ times to get variability estimates. Note that $CER$ and $MER$, although related, are distinct: the $MER$ evaluates the accuracy to predict that a sample will fall into (or outside) the highest-risk box/group found by a method/survival model, whereas the $CER$ evaluates the accuracy to predict a sample survival time.

Prediction performances are then assessed using the usual accuracy metrics and Receiver Operating Characteristics (ROC) for each method and compared between them. For binary classes, a common metric for assessing the prediction performance is prediction accuracy through the use of True- and False-Positive Rates $TPR$ and $FPR$, respectively, also known as $Sensitivity$ and $1-Specificity$. By definition, the True- and False-Positive Rates are defined as:
\begin{align*}
&TPR = Sensitivity = \frac{TP}{TP+FN}\\
&FPR = 1 - Specificity = \frac{FP}{FP+TN}
\end{align*}
where $TP$, $FP$, $TN$, $FN$ stands for True-Positive, False-Positive, True-Negative and False-Negative, respectively. The performance of classification is naturally assessed by measuring the accuracy of prediction, whereas the performance of ranking is commonly measured by taking $(AUC)$, the Area Under the Receiver Operating Characteristics (ROC) Curve ($TPR$ versus $FPR$) \cite{Klement_2008}. An $AUC = 1$ corresponds to a perfect classifier, while an $AUC = 0.5$ corresponds to all possible performances of a random classifier. Finally, we also report Pearson's $\chi^2$ contingency table test \emph{p}-values (after continuity correction) of independence between the observed versus predicted counts. Table \ref{Table04} reports the classification performance results of various survival models/methods in terms of contingency table test, area under the ROC curve and sensitivity/specificity.

\begin{table}[!hbt]
 \caption{\sl \footnotesize Empirical $\chi^2$ contingency table test p-values $(\widehat{P.val})$, Area Under the Curve $(\widehat{AUC})$, Sensitivity $(\widehat{1-FPR})$ and Specificity $(\widehat{TPR})$ for each method. Comparisons include (i) Survival Bump Hunting (SBH), (ii) Regression Survival Trees (RST), (iii) Random Survival Forest (RSF), (iv) Cox Proportional Hazard Regression (CPHR), (v) Survival Supervised PCA (SSPCA), (vi) Survival Supervised Clustering (SSC). Values are median estimates with standard errors of the sample mean in parenthesis. In parenthesis, next to the method, is also shown the criterion used for peeling or partitioning as it applies. Results are for the ``Replicated Combined CV'' (RCCV) technique and the LRT statistic used in the optimization criteria.}
 \begin{center}
    \footnotesize
    \begin{tabular}{lcccccc}
        \hline
        \hline
        Method                           &SBH (LRT)    &RST (LRT)    &RSF (LRT)    &CPHR         &SSPCA        &SSC         \\
        \hline
        $\widehat{P.val}$                &0.000 (0.32) &0.909 (0.42) &0.065 (0.24) &0.037 (0.29) &0.037 (0.34) &0.519 (0.33)\\
        $\widehat{1-FPR}$ ($Specificity$)&0.800 (0.11) &0.947 (0.04) &0.516 (0.01) &0.516 (0.01) &0.516 (0.01) &0.373 (0.05)\\
        $\widehat{TPR}$ ($Sensitivity$)  &1.000 (0.38) &0.125 (0.17) &1.000 (0.14) &1.000 (0.26) &1.000 (0.25) &0.833 (0.21)\\
        $\widehat{AUC}$                  &0.899 (0.24) &0.533 (0.07) &0.757 (0.07) &0.758 (0.14) &0.758 (0.13) &0.591 (0.11)\\
        \hline
    \end{tabular}
 \end{center}
 \label{Table04}
\end{table}

Table \ref{Table04} shows that SBH has a better prediction performance on all metrics than any other method/model under study. This directly results from its better trade-off of $Specificity$ and $Sensitivity$. Overall, this reflects the above results on comparative end-point estimates: SBH reaches out a more specific and smaller group of samples that is more extreme in survival hazards or risks. Conversely, other methods tend to be more sensitive, but way too un-specific. Note that this applies to Regression Survival Trees (RST) as well.

%=========================================================================================
\section{Real Data Analysis} \label{real}
%=========================================================================================
Finally, we applied our Survival Bump Hunting (SBH) procedure to a publicly available real clinical dataset from the Women's Interagency HIV cohort Study (WIHS) \cite{Bacon_2005}. It involves competing risks ``AIDS/Death (before HAART)'' and ``Treatment Initiation (HAART)'' during HIV-1 Infection in women. Here, for simplification purposes, only the first of the two competing events (the time to AIDS/Death) was used in our analysis. The data consists of $n = 485$ complete observations on the following $p = 4$ covariates in addition to the censoring indicator and (censored) time-to-event variables (Table \ref{Table05}).

\begin{table}[!hbt]
    \caption{\sl \footnotesize Women's Interagency HIV Study (WIHS). Clinical dataset used with covariates description.}
    \begin{center}
        \vskip -10pt \footnotesize
        \begin{tabular}{ll}
        \hline
        \hline
        \ Covariate Description                                            & Range\\
        \hline
        \ AIDS/Death Diagnosis Time                                        & $T$ (years)\\
        \ Event indicator variable                                         & $C \in \{\text{AIDS/Dead} = 1, \text{Censored} = 0\}$\\
        \ Patient age at time of FDA approval of first protease inhibitor  & $Age$ (years)\\
        \ Injection Drug Users (IDU) history                               & $IDU \in \{\text{No history} = 0, \text{History} = 1\}$\\
        \ Patient’s race                                                   & $Race \in \{\text{Other} = 0, \text{African-American} = 1\}$.\\
        \ CD4 count                                                        & $CD4 \in [0,+\infty] \slash 100 \ \text{cells} \slash \mu\text{l}$\\
        \hline
        \end{tabular}
    \end{center}
  \label{Table05}
\end{table}

\begin{figure}[!hbt]
    \centering\includegraphics[width=4.0in]{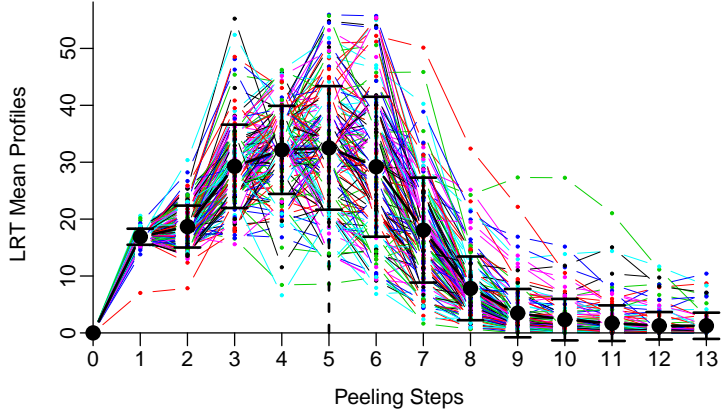}
    \caption{\sl \footnotesize Cross-validated tuning profile of the WIHS clinical dataset. The ``Replicated Combined CV'' cross-validated optimal peeling length ($\bar{L}^{rcv} = 5$) is shown with the vertical black dotted line. Each colored profile corresponds to one of the replications ($B = 128$). The cross-validated mean profile of the \emph{LRT} statistic is shown by the solid black line with standard error of the sample mean.}
    \vskip -10pt
    \label{Figure12}
\end{figure}

All results in the WIHS clinical dataset were achieved using the ``Replicated Combined CV'' (RCCV) technique and the \emph{LRT} statistic as peeling and optimization criterion, using $K = 5$, $A = 1024$ and $B = 128$. We show in Figure \ref{Figure12} the cross-validated tuning profile of $LRT$ as a function of the number of peeling steps. According to our optimization criterion, we determined that the resulting ``Replicated Combined CV'' optimal length of the peeling trajectory is $\bar{L}^{rcv} = 5$ (Figure \ref{Figure12}).

We show in Table \ref{Table06} the cross-validated decision rules and highest-risk box/group statistics at each step. Note that the box sample size in the final highest-risk box/group is $n(l=5) = 262$ out of a total sample size of $n=485$. Here, the cross-validated trace of covariate usage is: $CD4$, $Age$, $Age$, $Age$ (Table \ref{Table06}).

\begin{table}[!hbt]
 \caption{\sl \footnotesize Cross-validated decision rules (top) and highest-risk box/group statistics (bottom) of the WIHS clinical dataset. Values are sample mean estimates with corresponding standard errors in parenthesis. The box sample size at each step is also shown. Step \#0 corresponds to the situation where the starting box covers the entire test-set data $\mathcal{L}_{k}$ before peeling.}
 \footnotesize
 \begin{tabular}{lrccccccc}
    \hline
    \hline
    &Step $l$  &$Age$                     &$IDU$                    &$Race$                    &$CD4$                    & & & \\
    \hline
    &$ 0$      &$Age \ge 19.00 \ (0.00)$  &$IDU \ge 0.00 \ (0.00)$  &$Race \le 1.00 \ (0.00)$  &$CD4 \le 19.33 \ (0.00)$ & & & \\
    &$ 1$      &$Age \ge 19.00 \ (0.00)$  &$IDU \ge 0.00 \ (0.00)$  &$Race \le 1.00 \ (0.00)$  &$CD4 \le  8.64 \ (0.35)$ & & & \\
    &$ 2$      &$Age \ge 22.07 \ (1.94)$  &$IDU \ge 0.00 \ (0.00)$  &$Race \le 1.00 \ (0.00)$  &$CD4 \le  8.51 \ (0.35)$ & & & \\
    &$ 3$      &$Age \ge 23.30 \ (2.65)$  &$IDU \ge 0.00 \ (0.00)$  &$Race \le 1.00 \ (0.00)$  &$CD4 \le  8.46 \ (0.16)$ & & & \\
    &$ 4$      &$Age \ge 28.79 \ (0.51)$  &$IDU \ge 0.00 \ (0.00)$  &$Race \le 1.00 \ (0.00)$  &$CD4 \le  7.66 \ (0.83)$ & & & \\
    &$ 5$      &$Age \ge 29.22 \ (0.53)$  &$IDU \ge 0.00 \ (0.00)$  &$Race \le 1.00 \ (0.00)$  &$CD4 \le  6.79 \ (0.77)$ & & & \\
 \end{tabular}\\
 \begin{tabular}{lrccccccc}
        \hline
        \hline
        \\[-10pt]
        &Step $l$  &$n(l)$         &$\bar{\beta}^{rcv}(l)$  &$\widebar{T_{0}^{\prime}}^{rcv}(l)$  &$\widebar{P_{0}^{\prime}}^{rcv}(l)$  &$\bar{\lambda}^{rcv}(l)$  &$\bar{\chi}^{rcv}(l)$  &$\bar{\theta}^{rcv}(l)$\\
        \hline
        &$ 0$      &485 \  (0.00)  &1.00 \ (0.00)           &10.8 \ (0.00)                        &0.17 \ (0.00)                        & 0.00 \ (0.00)            &  0.00 \  (0.00)        &1.00 \ (0.00)          \\
        &$ 1$      &436 \  (0.00)  &0.90 \ (0.00)           &10.8 \ (0.00)                        &0.17 \ (0.00)                        & 0.61 \ (0.02)            & 16.90 \  (1.41)        &0.46 \ (0.00)          \\
        &$ 2$      &398 \  (4.85)  &0.82 \ (0.01)           &10.8 \ (0.00)                        &0.16 \ (0.01)                        & 0.54 \ (0.05)            & 18.69 \  (3.68)        &0.45 \ (0.01)          \\
        &$ 3$      &364 \  (9.70)  &0.75 \ (0.02)           &10.8 \ (0.00)                        &0.14 \ (0.01)                        & 0.61 \ (0.07)            & 29.28 \  (7.32)        &0.43 \ (0.01)          \\
        &$ 4$      &325 \ (19.40)  &0.67 \ (0.04)           &10.8 \ (0.00)                        &0.13 \ (0.01)                        & 0.61 \ (0.08)            & 32.17 \  (7.74)        &0.42 \ (0.01)          \\
        &$ 5$      &262 \ (33.95)  &0.54 \ (0.07)           &10.8 \ (0.00)                        &0.11 \ (0.01)                        & 0.61 \ (0.10)            & 32.51 \ (10.86)        &0.42 \ (0.01)          \\
        \hline
 \end{tabular}
 \label{Table06}
\end{table}

\begin{figure}[!hbt]
    \centering\includegraphics[width=6in]{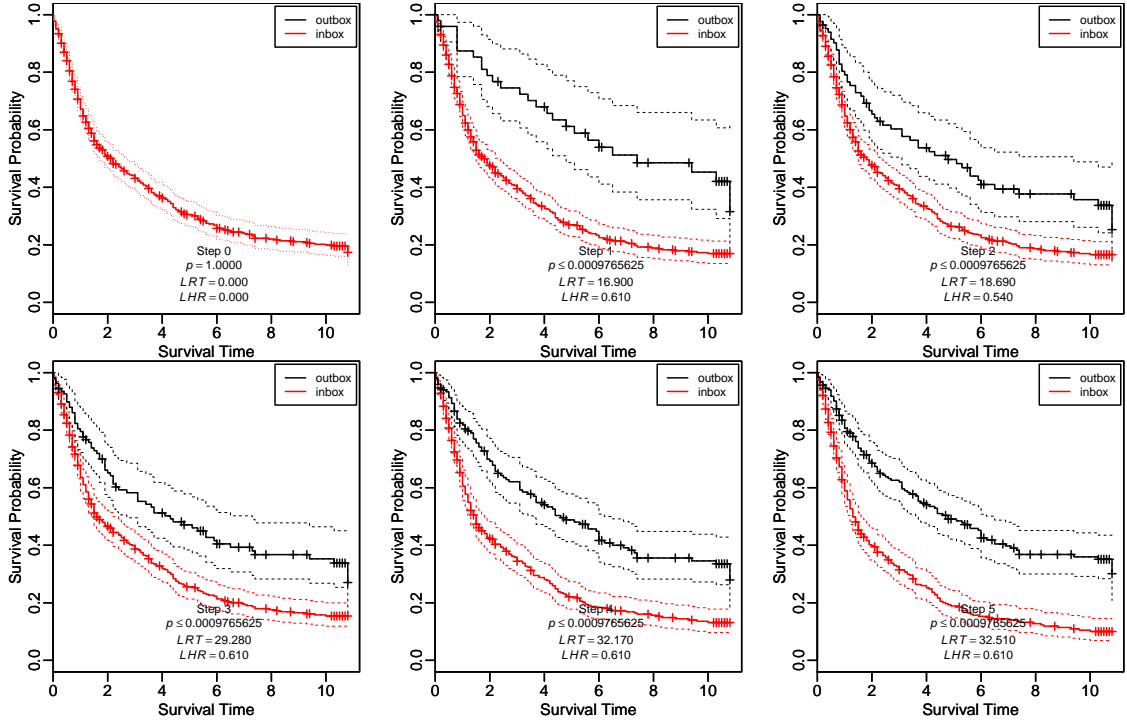}
    \caption{\sl \footnotesize Kaplan--Meier plots of RCCV survival probability curves of the WIHS clinical dataset. Each plot represents a step of the peeling sequence. Step \#0 corresponds to the situation where the starting box covers the entire test-set data $\mathcal{L}_{k}$ before peeling. The ``in-box'' legends (red) corresponds to the highest-risk box/group. Cross-validated LRT, LHR and permutation p-values of ``in-box'' samples are shown at the bottom of the plot with the corresponding peeling step for each method. P-values $\hat{p}^{cv}(l) \le 9.7e-5$ correspond to 1/10th of the precision limit (see section \ref{cvpval}). Notice the single survival curve at Step \#0 before peeling and how the survival curves of ``in-box'' and ``out-of-box'' samples separates as the peeling progresses.}
    \label{Figure13}
\end{figure}

Finally, we show in Figure \ref{Figure13} the cross-validated Kaplan--Meir survival probability curves of the highest-risk group vs. lower-risk group at each step with their corresponding permutation \emph{p}-values of separation. Notice how the curve separation increases with the peeling steps. The permutation \emph{p}-values at each step are respectively: $\tilde{p}^{cv}(l=0) = 1$, $\tilde{p}^{cv}(l=1-5) \le 9.7e-5$ (Figure \ref{Figure13}).

The question of this study was whether it is possible to achieve a stratification or prognostication of patients for AIDS and HAART by using e.g. the `Injection Drug Users` (IDU) history. Overall, SBH shows that it is possible to achieve a stratification and prognostication of patients that are more likely to be diagnosed or die of AIDS than others. The decision rule identifies a subgroup of $n(l=5) = 262$ such patients that should be treated more aggressively than others, e.g. by putting them on HAART treatment sooner. In addition, SBH reveals that these patients are characterized by a lower $CD4$ count of $CD4 \lessapprox 6.79 (\pm 0.77) \slash 100 \ \text{cells} \slash \mu\text{l}$ with an $Age \gtrapprox 29.22 (\pm 0.53)$. Moreover, SBH reveals that `Injection Drug Users` history ($IDU$) was actually \emph{not} the most useful covariate at this stage of determination.

%=========================================================================================
\section{Discussion and Conclusion} \label{discussion}
%=========================================================================================
To build a survival bump hunting model, fit by a recursive peeling procedure, we used two sets of criteria with different purposes: (i) the peeling criteria for model fitting by maximization of the rate of increase in \emph{LHR}, \emph{LRT} or \emph{CHS} statistics (section \ref{criterion}), and (ii) the optimization criterion used for model tuning by maximization of the \emph{LHR} or \emph{LRT} or by minimization of the \emph{CER} (section \ref{optimization}).

Despite being a well established concept as a splitting criterion in regression survival trees (section \ref{criterion}), one criticism about the \emph{LRT} is that it tends to favor continuous variables and causes some uneven splits (end-cut preference). In this study, the main difference that we observed between the three peeling criteria used so far (section \ref{peelcrit}) is in the conservativeness of the number of peeling steps or peeling sequence length (model complexity), and to a lower extent, of the number of used covariates (model size), regardless of the dimensionality of the data. In summary, we recommended using as peeling criterion primarily the Cumulative Hazard Summary \emph{CHS} and secondarily the Log-Rank Test \emph{LRT} to induce more conservative estimates and reduce the risk of over-fitting, especially in high-dimensional data.

In contrast, we noted that the overall effect of the peeling criteria is relatively marginal compared to that of the optimization criterion. In this study, we showed that the choice of the optimization criteria is crucial and dependent on the dimensionality of the data. In summary, we recommended using the Concordance Error Rate \emph{CER} as optimization criterion in every situation (low- and high-dimensional data), or alternatively the Log-Rank Test \emph{LRT} in low-dimensional situation only (section \ref{optcrit}).

Overall, both of our replicated cross-validation techniques, namely the ``Replicated Combined CV'' (RCCV) and ``Replicated Averaged CV'' (RACV) techniques, were found effective at controlling (at least in part) the overfitting and under-fitting issues, confirming that these techniques are appropriate to the task of survival bump hunting modeling by a recursive peeling procedure. However, we observed differences in the cross-validation techniques, which raised the questions whether RACV could lend to over-fitting more than RCCV, especially in high-dimensional data, and wether RACV performance could degrade faster than RCCV in situations with small sample sizes (see section \ref{techniques}). For these reasons, we recommended using RCCV preferably to RACV.

It is known that the stepwise covariate selection/usage procedure in the peeling loop of Algorithm \ref{algo} induces an inflation of variance estimates primarily because of the adaptive nature of the algorithm (each peeling step is conditional on the previous step). Using replicated cross-validation is therefore recommended to reduce the variability of cross-validation estimates in recursive peeling methods.

Survival estimates and model performance accuracy can be improved by the resampling technique used and resulting bias-variance trade-off. For instance, using a ``larger'' number of folds in $K$-fold CV in the presence of ``small'' number of events or samples is known to reduce bias, but also increase variance of estimates. Beside our cross-validation techniques, the Leave-one-Out Cross-Validation (LOOCV - Jackknife) or bootstrap-based resampling techniques are available, such as the ordinary bootstrap \cite{Efron_1993}, the 0.632 / Out-of-Bootstrap Cross-Validation (0.632 OOBCV - \cite{Efron_1983}) or its latest variant (0.632+ OOBCV \cite{Efron_1997}). As per Efron, ''\emph{the ordinary bootstrap gives an estimate of error with low variability, but with a possibly large downward bias, particularly in highly over-fitted situations}'' \cite{Efron_1983}. Conversely, cross-validation estimators are nearly unbiased (slightly upward), but have generally unacceptable high variability. For this reason, we chose to use a cross-validation procedure that could be replicated. The LOOCV and 0.632 OOBCV could be used instead of $K$-fold CV, but have larger variance estimators. An alternative may be the 0.632+ OOBCV estimator, which combines lower variance with a correction for bias \cite{Efron_1997}.

Collectively, our peeling strategy, combined with our model tuning/selection method along with cross-validation techniques support the claim that optimal survival bump hunting modeling can be done. Further, results also indicate that our strategies help our ``Patient Recursive Survival Peeling'' (PRSP) algorithm (\ref{algo}) use the most informative covariates in the decision rule. This suggests the possibility of carrying out a joint internal variable selection with our PRSP procedure.

Some interesting differences between decision-tree and decision-box models lie in the weaknesses and strengths of the estimated solutions and in their applications. Here are some: (i) {\it Stratification}: if multiple groups are of interest, an advantage of recursive partitioning is that they directly lead to multi-group stratifications of the data, instead of just presenting a rule for a single high (low) vs. low (high) response group; (ii) {\it Interpretability}: binary decision trees lead to an intuitive hierarchical interpretation of groups that facilitates their interpretation unlike peeling methods that are not constrained to a tree structure; (iii) {\it Patience vs. Greediness}: recursive partitioning methods are however notoriously ``greedy'' (exponential decrease of the data as the space undergo partitioning based on typically binary split), but recursive peeling methods can be made ``patient'' at will (quantile-controlled decrease of the data), eventually helping recursive top-down peeling algorithms such as ours to better learn from the data.

%=========================================================================================
\section{Acknowledgments} \label{acknowledgments}
%=========================================================================================
\normalsize
This research was made possible with the contribution of the National Institute of Health (NIH), National Cancer Institute (NCI). J-E. Dazard and J. Sunil Rao were supported in part by NIH grant NCI R01-CA160593A1. Additional support came from NIH grant NCI P30-CA043703 of the Comprehensive Cancer Center at Case Western Reserve University (J-E. D). This work made use of the High Performance Computing Cluster in the Core Facility for Advanced Research Computing at Case Western Reserve University.

%=========================================================================================
\section{Supporting Information} \label{supporting}
%=========================================================================================
\normalsize
\begin{itemize}
    \item Supporting\_Text; Supporting\_Tables 1, 2; Supporting\_Figures 1, 2, 3, 4, 5, 6.
    \item Supporting\_Code: R package \pkg{PRIMsrc}, available at:\\
     CRAN repository: \url{https://cran.r-project.org/web/packages/PRIMsrc/index.html}\\
     GitHub repository: \url{https://github.com/jedazard/PRIMsrc}
    \item Supporting\_Datasets: R package \pkg{PRIMsrc}, available at:\\
     CRAN repository: \url{https://cran.r-project.org/web/packages/PRIMsrc/index.html}\\
     GitHub repository: \url{https://github.com/jedazard/PRIMsrc/tree/master/data}
\end{itemize}

%=========================================================================================
\section{References} \label{references}
%=========================================================================================
\spacingset{0}
\footnotesize
\bibliographystyle{Survival_Bump_Hunting}
\bibliography{Survival_Bump_Hunting}
\spacingset{1}

\newpage
%=========================================================================================
{\center \section*{SUPPORTING INFORMATION}}
%=========================================================================================
\normalsize
\setcounter{page}{1}
\pagestyle{empty}
\setcounter{section}{1}
\renewcommand{\thefigurename}{Supporting\_Figure}
\renewcommand{\thetablename}{Supporting\_Table}
\setcounter{figure}{0}
\setcounter{table}{0}

%%%%%%%%%%%%%%%%%%%%%%%%%%%%%%%%%
% Supporting\_Figure 01
%%%%%%%%%%%%%%%%%%%%%%%%%%%%%%%%%
\begin{figure}[!hbt]
    \centering\includegraphics[width=6in]{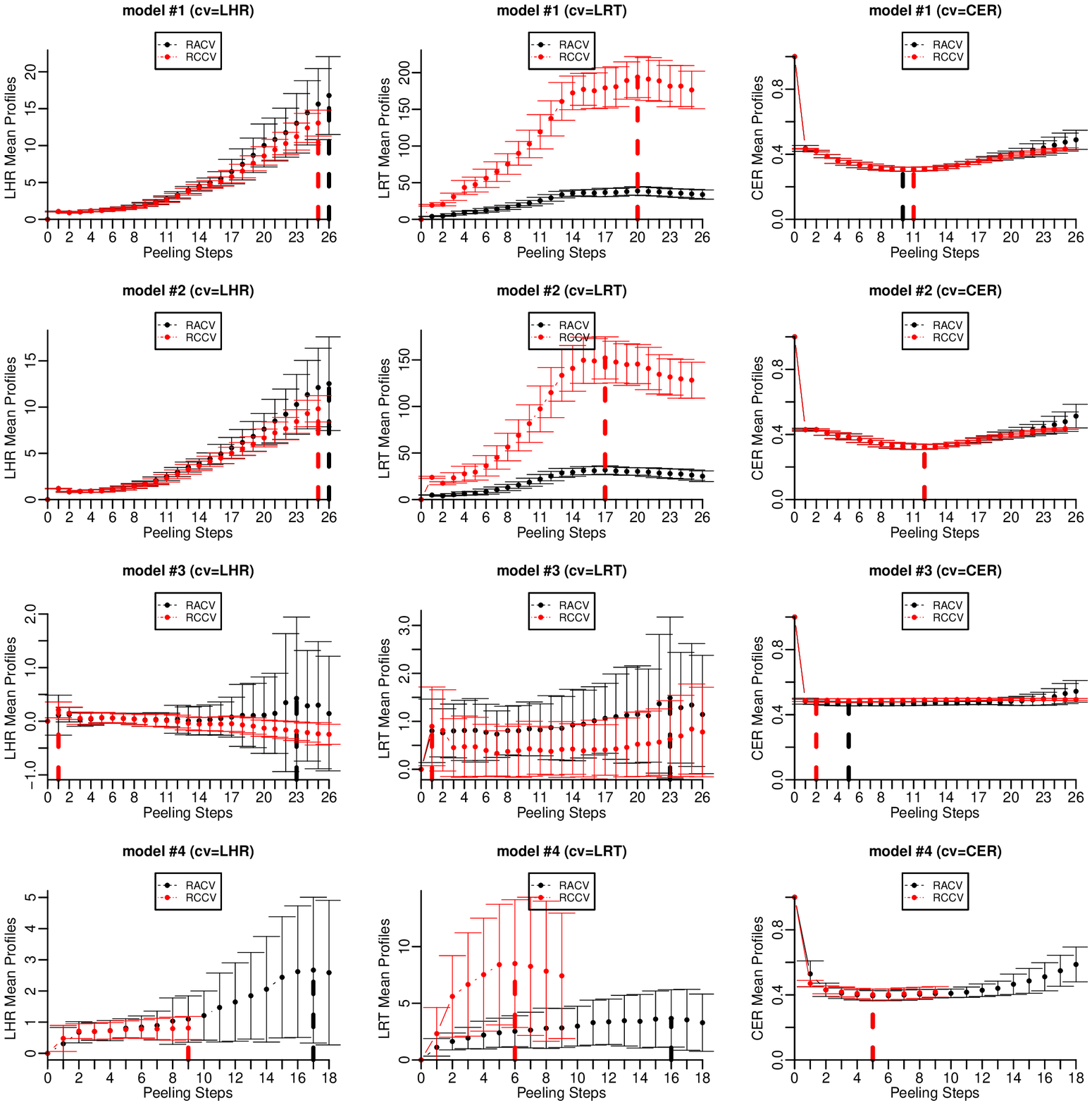}
    \caption{\sl \footnotesize Comparison of cross-validated tuning profiles of box end-point statistics between cross-validation techniques (overlaid: ``Replicated Averaged CV'' or RACV (black) vs. ``Replicated Combined CV'' or RCCV (red)) and optimization criteria (by rows: Log Hazard Ratio (LHR), Log-Rank Test (LRT) or Concordance Error Rate CER) in a given simulation model (by columns: simulation models \#1, \#2, \#3 or \#4). Results are for the Log Hazard Ratio (LHR) peeling criterion. The resulting ``Replicated CV'' optimal peeling length $\bar{L}^{rcv}$ (see eq. \ref{rep.optimizationeq}) of the peeling trajectory is shown in each case (vertical dashed lines). Each dotted line corresponds to a cross-validated mean profile of the statistic used in the optimization criterion with the corresponding standard error of the sample mean, both calculated over the replications ($B = 128$). Notice the situations of cross-validation success or failure as described in section \ref{optcrit} and Figure \ref{Figure04}. Also, notice the expected increase of variance of cross-validated point estimates towards the right-end of the profiles corresponding to an increase in model uncertainty and regions of risk of overfitting.}
    \label{SuppFigure01}
\end{figure}

%%%%%%%%%%%%%%%%%%%%%%%%%%%%%%%%%
% Supporting\_Figure 02
%%%%%%%%%%%%%%%%%%%%%%%%%%%%%%%%%
\newpage
\begin{figure}[!hbt]
    \centering\includegraphics[width=6in]{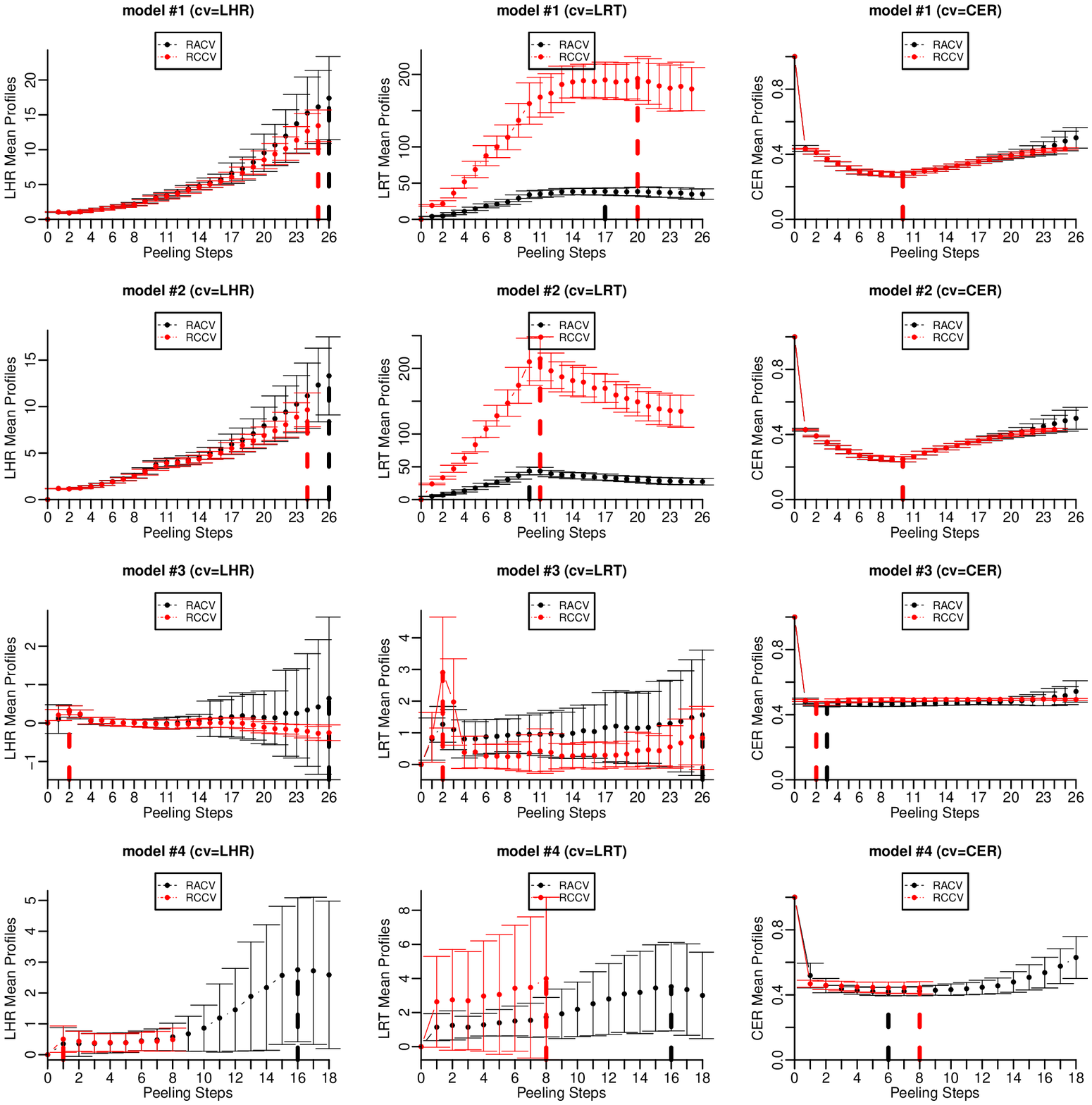}
    \caption{\sl \footnotesize Comparison of cross-validated tuning profiles of box end-point statistics between cross-validation techniques (overlaid: ``Replicated Averaged CV'' or RACV (black) vs. ``Replicated Combined CV'' or RCCV (red)) and optimization criteria (by rows: Log Hazard Ratio (LHR), Log-Rank Test (LRT) or Concordance Error Rate CER) in a given simulation model (by columns: simulation models \#1, \#2, \#3 or \#4). Results are for the Log-Rank Test (LRT) peeling criterion. The resulting ``Replicated CV'' optimal peeling length $\bar{L}^{rcv}$ (see eq. \ref{rep.optimizationeq}) of the peeling trajectory is shown in each case (vertical dashed lines). Each dotted line corresponds to a cross-validated mean profile of the statistic used in the optimization criterion with the corresponding standard error of the sample mean, both calculated over the replications ($B = 128$). Notice the situations of cross-validation success or failure as described in section \ref{optcrit} and Figure \ref{Figure04}. Also, notice the expected increase of variance of cross-validated point estimates towards the right-end of the profiles corresponding to an increase in model uncertainty and regions of risk of overfitting.}
    \label{SuppFigure02}
\end{figure}

%%%%%%%%%%%%%%%%%%%%%%%%%%%%%%%%%
% Supporting\_Figure 03
%%%%%%%%%%%%%%%%%%%%%%%%%%%%%%%%%
\newpage
\begin{figure}[!hbt]
    \centering\includegraphics[width=6in]{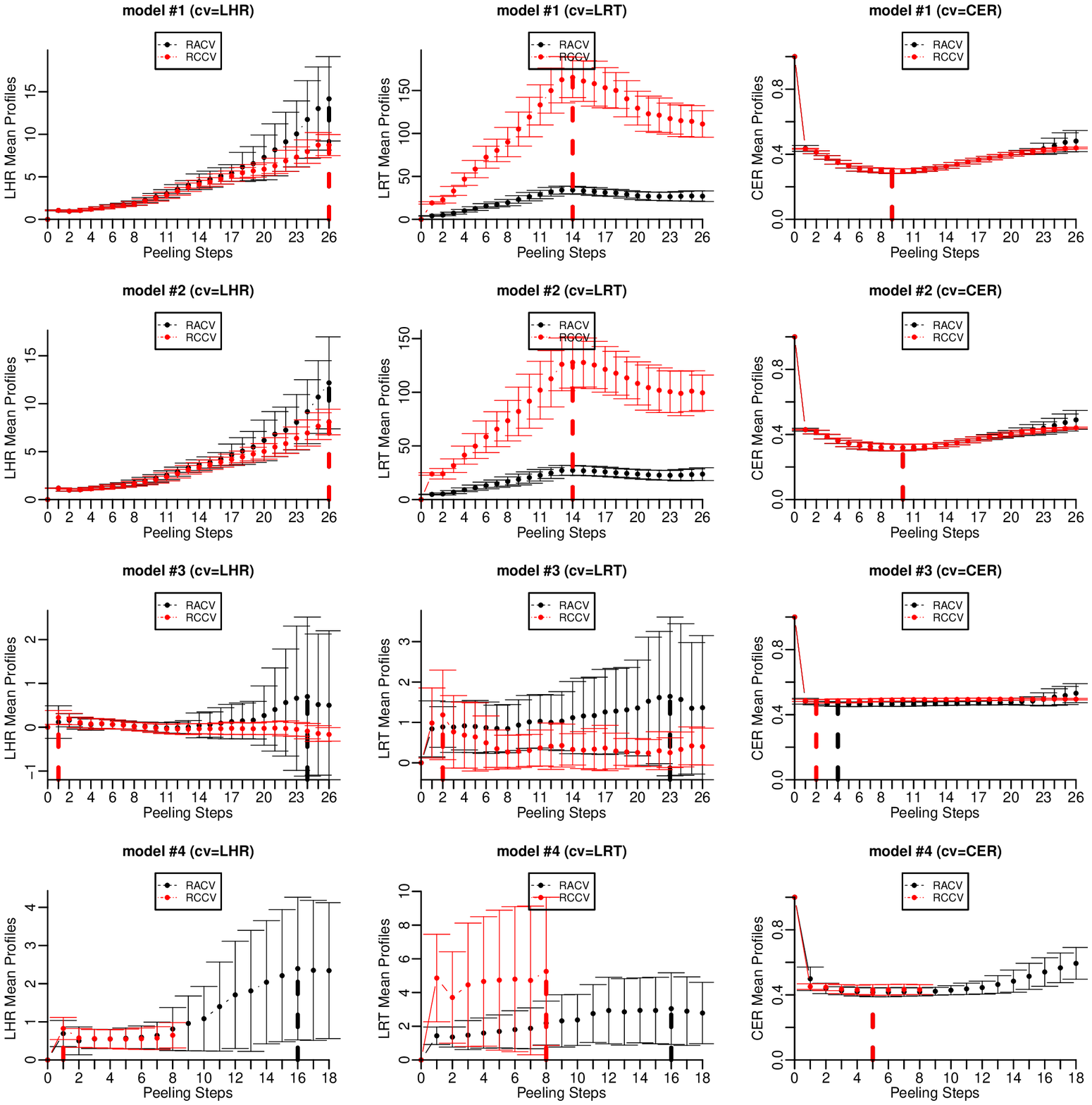}
    \caption{\sl \footnotesize Comparison of cross-validated tuning profiles of box end-point statistics between cross-validation techniques (overlaid: ``Replicated Averaged CV'' or RACV (black) vs. ``Replicated Combined CV'' or RCCV (red)) and optimization criteria (by rows: Log Hazard Ratio (LHR), Log-Rank Test (LRT) or Concordance Error Rate CER) in a given simulation model (by columns: simulation models \#1, \#2, \#3 or \#4). Results are for the Cumulative Hazard Summary (CHS) peeling criterion. The resulting ``Replicated CV'' optimal peeling length $\bar{L}^{rcv}$ (see eq. \ref{rep.optimizationeq}) of the peeling trajectory is shown in each case (vertical dashed lines). Each dotted line corresponds to a cross-validated mean profile of the statistic used in the optimization criterion with the corresponding standard error of the sample mean, both calculated over the replications ($B = 128$). Notice the situations of cross-validation success or failure as described in section \ref{optcrit} and Figure \ref{Figure04}. Also, notice the expected increase of variance of cross-validated point estimates towards the right-end of the profiles corresponding to an increase in model uncertainty and regions of risk of overfitting.}
    \label{SuppFigure03}
\end{figure}

%%%%%%%%%%%%%%%%%%%%%%%%%%%%%%%%%
% Supporting\_Figure 04
%%%%%%%%%%%%%%%%%%%%%%%%%%%%%%%%%
\newpage
\begin{figure}[!hbt]
    \centering \rotatebox{-90}{\includegraphics[width=6in]{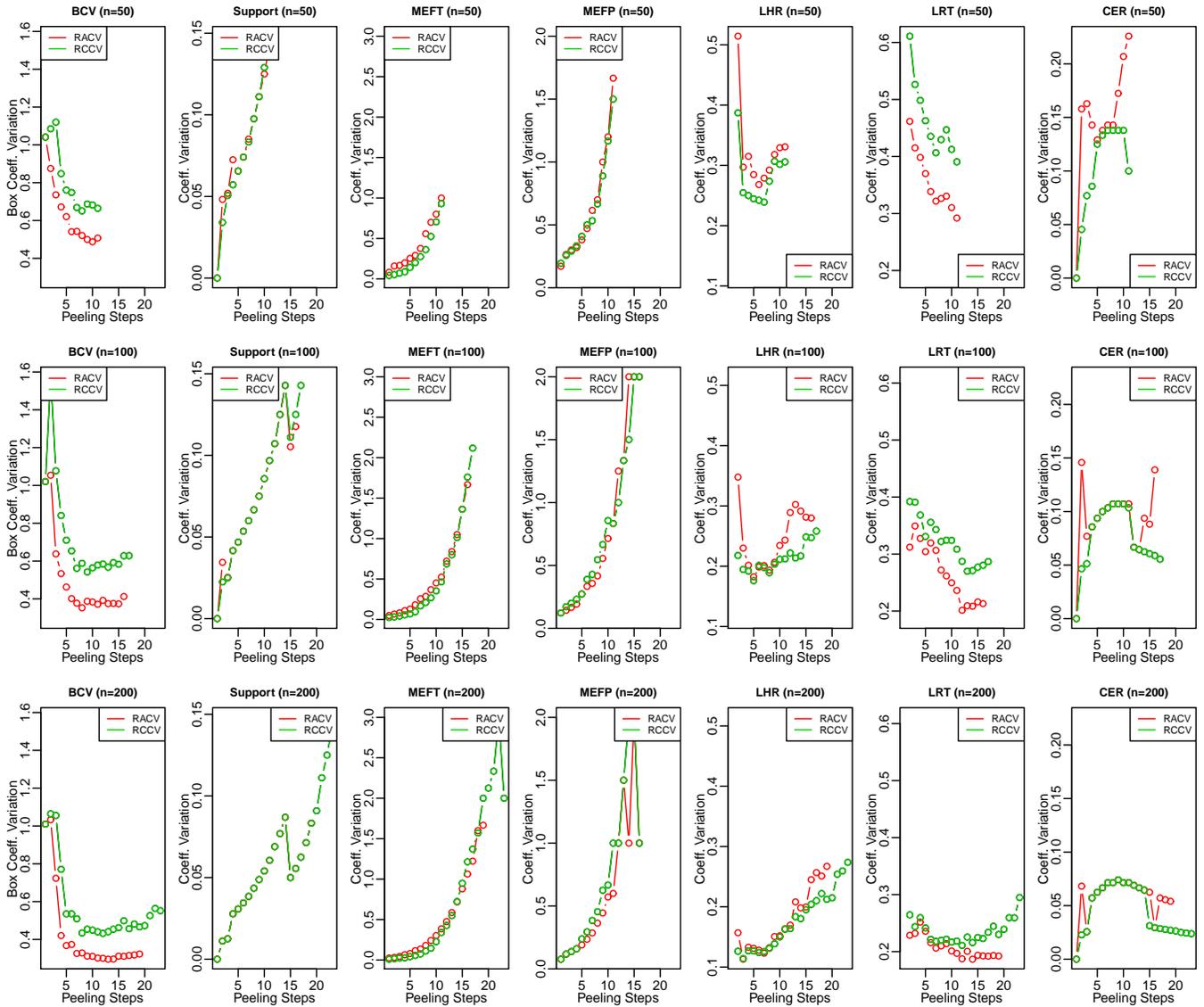}}
    \caption{\sl \footnotesize Profiles of coefficient of variation of Box Coefficient of Variation (BCV), survival end-points and prediction performance metrics. Comparative coefficient of variation profiles are shown for situations with decreasing sample sizes $n \in \{50, 100, 200\}$. Results are for simulated model \#1 and the LRT statistic used in both peeling and optimization criteria.}
    \label{SuppFigure04}
\end{figure}

%%%%%%%%%%%%%%%%%%%%%%%%%%%%%%%%%
% Supporting\_Table 01
%%%%%%%%%%%%%%%%%%%%%%%%%%%%%%%%%
\newpage
\begin{table}[!hbt]
    \caption{\sl \footnotesize Effect of peeling and optimization criteria as well as cross-validation techniques on the cross-validated numbers of used covariates by the PRSP algorithm (see Algorithm \ref{algo}) out of the total number of pre-selected ones (brackets). Numbers are reported for the combined effects of: (i) peeling criteria (by rows: Log Hazard Ratio (LHR), Log-Rank Test (LRT) or Cumulative Hazard Summary (CHS)), (ii) optimization criteria (by columns: Log Hazard Ratio (LHR), Log-Rank Test (LRT) or Concordance Error Rate (CER)), (iii) cross-validation techniques (by columns: ``Replicated Averaged CV'' or RACV and ``Replicated Combined CV'' or RCCV), and (iv) the four tested simulation models (by rows: Model \#1, \#2, \#3 or \#4).}
    \begin{center}
        \footnotesize
        \begin{tabular}{llrr@{}rrr@{}rrr}
            \hline
            \hline
            \ Model \#1                                                                                            \\
            \hline
            \hline
            \                & &\multicolumn{8}{c}{Optimization Criterion}                                         \\
            \cline{3-10}
            \                & &\multicolumn{2}{c}{$LHR$} & &\multicolumn{2}{c}{$LRT$} & &\multicolumn{2}{c}{$CER$}\\
            \cline{3-4} \cline{6-7} \cline{9-10}
            \                &            &RACV    &RCCV           & &RACV    &RCCV             & &RACV    &RCCV   \\
            \hline
            \Lower{Peeling}  &$LHR$       &3[  3]  &3[  3]         & &3[  3]  &3[  3]           & &3[  3]  &3[  3] \\
            \Lower{Criterion}&$LRT$       &3[  3]  &3[  3]         & &3[  3]  &3[  3]           & &2[  3]  &2[  3]\\
            \                &$CHS$       &3[  3]  &3[  3]         & &3[  3]  &3[  3]           & &3[  3]  &3[  3] \\
            \hline
            \\[-1pt]
            \hline
            \hline
            \ Model \#2                                                                                            \\
            \hline
            \hline
            \                & &\multicolumn{8}{c}{Optimization Criterion}                                         \\
            \cline{3-10}
            \                & &\multicolumn{2}{c}{$LHR$} & &\multicolumn{2}{c}{$LRT$} & &\multicolumn{2}{c}{$CER$}\\
            \cline{3-4} \cline{6-7} \cline{9-10}
            \                &            &RACV    &RCCV           & &RACV    &RCCV             & &RACV    &RCCV   \\
            \hline
            \Lower{Peeling}  &$LHR$       &3[  3]  &3[  3]         & &3[  3]  &3[  3]           & &3[  3]  &3[  3]  \\
            \Lower{Criterion}&$LRT$       &3[  3]  &3[  3]         & &2[  3]  &2[  3]           & &2[  3]  &2[  3]  \\
            \                &$CHS$       &3[  3]  &3[  3]         & &3[  3]  &3[  3]           & &3[  3]  &3[  3]  \\
            \hline
            \\[-1pt]
            \hline
            \hline
            \ Model \#3                                                                                            \\
            \hline
            \hline
            \                & &\multicolumn{8}{c}{Optimization Criterion}                                         \\
            \cline{3-10}
            \                & &\multicolumn{2}{c}{$LHR$} & &\multicolumn{2}{c}{$LRT$} & &\multicolumn{2}{c}{$CER$}\\
            \cline{3-4} \cline{6-7} \cline{9-10}
            \                &            &RACV    &RCCV           & &RACV    &RCCV             & &RACV    &RCCV   \\
            \hline
            \Lower{Peeling}  &$LHR$       &3[  3]  &1[  3]         & &3[  3]  &1[  3]           & &3[  3]  &2[  3] \\
            \Lower{Criterion}&$LRT$       &3[  3]  &2[  3]         & &3[  3]  &2[  3]           & &3[  3]  &2[  3] \\
            \                &$CHS$       &2[  3]  &1[  3]         & &2[  3]  &2[  3]           & &2[  3]  &2[  3] \\
            \hline
            \\[-1pt]
            \hline
            \hline
            \ Model \#4                                                                                            \\
            \hline
            \hline
            \                & &\multicolumn{8}{c}{Optimization Criterion}                                         \\
            \cline{3-10}
            \                & &\multicolumn{2}{c}{$LHR$} & &\multicolumn{2}{c}{$LRT$} & &\multicolumn{2}{c}{$CER$}\\
            \cline{3-4} \cline{6-7} \cline{9-10}
            \                &            &RACV    &RCCV           & &RACV    &RCCV             & &RACV    &RCCV   \\
            \hline
            \Lower{Peeling}  &$LHR$       &11[352] &6[352]         & &10[352] &3[352]           & &3[352]  &3[352] \\
            \Lower{Criterion}&$LRT$       &8[352]  &1[352]         & &8[352]  &5[352]           & &3[352]  &5[352] \\
            \                &$CHS$       &3[352]  &1[352]         & &3[352]  &3[352]           & &3[352]  &3[352] \\
            \hline
        \end{tabular}
    \end{center}
    \label{SuppTable01}
\end{table}

%%%%%%%%%%%%%%%%%%%%%%%%%%%%%%%%%
% Supporting\_Figure 05
%%%%%%%%%%%%%%%%%%%%%%%%%%%%%%%%%
\newpage
\begin{figure}[!hbt]
  \centering\includegraphics[width=4in]{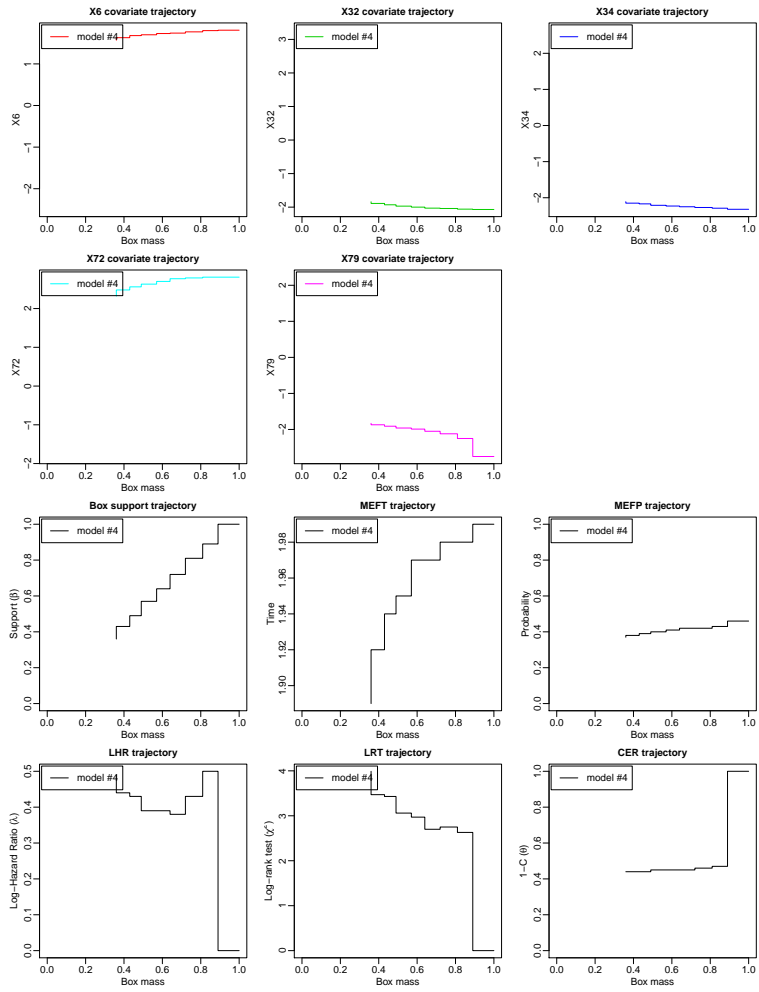}
  \caption{\sl \footnotesize Comparison of replicated combined cross-validated results for the peeling trajectories in simulated model \#4 for the ``Replicated Combined CV'' (RCCV) technique and the Cumulative Hazard Summary CHS as peeling criterion and the Concordance Error Rate CER as optimization criteria. Notice that covariates $({\bf x}_{6},{\bf x}_{32},{\bf x}_{34},{\bf x}_{72},{\bf x}_{79})$ were those effectively used by the PRSP algorithm (see Algorithm \ref{algo}) out of $p=1000$ total covariates and $p=100$ informative ones (see simulation design \ref{design}).}
  \label{SuppFigure05}
\end{figure}

%%%%%%%%%%%%%%%%%%%%%%%%%%%%%%%%%
% Supporting\_Figure 06
%%%%%%%%%%%%%%%%%%%%%%%%%%%%%%%%%
\newpage
\begin{figure}[!hbt]
  \centering\includegraphics[width=3in]{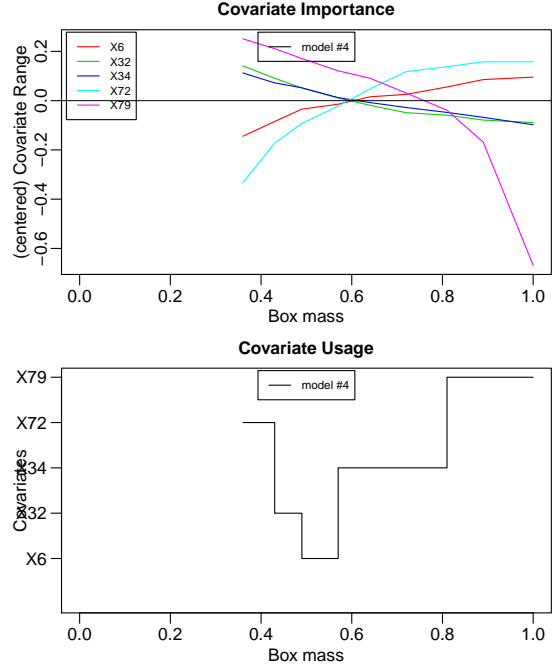}
  \caption{\sl \footnotesize Comparison of replicated combined cross-validated trace plots of covariate importance $\widebar{VI}(l)$ (top) and covariate usage $\widebar{VU}(l)$ (bottom) in simulated model \#4 for the ``Replicated Combined CV'' (RCCV) technique and the Cumulative Hazard Summary CHS as peeling criterion and the Concordance Error Rate CER as optimization criteria. Notice that covariates $({\bf x}_{6},{\bf x}_{32},{\bf x}_{34},{\bf x}_{72},{\bf x}_{79})$ were those effectively used by the PRSP algorithm (see Algorithm \ref{algo}) out of $p=1000$ total covariates and $p=100$ informative ones (see simulation design \ref{design}).}
  \label{SuppFigure06}
\end{figure}

%%%%%%%%%%%%%%%%%%%%%%%%%%%%%%%%%
% Supporting\_Table 02
%%%%%%%%%%%%%%%%%%%%%%%%%%%%%%%%%
\
\begin{table}[!hbt]
  \caption{\sl \footnotesize Comparison of cross-validated decision rules (upper Supporting Table) and box end points statistics of interest (lower Supporting Table) in simulated model \#4 for the ``Replicated Combined CV'' (RCCV) technique and the Cumulative Hazard Summary CHS as peeling criterion and the Concordance Error Rate CER as optimization criteria. For conciseness, only the initial and final decision rules ($\bar{L}^{rcv}$th step) are shown. Step \#0 corresponds to the situation where the starting box covers the entire test-set data $\mathcal{L}_{k}$ before peeling. Values are sample mean estimates with corresponding standard errors in parenthesis. Notice that covariates $({\bf x}_{6},{\bf x}_{32},{\bf x}_{34},{\bf x}_{72},{\bf x}_{79})$ were those effectively used by the PRSP algorithm (see Algorithm \ref{algo}) out of $p=1000$ total covariates and $p=100$ informative ones (see simulation design \ref{design}).}
 \footnotesize
 \begin{tabular}{lrccccc}
    \hline
    \hline
    \hspace{-0.2in}                    &Step $l$  &${\bf x}_{6}$                    &${\bf x}_{32}$                     &${\bf x}_{34}$                     &${\bf x}_{72}$                    &${\bf x}_{79}$\\
    \hline
                      \hspace{-0.2in}  &$0$       &${\bf x}_{6} \le 1.81 \ (0.00)$  &${\bf x}_{32} \ge -2.07 \ (0.00)$  &${\bf x}_{34} \ge -2.32 \ (0.00)$  &${\bf x}_{72} \ge 2.81 \ (0.00)$  &${\bf x}_{79} \ge -2.75 \ (0.00)$\\
    \Lower{model \#4} \hspace{-0.2in}  &$1$       &${\bf x}_{6} \le 1.80 \ (0.43)$  &${\bf x}_{32} \ge -2.06 \ (0.03)$  &${\bf x}_{34} \ge -2.29 \ (0.06)$  &${\bf x}_{72} \ge 2.81 \ (0.04)$  &${\bf x}_{79} \ge -2.25 \
                      (0.30)$\\
                      \hspace{-0.2in}  &$\vdots$  &$\vdots$                         &$\vdots$                           &$\vdots$                           &$\vdots$                          &$\vdots$\\
                      \hspace{-0.2in}  &$8$       &${\bf x}_{6} \le 1.57 \ (0.25)$  &${\bf x}_{32} \ge -1.84 \ (0.23)$  &${\bf x}_{34} \ge -2.11 \ (0.20)$  &${\bf x}_{72} \ge 2.32 \ (0.50)$  &${\bf x}_{79} \ge -1.83 \ (0.40)$\\
    \hline
 \end{tabular}\\
 \begin{tabular}{lrccccccc}
        \hline
        \hline
        \hspace{-0.2in}                    &Step $l$  &$n(l)$        &$\bar{\beta}^{rcv}(l)$  &$\widebar{T_{0}^{\prime}}^{rcv}(l)$  &$\widebar{P_{0}^{\prime}}^{rcv}(l)$  &$\bar{\lambda}^{rcv}(l)$  &$\bar{\chi}^{rcv}(l)$  &$\bar{\theta}^{rcv}(l)$\\
        \hline
                          \hspace{-0.2in}  &$0$       &100 \ (0.00)  &1.00 \ (0.00)           &1.99 \ (0.00)                        &0.46 \ (0.00)                        &0.00 \ (0.00)             &0.00 \ (0.00)
        &1.00 \ (0.00)        \\
        \Lower{model \#4} \hspace{-0.2in}  &$1$       & 89 \ (2.00)  &0.89 \ (0.02)           &1.98 \ (0.01)                        &0.43 \ (0.02)                        &0.50 \ (0.43)             &2.63 \ (2.66)    &0.47 \ (0.02)       \\
                          \hspace{-0.2in}  &$\vdots$  &$\vdots$      &$\vdots$                &$\vdots$                             &$\vdots$                             &$\vdots$                  &$\vdots$
        &$\vdots$               \\
                          \hspace{-0.2in}  &$8$       & 36 \ (6.00)  &0.36 \ (0.06)           &1.89 \ (0.16)                        &0.37 \ (0.07)                        &0.49 \ (0.37)             &3.99 \ (4.78)
        &0.44 \ (0.04)        \\
        \hline
 \end{tabular}
 \label{SuppTable02}
\end{table}

\end{document}